\pdfoutput=1
\documentclass[11pt,table]{article}

\usepackage[preprint]{acl}

\usepackage{times}
\usepackage{latexsym}
\usepackage{xcolor}

\usepackage[T1]{fontenc}
\usepackage[utf8]{inputenc}

\usepackage{microtype}

\usepackage{inconsolata}

\usepackage{graphicx}
\usepackage{tablefootnote}
\usepackage{microtype}
\usepackage{hyperref}
\usepackage{url}
\usepackage{booktabs}
\usepackage{amsmath}
\usepackage{graphicx}
\usepackage{lineno}
\usepackage{booktabs}       % professional-quality tables
\usepackage{pifont}
\usepackage{colortbl}
\usepackage{multirow}

\newcommand{\xmark}{\textcolor{red}{\ding{55}}}   % Red cross

\usepackage[disable]{todonotes}

\usepackage{subcaption}
\usepackage{makecell}
\usepackage{listings}
\usepackage{wrapfig}
\usepackage[normalem]{ulem}
\usepackage{hyperref}
\usepackage{cleveref}
\usepackage{amssymb}
\usepackage{pifont}

\usepackage{float}        % for [H] if you need it
\usepackage{caption}
\usepackage{makecell}
\usepackage{xurl} 

\usepackage{subcaption}
\usepackage{threeparttable}

\definecolor{bgprompt}{RGB}{250,250,250}
\usepackage{listings}
\usepackage{xcolor} % Required for background color

\definecolor{bgprompt}{gray}{0.95}

\lstdefinestyle{promptstyle}{
  backgroundcolor=\color{bgprompt},    
  basicstyle=\small\ttfamily,          
  frame=single,                        
  framerule=0.4pt,                     
  framesep=2mm,                        
  numbers=none,                        
  breaklines=true,                    
}

\lstnewenvironment{promptcode}
  {\lstset{style=promptstyle}}
  {}

\definecolor{HighlightGray}{gray}{0.92} % You can adjust 0.92 to be lighter (e.g., 0.95) or slightly darker (e.g., 0.90)
\definecolor{DimGrayText}{gray}{0.55}

\definecolor{darkblue}{rgb}{0, 0, 0.5}
\hypersetup{colorlinks=true, citecolor=darkblue, linkcolor=darkblue, urlcolor=darkblue}
\newcommand{\GitChameleon}{\textcolor{violet}{\textbf{GitChameleon 2.0}}}

\title{\GitChameleon{}: Evaluating AI Code Generation Against Python Library Version Incompatibilities}

\author{
  \textbf{\textcolor{violet}{Diganta Misra}\textsuperscript{1,2}}\thanks{Equal Contribution},
  \textbf{\textcolor{violet}{Nizar Islah}\textsuperscript{3,10}}\footnotemark[1],
  \textbf{\textcolor{violet}{Victor May}\textsuperscript{4}},
  \textbf{\textcolor{blue}{Brice Rauby}\textsuperscript{3, 5}},
\\
  \textbf{\textcolor{blue}{Zihan Wang}\textsuperscript{6}},
  \textbf{\textcolor{blue}{Justine Gehring}\textsuperscript{3,7,8}},
  \textbf{Antonio Orvieto\textsuperscript{1,2,9}},
  \textbf{\textcolor{blue}{Muawiz Chaudhary}\textsuperscript{3}},
\\
  \textbf{Eilif B. Muller\textsuperscript{3,10}},
  \textbf{Irina Rish\textsuperscript{3,10}},
  \textbf{Samira Ebrahimi Kahou\textsuperscript{3}},
  \textbf{Massimo Caccia\textsuperscript{11}}
\\
  \textcolor{violet}{Team Leads},
  \textcolor{blue}{Data and Core Contributors},
  Senior Advisors
\\
  \textsuperscript{1}ELLIS Institute Tübingen,
  \textsuperscript{2}MPI-IS Tübingen,
  \textsuperscript{3}Mila Quebec AI Institute,
  \textsuperscript{4}Google,
\\
  \textsuperscript{5}Polytechnique Montréal,
  \textsuperscript{6}McGill University, Montréal,
  \textsuperscript{7}Moderne,
  \textsuperscript{8}Gologic,
\\
  \textsuperscript{9}Tübingen AI Center,
  \textsuperscript{10}Université de Montréal,
  \textsuperscript{11}ServiceNow Research
\\
\small{
  \textbf{Correspondence:} \href{mailto:diganta.misra@tue.ellis.eu}{diganta.misra@tue.ellis.eu}, \href{mailto:nizar.islah@mila.quebec}{nizar.islah@mila.quebec}
}
}

\begin{document}
\maketitle
\begin{abstract}
The rapid evolution of software libraries poses a considerable hurdle for code generation, necessitating continuous adaptation to frequent version updates while preserving backward compatibility. While existing code evolution benchmarks provide valuable insights, they typically lack execution-based evaluation for generating code compliant with specific library versions. To address this, we introduce \GitChameleon{}, a novel, meticulously curated dataset comprising 328 Python code completion problems, each conditioned on specific library versions and accompanied by executable unit tests. \GitChameleon{} rigorously evaluates the capacity of contemporary large language models (LLMs), LLM-powered agents, code assistants, and RAG systems to perform version-conditioned code generation that demonstrates functional accuracy through execution. Our extensive evaluations indicate that state-of-the-art systems encounter significant challenges with this task; enterprise models achieving baseline success rates in the 48-51\% range, underscoring the intricacy of the problem. By offering an execution-based benchmark emphasizing the dynamic nature of code libraries, \GitChameleon{} enables a clearer understanding of this challenge and helps guide the development of more adaptable and dependable AI code generation methods. We make the dataset and evaluation code publicly available~\footnote{\url{https://github.com/mrcabbage972/GitChameleonBenchmark}}.
\end{abstract}
\section{Introduction}

% \begin{figure}[h!]
%     \centering
%     \includegraphics[width=0.9\linewidth]{figures/motivation.pdf}
%     \caption{Modern LLMs often struggle with generating version-accurate code, highlighting the need for benchmarks that specifically assess their ability to handle versioning.}
%     \label{fig:motivating_example}
% \end{figure}

\begin{figure}[ht]
    \centering
    \includegraphics[width=1\linewidth]{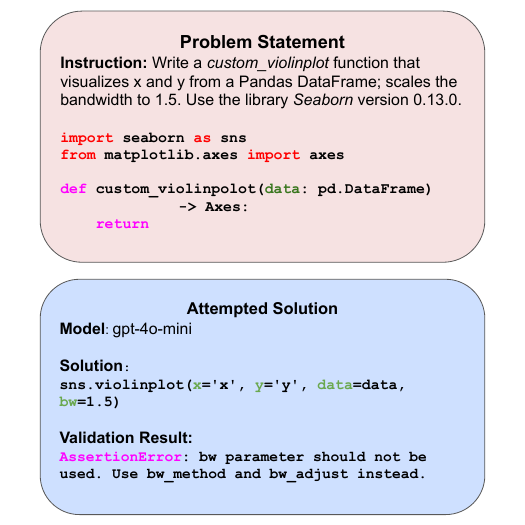}
    \caption{In this \GitChameleon{} problem, the \texttt{gpt-4o-mini} model produced an incorrect solution due for \texttt{seaborn.violinplot} by using the deprecated \texttt{bw} parameter, instead of the appropriate \texttt{bw\_method} and \texttt{bw\_adjust} required by the specified library version.}
    \label{fig:motivating_example}
\end{figure}

Large language models (LLMs) are increasingly integral to software development, being adopted for tasks like code generation and review~\citep{forbes_llms_swdev, lambiase2025exploringindividualfactorsadoption}.

Despite LLM advancements like larger context windows \citep{su2023roformerenhancedtransformerrotary}, faster inference \citep{dao2022flashattention}, and high performance on general coding benchmarks \citep{hendrycksapps2021, chen2021codex}, a critical capability remains under-evaluated: generating code that is compliant with a specific library version. This task of version-switching, which is essential for robust development in environments with fixed or legacy dependencies, is not well-verified in contemporary LLMs. 

%While Despite the availability benchmarks for code evolution are often non-executable \citep{pymigbench}, do not evaluate performance against real-world breaking changes in popular libraries \citep{liu2024codeupdatearenabenchmarkingknowledgeediting}, or fail to consider the practical scenario of updating projects with specific library version requirements \citep{wang2024llmsusedeprecatedapis}.

Existing benchmarks, while valuable, often focus on migrating codebases to newer versions (i.e., code evolution) or use non-executable evaluation methods. They do not fully address the challenge of generating new, functionally correct code for a static version constraint. For instance, PyMigBench~\citep{pymigbench} provides comprehensive datasets of real-world, inter-library migrations, rather than focusing on executable, intra-library tasks conditioned on specific versions. CodeUpdateArena~\citep{liu2024codeupdatearenabenchmarkingknowledgeediting} valuably assesses LLM knowledge editing using synthetically generated API updates for functions in popular libraries, a different approach from using documented historical breaking changes. Other relevant studies, such as~\citet{wang2024llmsusedeprecatedapis}, investigate the propensity of LLMs to generate code with deprecated APIs, which does not entirely cover the broader capability of generating software that adheres to precise, user-specified library versions involving various types of API changes.

\begin{figure}
  \begin{center}
    \includegraphics[width=1\linewidth]{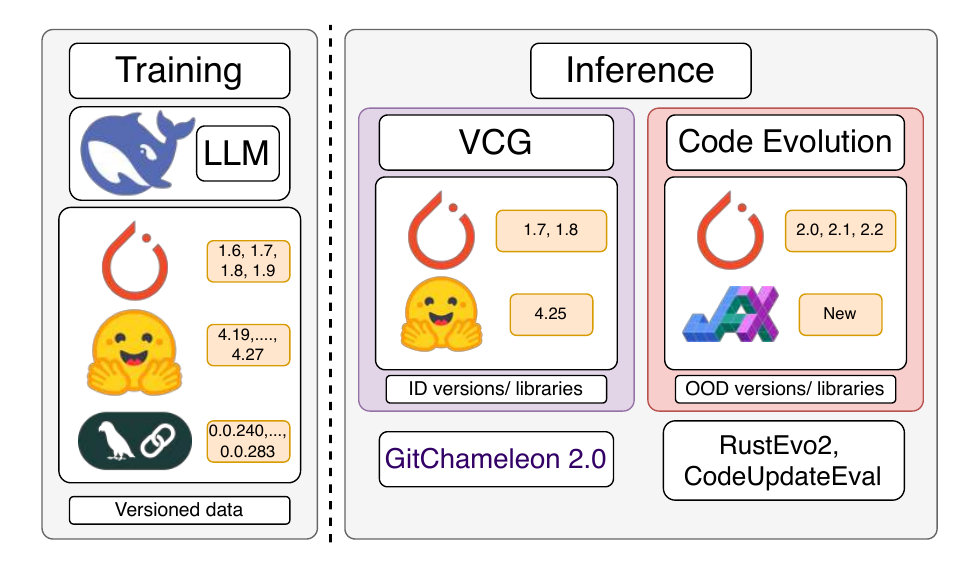}
  \end{center}
  \caption{\textit{An illustration of two evaluation paradigms for code generation models}. \textbf{Code Evolution} (right) assesses model capabilities on \emph{out-of-distribution (OOD)} data, using library versions or new libraries not encountered during training. In contrast, \textbf{Version-Conditioned Generation (VCG)} (left) focuses on the practical ability to generate code for specific, \emph{in-distribution (ID)} library versions that the model has seen before.}
  \label{fig:motiv_2}
\end{figure}

\paragraph{Code Evolution vs. Version Conditioned Generation (VCG).} 
Existing code evaluation benchmarks often focus on assessing the code evolution or migration capabilities of LLMs, where changes occur only in the forward direction and typically involve unseen library versions or entirely new libraries. This framing inherently makes the task out-of-distribution (OOD), as illustrated in Figure~\ref{fig:motiv_2}. In contrast, version-conditioned generation (VCG)—the ability of LLMs to produce code aligned with specific, previously seen library versions—is critical for practical deployment. It enables models to function reliably in real-world production environments or constrained settings where the libraries in use may not be the latest stable versions. To better evaluate this capability, a benchmark must pose problems that are strictly \emph{in-distribution (ID)} with respect to the relevant library version(s) required to solve them.

To bridge this gap, our work introduces \GitChameleon{}, an executable benchmark designed to assess the capability of LLMs and AI agents in generating version-aware Python code. \GitChameleon{} features problems centered on documented breaking changes from popular libraries, requiring models to produce solutions for explicitly specified versions (an illustrative example is shown in Figure~\ref{fig:motivating_example}). The development of such a benchmark faces challenges in meticulously curating version-specific breaking changes from library changelogs and crafting corresponding testable scenarios. Our comprehensive evaluation of diverse LLM-based tools on \GitChameleon{} reveals critical limitations in existing systems' ability to handle library versioning.

In summary, our contributions are highlighted as follows: 
\begin{itemize}
    \item We introduce a novel code completion benchmark \GitChameleon{} consisting of \textcolor{red}{328} Python-based version-conditioned problems, including visible tests for self-debugging and documentation references for Retrieval-Augmented Generation (RAG).

    \item We present a comprehensive empirical study on \GitChameleon{}, evaluating the capabilities of a diverse range of contemporary AI code generation systems, including AI agents, IDE-integrated and CLI-based coding assistants, and RAG-based LLM pipelines.
    \item We reveal critical limitations in the ability of current AI systems to adhere to specific versioning constraints and highlight factors impacting their performance, thereby providing insights to steer the development of more adaptable and dependable AI code generation methods.
\end{itemize}

\section{\GitChameleon{} Benchmark}

We introduce \GitChameleon{}, a manually authored benchmark that comprises \textcolor{red}{328} Python-based version-conditioned problems focused on popular code libraries. To evaluate performance on \GitChameleon{}, each problem is accompanied by a suite of assertion-based unit tests, enabling a thorough execution-based assessment of potential solutions. %This structured approach enables a thorough understanding and categorization of AI system failures in common scenarios involving version-specific code generation problems. The main challenge in constructing  the dataset was identifying breaking changes in the change logs of mature Python libraries.
In the following sections, we detail the dataset  structure, dataset statistics, evaluation metrics, and sample verification process.

\subsection{Dataset Structure}
Each dataset sample includes a problem related to a breaking change in a Python library. 

To validate a candidate solution, we provide a suite of tests, consisting of a comprehensive suite of \textbf{Hidden Tests} to be used for model performance evaluation and ranking and a concise \textbf{Visible Test} to provide execution feedback for Self-Debugging~\citep{chen2023teachinglargelanguagemodels} experiments.

\begin{figure}
\centering
\includegraphics[width=\columnwidth]{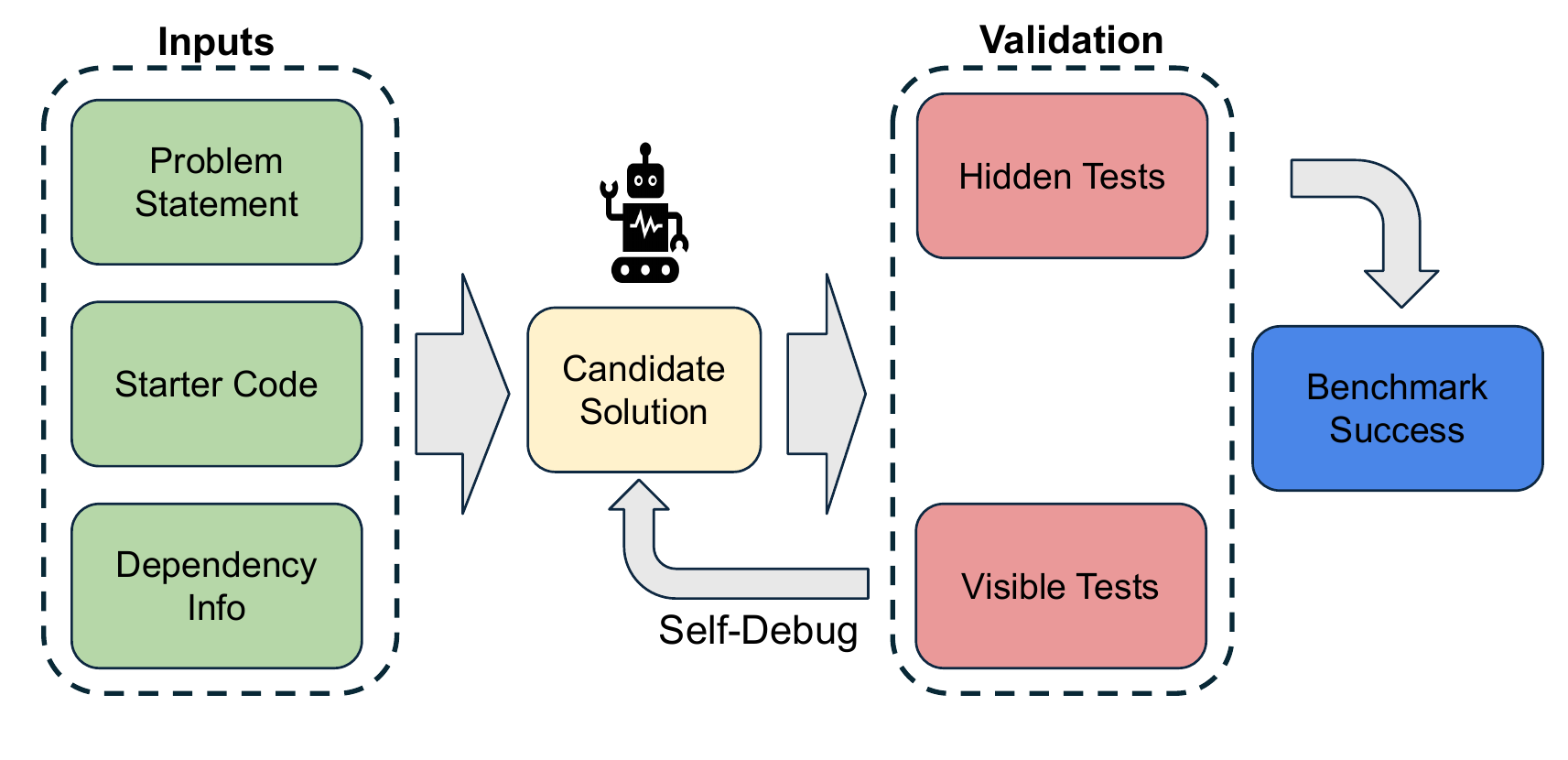}
\caption{An illustration of the workflow for a single example within \GitChameleon{}. The inputs, comprising the Problem Statement, Starter Code, and Dependency Info, are processed by an LLM or an AI agent to generate a Candidate Solution. This candidate solution then undergoes validation using the Hidden Tests to determine success on the benchmark. Results from the Visible Tests can be fed back into the solution method for self-debugging.}
\label{fig:schematic}
\end{figure}

The detailed structure of dataset samples is presented in Table~\ref{tab:column-defs}. For a schematic of the workflow for evaluating a method against a sample from \GitChameleon{}, see Figure~\ref{fig:schematic}. 

% Additionally, each problem has associated metadata to assist with analysis, as described in Table~\ref{tab:change-metadata}. Each problem is classified with a type of change among the categories defined in Table~\ref{tab:api_evolution_categories}. 

\subsection{Evaluation Metrics}
The benchmark metric is the success rate on hidden tests, which directly penalizes version mismatches that cause runtime errors during our execution-based validation. As a secondary metric, we use the API Hit Rate~\cite{wang2024systematicevaluationlargecode}: the percentage of generated solutions that correctly call all APIs specified in the ground-truth solution. Note that this hit rate can be lower than the success rate, as functionally correct alternative solutions may use different APIs.

\subsection{Statistics}

\begin{figure}
    \centering
    \includegraphics[width=1\linewidth]{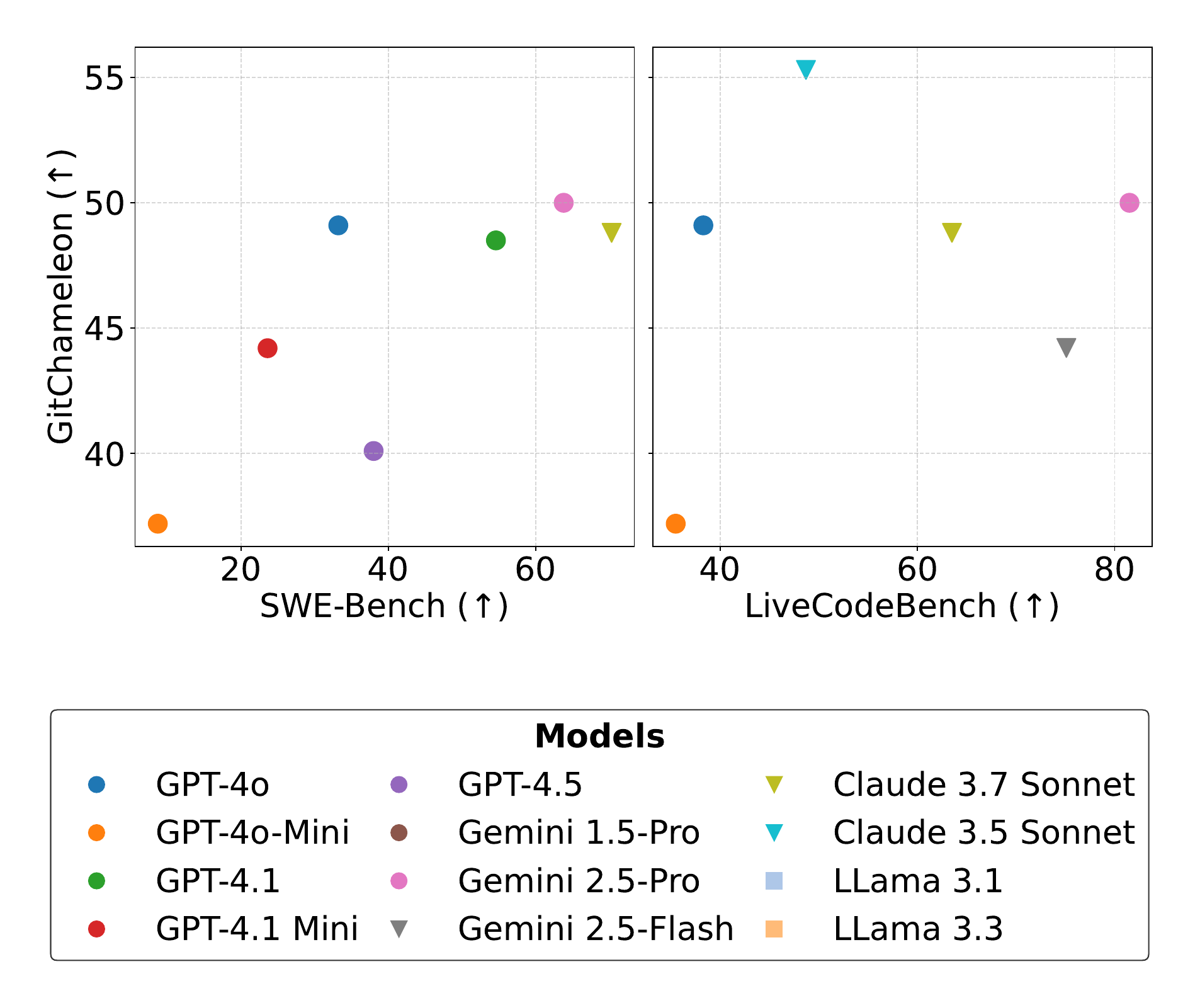} \caption{Can you predict \GitChameleon{} performance from other code generation benchmarks? Here we present the Spearman ($\rho$) and Pearson ($r$) correlations between \GitChameleon{}, SWE‐Bench~\citep{swebench}, and LiveCodeBench~\citep{livecodebench}. GitChameleon exhibits a moderate correlation with SWE‐Bench, with $\rho$ of 0.550 and $r$ of 0.675; and a weak correlation with LiveCodeBench, with $\rho$ of 0.214 and $r$ of 0.130.}
\end{figure}
\GitChameleon{} consists of 328 Python-based version conditioned problems based on 26 libraries spanning scientific computing, data science and web  development. The samples were collected from version releases over a period from the year 2014 to 2023 and exclude legacy and yanked version releases. %Among the 328 problems in the dataset, 9 have hidden test coverage between 70\% and 90\%, while the rest have 90\%+.

% Using the \texttt{cl100k\_base} tokenizer, we analyzed the token counts of the \GitChameleon{} samples. The problem statements average 32.1 tokens, and the starter code averages 55.3 tokens, leading to a combined average of 87.4 tokens per sample. Including the prompt template utilized for evaluating instruction-tuned LLMs, the total token count across all samples is 48,571 tokens.

\begin{figure}[ht]
  \centering
  % \begin{subfigure}{0.9\linewidth}
  %   \centering
  %   \includegraphics[width=\textwidth]{figures/version_year.pdf}
  %   \caption{Number of samples by version release year.}
  %   \label{fig:version_year}
  % \end{subfigure}
  % \vspace{2mm}
  % \begin{subfigure}{0.9\linewidth}
  %   \centering
  %   \includegraphics[width=\textwidth]{figures/type_change.pdf}
  %   \caption{Number of samples by change category.}
  %   \label{fig:change_type}
  % \end{subfigure}
  \includegraphics[width=\linewidth]{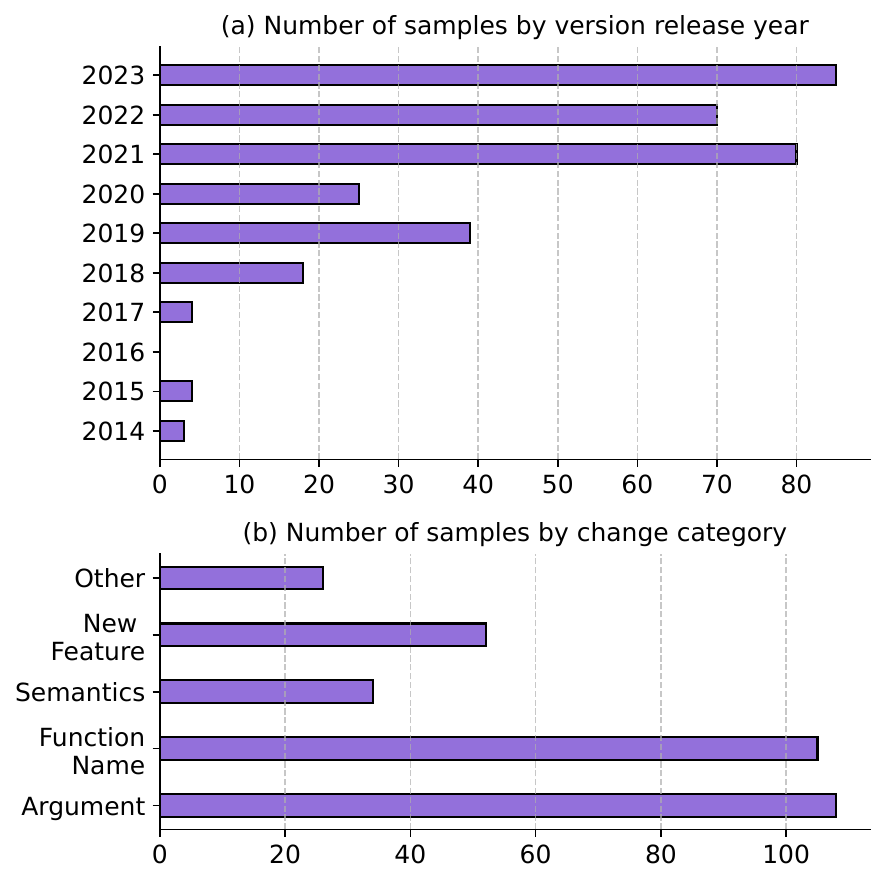}
  \caption{(a) Most versions in \GitChameleon{} were released between 2021–2023, with a few in earlier years.
    (b) The most common type of change between versions was an argument or attribute change, while semantic or functional changes were least common.}
    \label{fig:version_year}
\end{figure}

As demonstrated in Fig.~\ref{fig:version_year}(a), most of the samples in \GitChameleon{} are from versions of libraries released in the years 2021-2023. We intentionally use versions that fall within the training window of most evaluated models. The challenge is therefore not one of data contamination, but of \textbf{control and disambiguation}: when a model has been exposed to multiple library versions, can it correctly generate code for the specific version required by the prompt?

% \begin{figure*}[!htb]
%   \centering
%   \includegraphics[width=\textwidth]{figures/example.png}
%   \caption{%
%     Example of a problem statement derived from a changelog entry from Scipy from 1.11.1. We used this change to craft a problem statement that tests an LLM's ability to recognize and adapt to version-specific updates.
%   }
%   \label{fig:example_fig}
% \end{figure*}

The dataset was constructed through careful manual effort, with over 350 hours invested in identifying historical breaking changes, crafting problem statements, and validating unit tests. Further details about the benchmark and its construction process are presented in Appendix~\ref{app:bench_deets}.

\section{Empirical Study}
\label{sec:exp-study}
We evaluate \GitChameleon{} in a comprehensive selection of settings, including Greedy Decoding, Chain-of-Thought~\citep{wei2023chainofthoughtpromptingelicitsreasoning}, Self-Debugging~\citep{chen2023teachinglargelanguagemodels}, RAG~\citep{lewis2020retrieval}, Multi-Step Agents~\citep{yao2023react} and enterprise Coding Assistant software products, to assess their ability to generate version-specific executable code. %This study highlights how well current methods adapt to dynamic library versions and produce functionally correct code that passes the provided unit tests.

This section first presents the experimental setup, then reports the experiment results in each setting, and finally shows a breakdown of the observed results along a few key dimensions.

\subsection{Experimental Setup}
\label{sec:exp-setup}
In this section, we present the experimental setup used for each of our settings. To ensure version compliance, we use a dual control mechanism: the target version is explicitly included in the model’s prompt, and the validation environment is configured with that exact library version. All prompts are shown in Appendix~\ref{sec:prompt_template}. For prompt optimization, we used the Anthropic Prompt Improver~\footnote{\url{https://docs.anthropic.com/en/docs/build-with-claude/prompt-engineering/prompt-improver}}. Further automated prompt optimization efforts did not make a significant change, as described in Table~\ref{tab:round-prompt-optim}.
\subsubsection{Greedy Decoding}
We configured the generation parameters with a sampling temperature of 0 and a \texttt{top\_p} value of 0.95. We had specified a structured output schema that specifies the fields \texttt{Answer} and \texttt{Explanation}, where both are of type \texttt{string}. %The prompt template is given in Figure~\ref{fig:greedy-prompts} (a).

\subsubsection{Zero-Shot Chain-Of-Thought (CoT)}
We had used the same generation parameters as for Greedy Decoding and an output schema that specifies the fields \texttt{Answer} and \texttt{Steps}, where the former is a of type \texttt{string} and the latter is a list of \texttt{string}. %Figure~\ref{fig:greedy-prompts} (b).

\subsubsection{Self-Debugging}
On examples that failed with Greedy Decoding, we employed the method described in ~\cite{chen2023teachinglargelanguagemodels} to feed the visible test error trace along with the model's explanation of its output back to the model. %The prompt template is given in Figure~\ref{fig:self-debugging-prompts}.

\subsubsection{Retrieval-Augmented Generation} 
We designed a RAG~\cite{lewis2020retrieval} pipeline where we first constructed a vectorized database (VectorDB) by embedding each sample's relevant API documentation with the OpenAI \texttt{text-embedding-3 large} model \footnote{\url{https://openai.com/index/new-embedding-models-and-api-updates/}}. The corpus used for constructing the VectorDB included 536 documents, with 140 samples having 1 associated document, 168 having 2 associated documents and 20 having 3 documents. 

Subsequently, we used DocPrompting ~\cite{zhou_docprompting_2023} to query the VectorDB to generate solutions. %The detailed query template is presented in Figure~\ref{app:prompt_rag}.

\subsubsection{Multi‐Step Agent} We conducted experiments with a tool-calling agent, as implemented by the smolagents~\cite{smolagents}~\footnote{\url{https://huggingface.co/learn/agents-course/en/unit2/smolagents/tool_calling_agents}} framework. This agent implementation mostly follows the ReAct~\cite{yao2023react} method, but, it alternates between acting and planning~\cite{react_vs_plan_execute} steps. 

Following the Agentic RAG approach~\citep{singh2025agenticretrievalaugmentedgenerationsurvey}, we had equipped the agent with a grounding tool in order to assess its capability to independently fetch relevant info for solving the benchmark problems. To this end, we had  experimented with the following grounding tools:  DuckDuckGo Search~\cite{duckduckgo}, Perplexity~\cite{perplexity_ai}, and Gemini with Grounding~\cite{gemini_search_grounding_google}. %An agent equipped with one of these tools can be viewed as implementation of Agentic RAG ~\cite{singh2025agenticretrievalaugmentedgenerationsurvey}. 

Additionally, we examined agentic multi-step self-debugging~\cite{jin2024rgdmultillmbasedagent} by including or omitting a code execution sandbox tool~\cite{rabin2025sandboxevalsecuringtestenvironment}, which provides the needed dependencies for each example. The sandbox takes a Python program as input and outputs the standard output from the program. 

%We had experimented with agents that leverage each of three major model families: Claude, Gemini, and GPT.

%The prompt template is given in Figure~\ref{fig:agent-prompt}.

\subsubsection{AI Coding Assistants}
In addition to evaluating a generic agentic framework endowed with basic tools, we also analyze the performance of specialized AI coding assistant software. 

For this setting, we examine both Command-Line Interface (CLI), such as \texttt{Claude Code}\footnote{\url{https://docs.anthropic.com/en/docs/claude-code/overview}} coding assistants and Integrated Development Environment (IDE) coding assistants, such as \texttt{Cline}\footnote{\url{https://cline.bot/}}. 

Specifically, in this evaluation we aimed to evaluate the code completion functionality of code assistants in an IDE or terminal environment wherein the goal was to complete the starter code of each \GitChameleon{} problem with the generated solution. 

The input to the assistants is given as a Python file which consists of the required library, version and extra dependencies as in-line comments and subsequently the starter code. \textsc{Note:} All assistants had internet and terminal commands execution access.

We had furthermore ablated this setting versus giving the full problem statement as input.

\subsection{Experiment Results}
This section presents the benchmark results in each setting, as described in the \textbf{Experimental Setup} section~(\ref{sec:exp-setup}). Table~\ref{tab:base-model-performance} contains the results for Greedy Decoding, Self-Debug and Zero-Shot CoT.
\subsubsection{Greedy Decoding}
 We observe that the largest Enterprise-grade models, including \texttt{Claude 3.7 Sonnet}, \texttt{Gemini 2.5 Pro}, \texttt{GPT-4.1}, \texttt{GPT-4o}, and \texttt{o1}, exhibit comparable hidden success rates, generally falling within the 48--51\% range. Among these \texttt{o1} (51.2\% hidden) achieves the highest hidden success rate. 

The open-weight Llama models are notably behind, even the recently released \texttt{Llama 4 Maverick FP8} (40.8\% hidden success rate). 

Model size clearly impacts performance: for instance, \texttt{Gemini 2.5 Flash} trails its \texttt{Pro} counterpart by nearly 12\% on hidden tests (38.1\% vs. 50.0\%). Similarly, the \texttt{mini} and \texttt{nano} series within the GPT family (e.g., \texttt{GPT-4.1-mini}, \texttt{GPT-4.1-nano}, \texttt{GPT-4o-mini}) consistently show lower performance than their larger full-size siblings, with differences on hidden tests ranging from approximately 4 to 15 points.

\subsubsection{Zero-Shot Chain-Of-Thought}
This approach does not uniformly improve LLM performance across all models. While some models demonstrate significant gains in hidden success rates, a substantial number of enterprise-grade models and their smaller variants experience performance degradation.

For instance, notable improvements in hidden success rates are observed in models such as \texttt{Llama 3.1 Instruct Turbo} (from 30.2\% to 36.6\%, a +6.4 point increase) and \texttt{o3-mini} (from 45.1\% to 50.9\%, a +5.8 point increase).

Conversely, several models exhibit a decrease in performance with CoT. Prominent examples include \texttt{Gemini 2.0 Flash} (from 44.2\% to 36.0\%) and even the top-performing \texttt{o1} (from 51.2\% to 41.2\%).

\begin{table*}[!hbt]
  \centering
  \setlength{\tabcolsep}{8pt} % Adjust column spacing
  \renewcommand{\arraystretch}{1.2} % Optional: increase row spacing a bit
  \resizebox{0.85\linewidth}{!}{%
    \begin{tabular}{@{}
                     l
                     >{\columncolor{HighlightGray}}c c c                    
                     >{\columncolor{HighlightGray}}c c c
                     >{\columncolor{HighlightGray}}c c
                     @{}}
      \toprule
      \multirow{3}{*}{\textbf{Model}}
      & \multicolumn{3}{c}{\textbf{Greedy Decoding}}
      & \multicolumn{3}{c}{\textbf{Greedy with Self-Debug}}
      & \multicolumn{2}{c}{\textbf{Zero-shot CoT}} \\
      \cmidrule(lr){2-4} \cmidrule(lr){5-7} \cmidrule(lr){7-9}
      & \multicolumn{2}{c}{\shortstack[l]{\textbf{Success}\\\textbf{Rate (\%)}}}
      & \multirow{2}{*}{\shortstack[l]{\textbf{API} \\ \textbf{Hit} \\ \textbf{Rate (\%)}}}
      & \multicolumn{2}{c}{\shortstack[l]{\textbf{Success}\\\textbf{Rate (\%)}}}
      & \multirow{2}{*}{\shortstack[l]{\textbf{API} \\ \textbf{Hit} \\ \textbf{Rate (\%)}}}
      & \multicolumn{1}{c}{\shortstack[l]{\textbf{Success}\\\textbf{Rate (\%)}}}
      & \multirow{2}{*}{\shortstack[l]{\textbf{API} \\ \textbf{Hit} \\ \textbf{Rate (\%)}}} \\
      \cmidrule(lr){2-3} \cmidrule(lr){5-6} \cmidrule(lr){8-8}
      & \cellcolor{HighlightGray}\shortstack[c]{\textbf{Hidden}}
      & \textcolor{DimGrayText}{\shortstack[c]{\textbf{Visible}}}
      &
      & \cellcolor{HighlightGray}\shortstack[c]{\textbf{Hidden}}
      & \textcolor{DimGrayText}{\shortstack[c]{\textbf{Visible}}}
      &
      & \cellcolor{HighlightGray}\shortstack[c]{\textbf{Hidden}}     
      
      & \\
      \midrule
      \addlinespace[0.5em]
      \multicolumn{9}{@{}l}{\textbf{Open-Weights Models}} \\
Llama 3.1 Instruct Turbo        & 30.2\tiny$\pm$2.5  & 38.1\tiny$\pm$2.7  & 39.7\tiny$\pm$2.7  & 52.1\tiny$\pm$2.8  & 69.2\tiny$\pm$2.5  & 41.5\tiny$\pm$2.7  & 36.6\tiny$\pm$2.7  & 35.3\tiny$\pm$2.6  \\
Llama 3.3 Instruct Turbo 70B    & 36.3\tiny$\pm$2.7  & 43.3\tiny$\pm$2.7  & 36.4\tiny$\pm$2.7  & 53.0\tiny$\pm$2.8  & 70.1\tiny$\pm$2.5  & 37.4\tiny$\pm$2.7  & 37.5\tiny$\pm$2.7  & 37.2\tiny$\pm$2.7  \\
Llama 4 Maverick 400B           & 40.8\tiny$\pm$2.7  & 46.6\tiny$\pm$2.8  & \textbf{49.5}\tiny$\pm$2.8  & 58.5\tiny$\pm$2.7  & 72.3\tiny$\pm$2.5  & \textbf{46.8}\tiny$\pm$2.8  & \textbf{46.6}\tiny$\pm$2.8  & 41.3\tiny$\pm$2.7  \\
Qwen 2.5-VL Instruct 72B        & \textbf{48.2}\tiny$\pm$2.8  & \textbf{55.5}\tiny$\pm$2.7  & 43.8\tiny$\pm$2.7  & \textbf{64.6}\tiny$\pm$2.6  & \textbf{77.4}\tiny$\pm$2.3  & 45.3\tiny$\pm$2.7  & 45.1\tiny$\pm$2.7  & \textbf{43.0}\tiny$\pm$2.7  \\
\midrule
\addlinespace[0.5em]
\multicolumn{9}{@{}l}{\textbf{Enterprise Models}} \\
\midrule    
Claude 3.7 Sonnet               & 48.8\tiny$\pm$2.8  & 55.8\tiny$\pm$2.7  & 46.0\tiny$\pm$2.8  & 65.9\tiny$\pm$2.6  & 75.9\tiny$\pm$2.4  & 47.6\tiny$\pm$2.8  & 45.1\tiny$\pm$2.7  & 43.4\tiny$\pm$2.7  \\
Gemini 1.5 Pro                  & 45.1\tiny$\pm$2.7  & 51.5\tiny$\pm$2.8  & 46.8\tiny$\pm$2.7  & 62.5\tiny$\pm$2.8  & 72.6\tiny$\pm$2.4  & 48.6\tiny$\pm$2.7  & 43.3\tiny$\pm$2.7  & 44.6\tiny$\pm$2.8  \\
Gemini 2.0 Flash                & 44.2\tiny$\pm$2.7  & 50.6\tiny$\pm$2.8  & 43.8\tiny$\pm$2.7  & \textbf{70.4}\tiny$\pm$2.7  & 79.0\tiny$\pm$2.4  & 49.4\tiny$\pm$2.7  & 36.0\tiny$\pm$2.6  & 41.8\tiny$\pm$2.7  \\
Gemini 2.5 Pro                  & \textbf{50.0}\tiny$\pm$2.8  & \textbf{61.0}\tiny$\pm$2.8  & 47.7\tiny$\pm$2.7  & 61.3\tiny$\pm$2.8  & 73.8\tiny$\pm$2.2  & 49.2\tiny$\pm$2.7  & 49.4\tiny$\pm$2.8  & 49.1\tiny$\pm$2.8  \\
Gemini 2.5 Flash                & 38.1\tiny$\pm$2.6  & 41.8\tiny$\pm$2.7  & 45.4\tiny$\pm$2.7  & 65.9\tiny$\pm$2.8  & 73.2\tiny$\pm$2.4  & 45.8\tiny$\pm$2.7  & 30.8\tiny$\pm$2.5  & \textbf{49.8}\tiny$\pm$2.8  \\
GPT-4.1                         & 48.5\tiny$\pm$2.8  & 49.1\tiny$\pm$2.8  & 46.8\tiny$\pm$2.7  & 63.4\tiny$\pm$2.8  & 76.8\tiny$\pm$2.1  & 48.3\tiny$\pm$2.7  & 47.9\tiny$\pm$2.8  & 44.5\tiny$\pm$2.7  \\
GPT-4.1-mini                    & 44.2\tiny$\pm$2.7  & 50.0\tiny$\pm$2.8  & 44.5\tiny$\pm$2.7  & 68.0\tiny$\pm$2.8  & \textbf{79.3}\tiny$\pm$2.3  & 46.3\tiny$\pm$2.7  & 24.1\tiny$\pm$1.8  & 41.3\tiny$\pm$2.7  \\
GPT-4.1-nano                    & 33.8\tiny$\pm$2.6  & 35.1\tiny$\pm$2.6  & 43.1\tiny$\pm$2.7  & 67.7\tiny$\pm$2.7  & 74.4\tiny$\pm$2.6  & 45.8\tiny$\pm$2.7  & 11.9\tiny$\pm$1.8  & 32.1\tiny$\pm$2.5  \\
GPT-4o                          & 49.1\tiny$\pm$2.8  & 54.0\tiny$\pm$2.8  & 46.5\tiny$\pm$2.7  & 64.9\tiny$\pm$2.8  & 72.3\tiny$\pm$2.5  & 48.0\tiny$\pm$2.7  & \textbf{50.3}\tiny$\pm$2.8  & 42.5\tiny$\pm$2.7  \\
GPT-4o-mini                     & 37.2\tiny$\pm$2.6  & 46.3\tiny$\pm$2.7  & 38.4\tiny$\pm$2.6  & 60.4\tiny$\pm$2.7  & 71.6\tiny$\pm$2.6  & 40.6\tiny$\pm$2.7  & 36.0\tiny$\pm$2.6  & 37.3\tiny$\pm$2.6  \\
GPT-4.5                         & 40.8\tiny$\pm$2.7  & 46.0\tiny$\pm$2.7  & \textbf{52.8}\tiny$\pm$2.8  & 66.2\tiny$\pm$2.8  & 74.4\tiny$\pm$2.4  & \textbf{54.4}\tiny$\pm$2.7  & 39.9\tiny$\pm$2.6  & 48.8\tiny$\pm$2.8  \\
Grok 3                           & 48.2\tiny$\pm$2.8  & 53.7\tiny$\pm$2.8  & 44.8\tiny$\pm$2.7  & 67.1\tiny$\pm$2.8  & 77.1\tiny$\pm$2.3  & 46.3\tiny$\pm$2.8  & 49.4\tiny$\pm$2.8  & 44.2\tiny$\pm$2.7  \\
Mistral Medium 3                & 43.6\tiny$\pm$2.7  & 49.1\tiny$\pm$2.8  & 44.2\tiny$\pm$2.7  & 61.3\tiny$\pm$2.8  & 71.3\tiny$\pm$2.5  & 45.4\tiny$\pm$2.7  & 44.2\tiny$\pm$2.7  & 44.1\tiny$\pm$2.7  \\

      \bottomrule
    \end{tabular}%
  }
  \caption{Success rate on visible and hidden tests and API hit rate under the Greedy, Self-Debug, and Zero-shot CoT settings, grouped by OSS vs.\ Enterprise models. Model ranking on the benchmark is determined by \textbf{Hidden Success Rate}. Visible Success Rate figures are for context on Self-Debugging. The best result in each column is in bold. For full model details and citations, please refer to Appendix~\ref{app:artifacts}.}
  \label{tab:base-model-performance}
\end{table*}

\subsubsection{LLM Self-Debugging}
\iffalse
\textbf{Hidden Success Rate}: Across models, Self-Debugging improves the success rates by approximately 10 points (e.g., Llama 3.1: 30→48 \%) up to nearly 20 pp (GPT-4.1-mini: 44→60 \%). This demonstrates the ability of modern LLM's to  inspect failures and re-generate correct code in the context of code evolution.
\fi

\textbf{Hidden Success Rate}: Across models, Self-Debugging significantly improves the hidden success rates. Observed gains range from approximately 10\% to 20\%. For instance, \texttt{Llama 3.1}'s hidden success rate increases from 30\% to 52.1\%, and \texttt{GPT-4.1-mini} shows an improvement from 44\% to 68\%. This demonstrates the strong capability of modern LLMs to diagnose failures and generate corrected code.

\textbf{Visible Success Rate}: As expected, the improvement is even more pronounced on visible tests, ranging from 13 to 37 points. For instance, \texttt{GPT-4.1}'s success rate improves from 49\% to 69\%, \texttt{Claude 3.7 Sonnet}'s success rate improves from 56\% to 83\% and \texttt{Gemini 2.0 Flash} improves from  50\% to 75\%.

\begin{figure}[ht]
    \includegraphics[width=\linewidth]{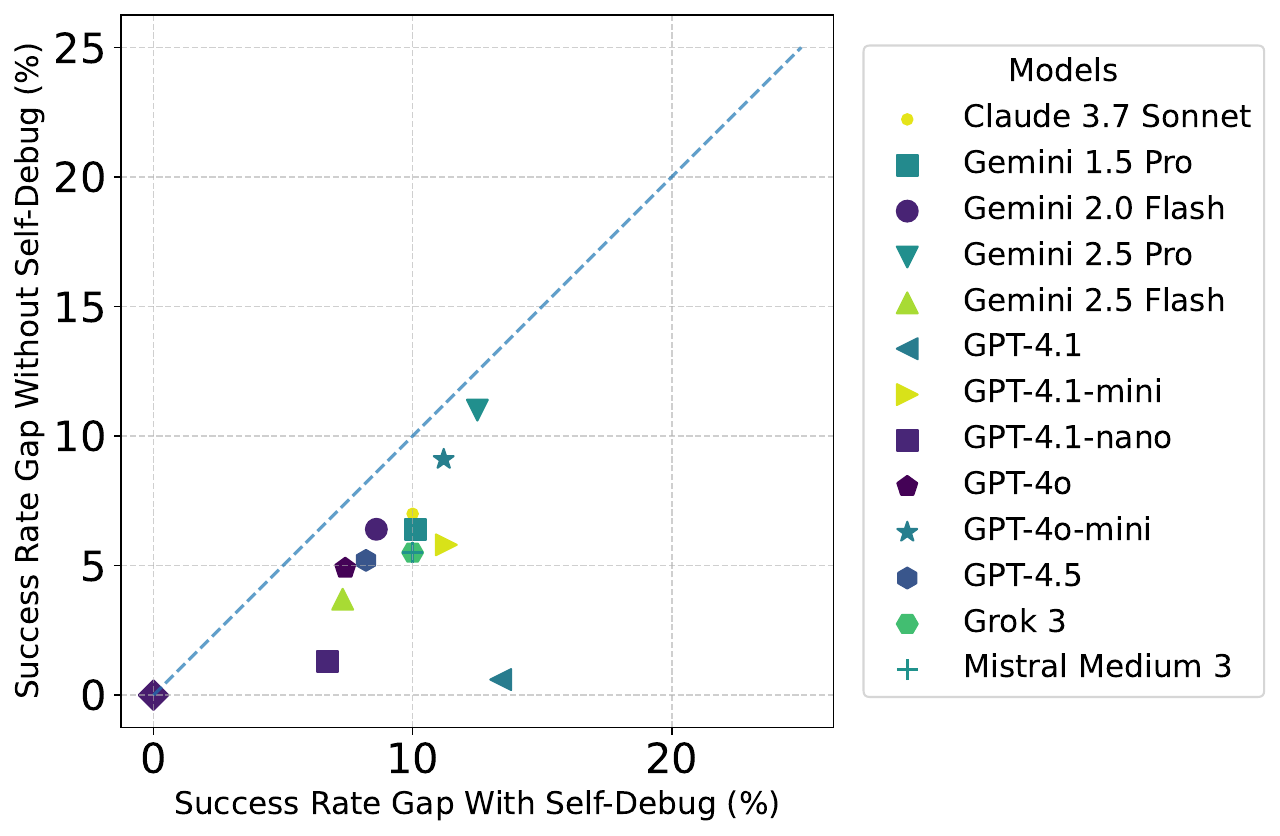}   
    \caption{\textbf{Analysis of the Visible-Hidden Gap Before and After Self-Debugging}. We analyze how self-debugging affects the gap between the success rate on visible and hidden tests. We can see that for all models, the gap increases after self-debugging. This shows that self-debugging on visible tests has a limited ability to improve on the hidden tests.}
    \label{fig:visible_hidden_gap}
\end{figure}
\textbf{Visible-Hidden Gap Analysis}: In Figure~\ref{fig:visible_hidden_gap}, we present the effect of self-debugging on the size of the gap between the success rate on visible tests and the success rate on hidden tests.

\subsubsection{Multi‐Step Agent}

\begin{table}[!htbp]
  \centering
  \resizebox{\columnwidth}{!}{%
  \begin{tabular}{ l l cc cc } % Column setup: Model | Grounding | SR (No SB) | SR (SB) | API (No SB) | API (SB)
    \toprule
    % Header Row 1: Model & Grounding span two rows, Metrics span one row
    \multirow{2}{*}{\textbf{Model}} & \multirow{2}{*}{\textbf{{\shortstack[l]{Grounding\\Method}}}} & \multicolumn{2}{c}{\shortstack[l]{\textbf{Success}\\\textbf{Rate (\%)}}} & \multicolumn{2}{c}{\shortstack[l]{\textbf{API Hit}\\\textbf{Rate (\%)}}} \\
    \cmidrule(lr){3-4} \cmidrule(lr){5-6} % Rules under "Success Rate (%)" and "API Hit Rate (%)"
    % Header Row 2: Specific sandbox states under each metric
                                      &                                       & \textbf{No Sandbox} & \textbf{Sandbox} & \textbf{No Sandbox} & \textbf{Sandbox} \\
    \midrule
    \multirow{3}{*}{\shortstack[l]{Claude\\Sonnet\\3.5}}
      & DuckDuckGo      & 41.7\tiny$\pm$2.7  & \textbf{55.3}\tiny$\pm$2.7  & 42.2\tiny$\pm$2.7  & 48.9\tiny$\pm$2.8  \\
      & Perplexity      & 44.1\tiny$\pm$2.7  & 51.4\tiny$\pm$2.8           & 41.8\tiny$\pm$2.7  & 46.0\tiny$\pm$2.8  \\
      & Grounded Gemini & 40.0\tiny$\pm$2.7  & 53.7\tiny$\pm$2.8           & 41.0\tiny$\pm$2.7  & 45.2\tiny$\pm$2.7  \\
\midrule
\multirow{3}{*}{\shortstack[c]{Gemini\\1.5 Pro}}
      & DuckDuckGo      & 46.0\tiny$\pm$2.8  & 49.8\tiny$\pm$2.8           & 47.4\tiny$\pm$2.8  & 50.3\tiny$\pm$2.8  \\
      & Perplexity      & \textbf{46.5}\tiny$\pm$2.8  & 44.4\tiny$\pm$2.7           & 47.2\tiny$\pm$2.8  & 46.6\tiny$\pm$2.8  \\
      & Grounded Gemini & 44.1\tiny$\pm$2.7  & 49.2\tiny$\pm$2.8           & \textbf{49.7}\tiny$\pm$2.8  & \textbf{51.2}\tiny$\pm$2.8  \\
\midrule
\multirow{3}{*}{GPT-4o}
      & DuckDuckGo      & 23.9\tiny$\pm$2.4  & 33.2\tiny$\pm$2.6           & 44.2\tiny$\pm$2.7  & 48.1\tiny$\pm$2.8  \\
      & Perplexity      & 33.5\tiny$\pm$2.6  & 41.5\tiny$\pm$2.7           & 43.2\tiny$\pm$2.7  & 44.7\tiny$\pm$2.7  \\
      & Grounded Gemini & 25.4\tiny$\pm$2.4  & 50.0\tiny$\pm$2.8           & 46.5\tiny$\pm$2.8  & 44.2\tiny$\pm$2.7  \\

    \bottomrule
  \end{tabular}%
  }
  \caption{Multi-Step Agent performance with different models, grounding methods, and sandbox states. The best result in each column is in bold.} % Original caption
  \label{tab:agents} % Original label
\end{table}

We report the performance of Multi-Step Agents on \GitChameleon{} in Table \ref{tab:agents}. A clear and significant trend is the substantial increase in success rates for all models and grounding methods when giving the agent a sandbox tool. Overall, \texttt{Claude Sonnet 3.5} demonstrated the highest success rates with a sandbox, across all grounding methods, while \texttt{Gemini 1.5 Pro} demonstrated the best results without a sandbox.

\subsubsection{AI Coding Assistants}
\begin{table}[!ht]
  \centering
  \resizebox{\columnwidth}{!}{%
  \begin{tabular}{@{} l l cc cc @{}}
    \toprule
    \multirow{2}{*}{\textbf{Name}} & \multirow{2}{*}{\textbf{Model}}
      & \multicolumn{2}{c}{\shortstack[c]{\textbf{Success Rate}\\\textbf{(\%)}}}
      & \multicolumn{2}{c}{\shortstack[c]{\textbf{API Hit Rate}\\\textbf{(\%)}}} \\
    \cmidrule(lr){3-4}\cmidrule(lr){5-6}
    & & \textbf{No-prob} & \textbf{Prob} & \textbf{No-prob} & \textbf{Prob} \\
    \midrule
    \multicolumn{6}{@{}l}{\textbf{CLI Assistants}} \\
    \midrule
 Claude Code   & Claude 3.7 Sonnet & 32.0\tiny$\pm$2.6  & 48.8\tiny$\pm$2.8  & \textbf{44.2}\tiny$\pm$2.7  & 45.5\tiny$\pm$2.7    \\
\midrule
\multirow{2}{*}{Goose}
             & GPT-4o            & \textbf{36.3}\tiny$\pm$2.7  & 36.9\tiny$\pm$2.7  & 43.9\tiny$\pm$2.7  & \textbf{54.5}\tiny$\pm$2.7    \\
             & GPT-4.1           & 19.2\tiny$\pm$2.2  & \textbf{55.5}\tiny$\pm$2.7  & 41.7\tiny$\pm$2.7  & 53.0\tiny$\pm$2.8    \\
\addlinespace[0.5em]
\multicolumn{6}{@{}l}{\textbf{IDE Assistants}} \\
\midrule
\multirow{5}{*}{Cline}
             & Claude 3.7 Sonnet & 32.9\tiny$\pm$2.6  & 44.8\tiny$\pm$2.7  & 40.5\tiny$\pm$2.7  & 50.2\tiny$\pm$2.8    \\
             & GPT-4.1           & 38.4\tiny$\pm$2.7  & \textbf{54.6}\tiny$\pm$2.7  & 42.4\tiny$\pm$2.7  & 48.8\tiny$\pm$2.8    \\
             & GPT-4.1-mini      & 27.1\tiny$\pm$2.5  & 42.1\tiny$\pm$2.7  & 32.9\tiny$\pm$2.6  & \textbf{52.4}\tiny$\pm$2.8    \\
             & GPT-4.1-nano      & 38.1\tiny$\pm$2.7  & 54.6\tiny$\pm$2.7  & 42.4\tiny$\pm$2.7  & 48.8\tiny$\pm$2.8    \\
             & GPT-4o            & \textbf{41.5}\tiny$\pm$2.7  & --                 & 42.7\tiny$\pm$2.7  & --                     \\
\midrule
Kilocode     & Claude 3.7 Sonnet & 30.2\tiny$\pm$2.5  & --                 & \textbf{43.3}\tiny$\pm$2.7  & --                     \\
\midrule
Roocode      & Claude 3.5 Sonnet & 12.5\tiny$\pm$1.8  & --                 & 41.2\tiny$\pm$2.7  & --                     \\

    \bottomrule
  \end{tabular}%
  }
  \caption{Success and API-hit rates for CLI and IDE coding assistants, under the setting where the problem statement is  given (\textbf{Prob}) and where it is not (\textbf{No-prob}), in which case we evaluate a scenario akin to tab code-completion. The results show that including the problem statement improves success rate by double-digit margins for 4 out of 5 cases evaluated.}
  \label{tab:cli-ide-assistants}
\end{table}

Table~\ref{tab:cli-ide-assistants} presents the success rates of various CLI and IDE assistants on the visible and hidden tests in \GitChameleon{}. When the problem statement is not given, \texttt{Cline} with GPT-4.1 achieves the best result, with a success rate of 38.4\%. All assistants besides for \texttt{Goose} on GPT-4o demonstrate significant gains, ranging from 12 to 35 points, from including the problem statement.

\subsubsection{Retrieval-Augmented Generation}

\begin{table}[!ht]
\centering
\resizebox{\linewidth}{!}{%
\begin{tabular}{@{}lccccc@{}}
\toprule
\textbf{Model} & \shortstack[c]{\textbf{Success}\\\textbf{Rate (\%)}} & \shortstack[c]{\textbf{API Hit}\\\textbf{Rate (\%)}} & \shortstack[c]{\textbf{Precision}\\\textbf{(\%)}} & \shortstack[c]{\textbf{Recall}\\\textbf{(\%)}} & \shortstack[c]{\textbf{MRR}\\\textbf{}} \\
\midrule
\addlinespace[0.5em]
\multicolumn{5}{@{}l}{\textbf{Open-Weights Models}} \\
\midrule
Deepseek V3          & \textbf{48.9}\tiny$\pm$2.8  & 48.5\tiny$\pm$2.8  & 41.6\tiny$\pm$2.2  & 50.4\tiny$\pm$2.8  & 0.62\tiny$\pm$0.03 \\
Llama 4 Maverick\tablefootnote{This version of the model is not FP8-quantized, unlike the one presented in Table~\ref{tab:base-model-performance}}
                     & 45.1\tiny$\pm$2.7  & \textbf{50.5}\tiny$\pm$2.8  & 41.2\tiny$\pm$2.2  & 49.8\tiny$\pm$2.8  & 0.61\tiny$\pm$0.03 \\
Qwen3                & 41.8\tiny$\pm$2.7  & 39.6\tiny$\pm$2.7  & 36.3\tiny$\pm$2.0  & 46.9\tiny$\pm$2.8  & 0.56\tiny$\pm$0.03 \\
Jamba 1.6 Large      & 41.8\tiny$\pm$2.7  & 47.1\tiny$\pm$2.8  & \textbf{41.9}\tiny$\pm$2.2  & \textbf{50.7}\tiny$\pm$2.8  & \textbf{0.62}\tiny$\pm$0.03 \\

\midrule
\addlinespace[0.5em]
\multicolumn{6}{@{}l}{\textbf{Enterprise Models}} \\
\midrule
Claude 3.7 Sonnet    & 56.1\tiny$\pm$2.7  & 53.0\tiny$\pm$2.8  & 41.9\tiny$\pm$2.2  & 50.7\tiny$\pm$2.8  & 0.62\tiny$\pm$0.03 \\
Claude 4 Sonnet      & \textbf{59.4}\tiny$\pm$2.8  & \textbf{55.8}\tiny$\pm$2.8  & \textbf{41.9}\tiny$\pm$2.2  & \textbf{50.7}\tiny$\pm$2.8  & \textbf{0.62}\tiny$\pm$0.03 \\
Gemini 2.5 Pro       & 56.7\tiny$\pm$2.7  & 51.1\tiny$\pm$2.8  & 41.9\tiny$\pm$2.2  & 50.7\tiny$\pm$2.8  & 0.62\tiny$\pm$0.03 \\
GPT-4.1              & 58.5\tiny$\pm$2.7  & 51.8\tiny$\pm$2.8  & 41.2\tiny$\pm$2.2  & 50.1\tiny$\pm$2.8  & 0.61\tiny$\pm$0.03 \\
Grok3                & 54.3\tiny$\pm$2.7  & 55.2\tiny$\pm$2.8  & 41.6\tiny$\pm$2.2  & 50.4\tiny$\pm$2.8  & 0.62\tiny$\pm$0.03 \\
Mistral Medium 3     & 52.4\tiny$\pm$2.7  & 51.2\tiny$\pm$2.8  & 41.6\tiny$\pm$2.2  & 50.4\tiny$\pm$2.8  & 0.62\tiny$\pm$0.03 \\
Devstral Small       & 43.3\tiny$\pm$2.7  & 45.1\tiny$\pm$2.8  & 41.6\tiny$\pm$2.2  & 50.4\tiny$\pm$2.8  & 0.62\tiny$\pm$0.03 \\
Nova Pro             & 44.2\tiny$\pm$2.7  & 42.4\tiny$\pm$2.7  & 40.7\tiny$\pm$2.2  & 49.6\tiny$\pm$2.8  & 0.60\tiny$\pm$0.03 \\

\bottomrule
\end{tabular}%
}
\caption{RAG performance for a subset of models when retrieving $k=3$ most relevant documents. The best success rate and API hit rate results for each model group are in bold. An extended version of the RAG experiment results is presented in \Cref{tab:rag_performance}.}
\label{tab:rag_performance_short}
\end{table}
% the code for precision is in compute_rag_interference.py

Table~\ref{tab:rag_performance_short} presents the performance of various models with RAG. Many models exhibit a significant (up to 10\%) boost in success rate with RAG compared to greedy decoding alone. Notably, \texttt{GPT-4.1}, the best performing model achieves a success rate of 58.5\%, up from 48.5\% with greedy decoding. These results demonstrate that the benchmark is still challenging even with access to the library documentation, with over 40\% of the problems remaining unsolved in the best case.

% \subsubsection{Prompt Optimization}
% \input{tables/prompt_optim}
% Table~\ref{tab:prompt_optim} presents the performance of various models with system prompt optimization 5 rounds.

\subsection{In-Depth Analysis of Findings} 
\label{ssec:ModelPerformanceMetrics}%

This section provides a detailed analysis of the experimental results, focusing on model performance across several key dimensions. These dimensions include the impact of different API change types, a comparison between success rate and API hit rate, and the effectiveness of self-debugging across various error types.

\begin{figure*}[t]
    \centering    
    \includegraphics[width=\linewidth]{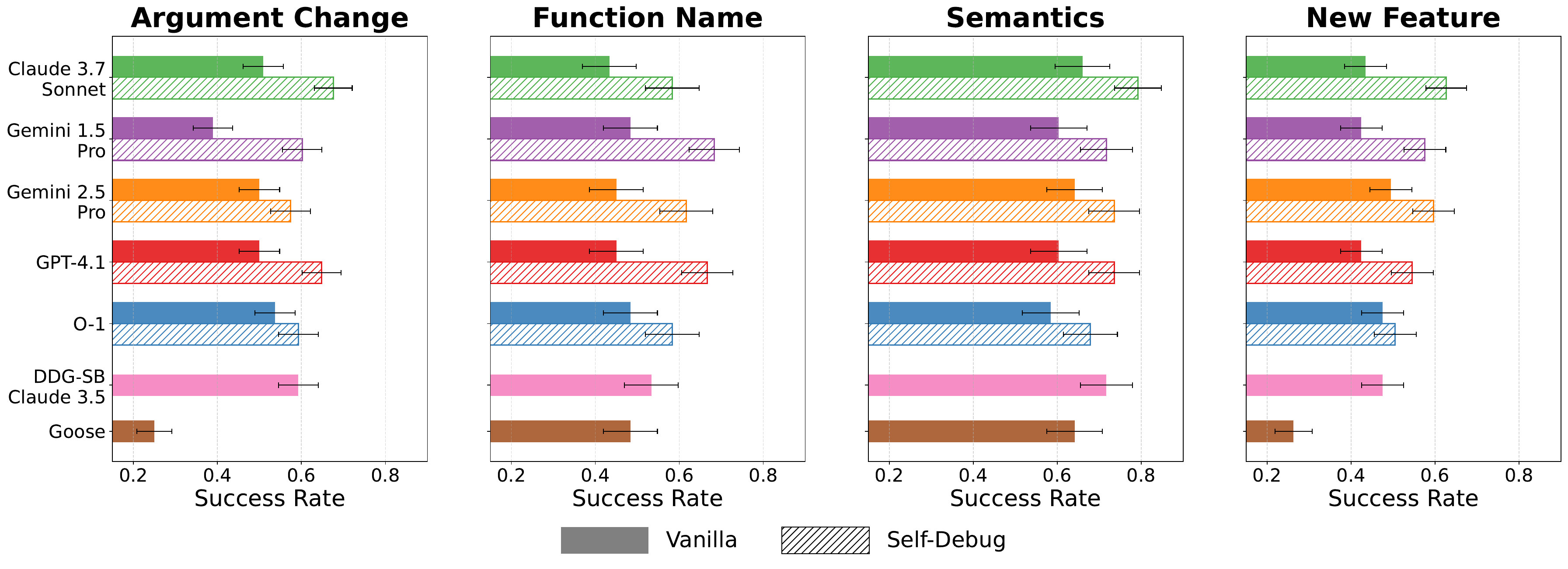}
        \label{fig:model_change_perf}
    \caption{\textbf{Success Rate Breakdown by Type of Change}: We analyze success rates with and without self-debugging, grouped by the type of change.   
    Light shaded bars represent values obtained from self-debugging. Standard error is drawn as a black line. We include ~\texttt{DDG-SB}, a Multi-Step Agent variant where DuckDuckGo is used for grounding and access to a sandbox is enabled. and the Coding Assistant \texttt{Goose}. Self-Debug results for these are omitted. }
    \label{fig:combined_model_perf}
\end{figure*}

\paragraph{Comparison of Success Rate and API Hit Rate}

API hit rate shows a moderate positive Pearson correlation with hidden‐test success under Greedy Decoding with the Pearson correlation coefficient (\(r = 0.392\), \(p = 0.097\), $N=19$), indicating that models which invoke the ground truth APIs more often tend to perform better on hidden tests in the Greedy setting, but falls just short of statistical significance at 5\% level. Under Zero‐Shot CoT, the correlation remains similar in magnitude (\(r = 0.483\)) and is statistically significant (\(p = 0.036\), $N=19$). In the Self‐Debug regime, however, the association becomes both stronger and highly significant (\(r = 0.615\), \(p = 0.011\), $N=16$), demonstrating that when models can iteratively refine their outputs, invoking ground truth APIs becomes an especially reliable predictor of hidden‐test performance.

% \paragraph{Analysis of Performance by Release Date}

% % 2019, 2020, 2021, 2022, 2023
% % 87.66962963 79.12814815 45.1496296	21.33925926	28.28148148

% At the top of Figure \ref{fig:combined_model_perf}, we present the year-over-year performance of a subset of the solution approaches. In each of the 2021, 2022, and 2023 panels, visible‐test success rates (lighter bars) exceed hidden‐test rates (darker bars) by roughly 10–15 points, illustrating the difficulty of the hidden tests. 

% \begin{itemize}
% \item overall trend is models are better at older than newer 2021 $>$ 2022 $\simeq$ 2023. this is likely due to the training data cutoff including at most only part of 2023.
% % \item Feedback improves 2022 significantly, no specific reason why...
% \end{itemize}

\paragraph{Analysis of Performance by Type of API Change}
% feedback_pass_at_10	feedback_pass_at_10	feedback_pass_at_10	feedback_pass_at_10	
% argument or attribute change	name change	other library or new feature	output change
% 22.08148148	52.67333333	55.53703704	9.327407407	
% pass_at_10	pass_at_10	pass_at_10	pass_at_10
% argument or attribute change	name change	other library or new feature	output change	
% 18.50814815	50.47259259	48.58444444	7.340740741

Figure \ref{fig:combined_model_perf} illustrates the performance of models across various API change types within the \GitChameleon{} benchmark, revealing notable variations in success rates. Semantic changes were the most tractable, with success rates ranging from 60–80\% with Self-Debug and 55–65\% without. New-feature additions proved to be the most challenging, with success rates between 25–50\% for Greedy Decoding and 50–65\% for Self-Debug. Notably, the Code Assistant \texttt{Goose} exhibited a substantial discrepancy in its performance on semantic and function-name changes compared to argument changes and new features. This suggests a heightened sensitivity to change category for \texttt{Goose}, a characteristic not observed in the enterprise models or the \texttt{Claude}-powered tool-calling agent.

\begin{figure}[!htbp]
  \centering
    \includegraphics[width=\linewidth]{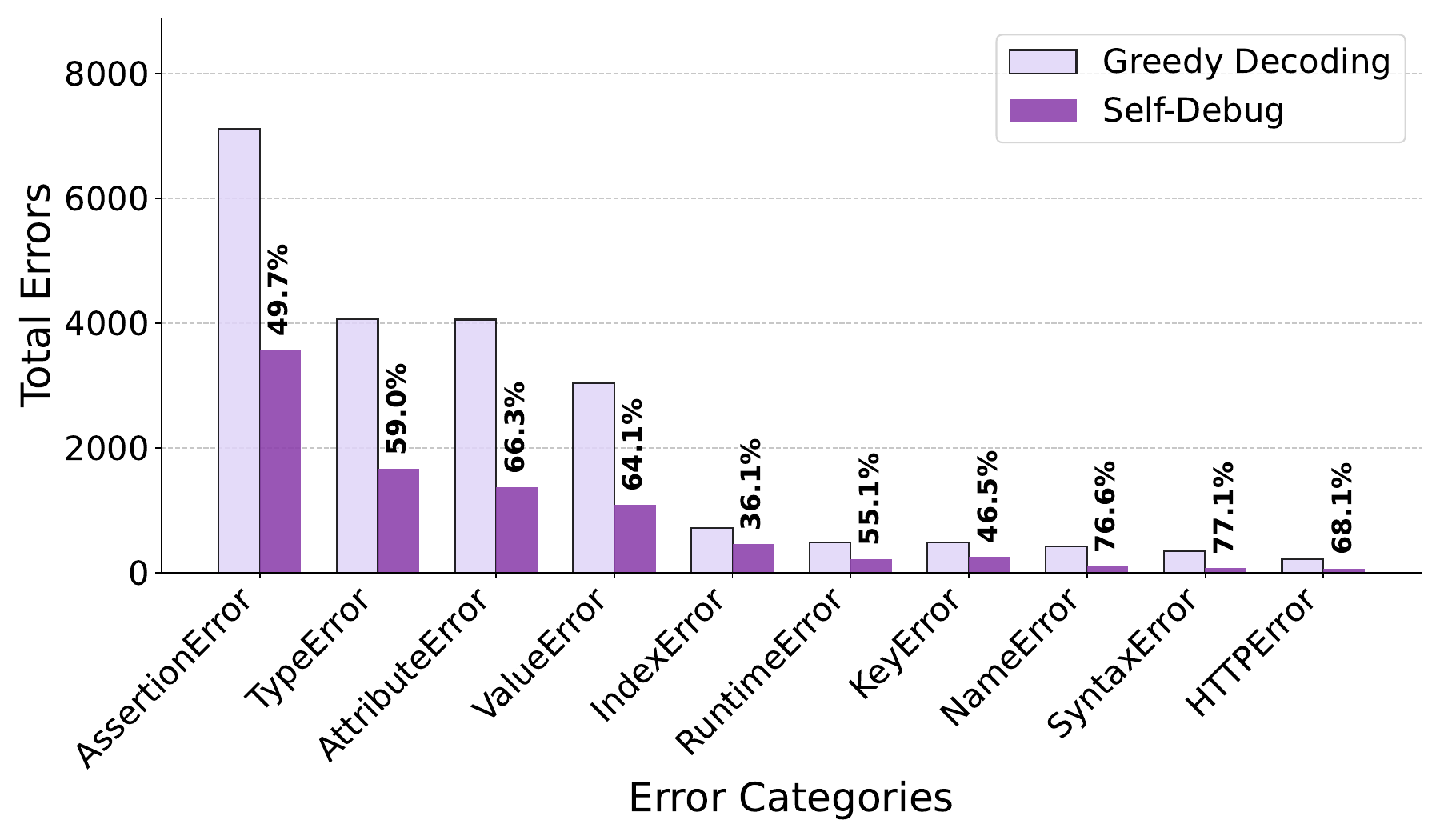}
  \label{fig:error-trace}
\caption{\textbf{Total error count for each category under Greedy decoding versus Self‐Debug}. Self‐Debug yields substantial decreases all types of errors.
}
    \label{fig:self-debug-error-cats}
  % \vspace{-10mm}
\end{figure}

\vspace{-3mm}

\paragraph{Self-Debug Error Categorization} Figure~\ref{fig:self-debug-error-cats} shows that self-debugging consistently lowers the rate of every class of traceback error, both in absolute numbers and relative terms:

(a) \textbf{Raw Counts:} We observe that for all error categories—from the most common (\texttt{AssertionError} and \texttt{TypeError}) down to the rarest (\texttt{RuntimeError})—applying Self-Debugging significantly lowers the total number of failures.

(b) \textbf{Percentage Reduction:} When normalized by the \texttt{Greedy Decoding} baseline, reductions span roughly 50\% up to about 90\%. The biggest relative improvements appear in the infrequent categories—such as \texttt{RuntimeError} and \texttt{SyntaxError}—while the common \texttt{AssertionError} and \texttt{TypeError} still see decrease in the range of 60-70\%.
\section{Related Work}
The continuous evolution of software libraries presents significant challenges for AI-driven code generation. This section reviews existing benchmarks designed to evaluate model performance in this context. Specialized frameworks developed to address the challenge are presented in \cref{app:ext_related_work}

% \subsection{Code Evolution Benchmarks}
The challenge of evaluating large language models (LLMs) in the context of evolving software libraries and their versions has been approached by several benchmarks. These benchmarks, while valuable, often differ in scope, methodology, or evaluation techniques compared to \GitChameleon{}.

%\textbf{RustEvo / EvoEval}
%This benchmark addresses Rust's rapid API evolution by automatically creating dynamic benchmarks from identified API changes, complete with natural language tasks, reference solutions, and executable tests\cite{arxiv_rustevo2_2025}. Experiments highlight that LLM performance varies by API change type and knowledge cutoff, with Retrieval-Augmented Generation (RAG) helping to mitigate these cutoff limitations.

\textbf{PyMigBench}
Focusing on Python library migration, this benchmark uses 321 real-world instances, evaluating both individual code transformations and the functional correctness of entire migrated segments via unit tests~\cite{pymigbench}. PyMigBench revealed that LLMs often handle individual changes well but struggle with achieving full functional correctness, especially for complex argument transformations.

%\textbf{RustEvo / EvoEval}
%This line of work addresses the challenges posed by Rust's particularly rapid API evolution. The framework automates the construction of dynamic benchmarks by systematically identifying API changes between Rust versions from sources like changelogs and documentation\cite{arxiv_rustevo2_2025}. These changes are categorized and automatically synthesized into programming tasks, complete with natural language descriptions, reference solutions, and executable tests. Experiments using RustEvo demonstrate significant variations in LLM performance depending on the type of API change and whether the API was released before or after the model's knowledge cutoff date. It also provides evidence for the effectiveness of RAG in mitigating knowledge cutoff limitations.

%\textbf{PyMigBench}
%This benchmark focuses specifically on the task of library migration within the Python ecosystem. It consists of 321 real-world library migration instances extracted from software repositories, encompassing nearly 3,000 individual migration-related code changes~\cite{pymigbench}. PyMigBench allows evaluation at two granularities: the correctness of individual code transformations and the functional correctness of the entire migrated code segment via unit tests from the original client repositories. PyMigBench demonstrated that while LLMs can correctly handle a high percentage of individual changes, achieving full functional correctness is less frequent, and they particularly struggle with certain types of changes, such as those requiring complex argument transformations.

\textbf{VersiCode} \cite{wu2024versicodeversioncontrollablecodegeneration} and the dataset by Wang et al.~\cite{wang2024llmsusedeprecatedapis} address library evolution but primarily depend on string matching for evaluation. 

\textbf{CodeUpdateArena} \cite{liu2024codeupdatearenabenchmarkingknowledgeediting} investigates model adaptation to synthetically generated API updates for functions in popular libraries. 

\textbf{GitChameleon} \cite{islah2024gitchameleon} serves as the primary predecessor to our proposed \GitChameleon{} benchmark, establishing the foundation for version-conditioned evaluation. However, it suffers from limited dataset coverage, comprising only 116 problems with a single manually crafted test per instance. Moreover, its experimental scope is narrow—lacking evaluations on agentic frameworks, retrieval-augmented generation (RAG), code assistants, and the deeper analyses that our work contributes. Building upon GitChameleon, we significantly enhance both the dataset and evaluation pipeline, offering broader problem coverage and a more comprehensive experimentation suite.

\GitChameleon{} distinguishes itself by focusing on the real-world scenario where developers are often constrained to specific library versions due to technical debt. Unlike CodeUpdateArena's synthetic changes, \GitChameleon{} evaluates LLMs on their ability to generate code for actual, documented historical breaking changes within library versions they were likely exposed to during training. Furthermore, diverging from the string-matching evaluations of VersiCode and Wang et al.~\cite{wang2024llmsusedeprecatedapis}, \GitChameleon{} is based on executable tests. This provides a more practical and rigorous assessment of functional accuracy in version-specific code generation. For an extended discussion of how \GitChameleon{} is differentiated from existing work, please see Appendix~\ref{app:ext_related_work}. %Our benchmark includes coding problems that are conditioned on specific library versions and accompanied by  executable unit tests, additional visible tests for self-debugging, and documentation references for RAG experiments, therefore enabling the evaluation of a variety of important settings. 

% \subsection{Specialized Frameworks and Repair Techniques}

% Specialized frameworks are also emerging to tackle library evolution challenges. \textbf{DepsRAG} \cite{arxiv_depsrag_v2} uses a multi-agent system with RAG and Knowledge Graphs for reasoning about software dependencies. \textbf{Dr.Fix} \cite{behrang2025drfixautomaticallyfixingdata} combines LLMs with program analysis and RAG for automated program repair, focusing on API misuse in LLM-generated code. These indicate a trend towards hybrid and agentic systems, moving beyond single LLM calls to more sophisticated architectures that integrate LLMs with other methods for handling library evolution. \GitChameleon{} serves as an essential resource for evaluating such systems.
\section{Conclusion}
The rapid evolution of software libraries presents a critical challenge for LLM-powered AI systems in generating functionally correct, version-conditioned code. To address this, we introduce \GitChameleon{}, a novel Python-based benchmark meticulously curated with version-conditioned problems and executable tests. Our extensive evaluation reveals that state-of-the-art LLMs, agents and code assistants currently struggle significantly with this task, achieving modest success rates.

%A key finding is the substantial effectiveness of self-debugging, which markedly improves performance and reduces diverse error types, underscoring LLMs' capacity for self-correction. %Notably, the simple application of self-debugging proved more effective in improving success rates than the open-ended capabilities of multi-step AI agents and even specialized coding assistants.  

By shedding light on current limitations and facilitating execution-based evaluation, \GitChameleon{} aims to foster the development of more robust and adaptable code generation models for evolving software environments. 

\section*{Acknowledgements}
The authors thank the International Max Planck Research School for Intelligent Systems (IMPRS-IS) for supporting Diganta Misra. This work was partially enabled by compute resources provided by Mila\footnote{\url{https://mila.quebec}} and was funded by the Max Planck \& Amazon Science Hub. 

\section*{Limitations}
%We consider the lack of prompt optimization done for the instruct models as a limitation of our analysis. 
While we aim to provide a comprehensive and holistic evaluation of LLMs on the task of version-conditioned generation, our benchmark is currently limited to Python and a small set of libraries. Moreover, we focus solely on code generation from natural language instructions, and do not evaluate version-to-version translation—i.e., converting code from one library version to another—even when both versions are in-distribution relative to the model’s training. For instance, if a model has been trained on PyTorch versions 1.7, 1.8, and 1.9, it would be valuable to assess whether it performs better when given a solution in 1.8 and asked to upgrade to 1.9 or downgrade to 1.7. Finally, we do not include human evaluations, which could provide a baseline for estimating average human performance on this task.

% Finally, we do not explore approaches such as  finetuning on a split of our benchmark to observe an upper bound of performance on this task. Future work could explore such approaches using our dataset.
\clearpage
\newpage
\bibliography{anthology, custom}

% % Bibliography entries for the entire Anthology, followed by custom entries
% %\bibliography{anthology,custom}
% % Custom bibliography entries only
% % \bibliography{arxiv}
\clearpage
\newpage
\appendix

%\section{Differentiating Factor}
\label{sec:appendix}
\section{Benchmark Details}
\label{app:bench_deets}
 
This appendix provides additional details on the \GitChameleon{} benchmark. We provide details on the dataset construction process, the structure of the dataset samples, on the processes for validating the examples and constructing the hidden tests, and finally present additional statistics regarding the dataset.

\subsection{Dataset Construction Process}
The examples were created by the authors, which took roughly 350 human hours. To construct that dataset, we compiled a list of popular Python libraries, focusing on those that had more than 1000 stars on Github as well as detailed documentation of changes between versions. For each library, we reviewed the change logs to identify breaking changes: deprecated functions, argument changes, alterations in behavior, and newly introduced functions.

For each identified change, we wrote a concise problem statement, starter code, expected solution and a suite of tests, consisting of a comprehensive suite of hidden tests to be used for model performance evaluation and ranking and a manually written concise visible test to be used for self-debugging experiments. We also added a ground-truth set of relevant documents for RAG experiments.

\textbf{\textsc{Note:}} Low-level changes—such as backend optimizations that do not alter the surface-level API—are not considered valid changes for our benchmark. For example, if between \texttt{Torch 1.7} and \texttt{Torch 1.8} the \texttt{torch.nn.Softmax()} function received a CUDA-based numerical stability improvement, this does not modify the API usage of \texttt{Softmax()} and is therefore not labeled as a change in our benchmark. Since most changes in mature libraries primarily impact backend functionality, collecting 328 valid samples required significant effort.

\subsection{Structure of Dataset Samples}
\begin{table}
  \centering
  \resizebox{\linewidth}{!}{%
    \begin{tabular}{l p{7cm}}
      \toprule
      \textbf{Library}             & The software library under test. \\
      \textbf{Library Version}     & The exact version of that library. \\
      \textbf{Task Description}    & A problem centered on a particular library change. \\
      \textbf{Initial Code}        & The Python snippet provided as a starting point. \\
      \textbf{Extra Dependencies}  & Any additional packages required to solve the task. \\
      \textbf{Hidden Tests}        & Comprehensive unit tests designed to maximize coverage. The success rate on these is the benchmark metric. \\
      \textbf{Visible Test}       & A concise test that validates the specific target behavior, intended to be used for Self-Debugging experiments. \\
      \textbf{Reference Solution}  & A correct, ground-truth implementation. \\
      \textbf{Reference Documents}  & A set of version-specific reference documents, to be used for RAG experiments. \\
      \bottomrule
    \end{tabular}%
  }
  \caption{Problem column definitions for the \GitChameleon{} dataset.}
  \label{tab:column-defs}
\end{table}

\begin{table}[ht]
  \centering
  \resizebox{\linewidth}{!}{%
    \begin{tabular}{lp{7cm}}
      \toprule
      \textbf{Change Category}       & The type of library‐evolution changes, as defined in table~\ref{tab:api_evolution_categories}. \\
      \textbf{Target Entity}         & The specific function or class under test. \\
      \textbf{Solution Style}        & “Functional” if only a function body is expected, or “Full” for a general code completion. \\
      \textbf{Web Framework Task}    & “Yes” if the problem exercises a web‐development framework, otherwise “No.” \\
      \bottomrule
    \end{tabular}%
  }
  \caption{Metadata column definitions.}
  \label{tab:change-metadata}
\end{table}

\begin{table}[ht] 
\centering
\resizebox{\linewidth}{!}{
\begin{tabular}{p{4cm}p{7cm}} 
\toprule
\textbf{Change Category} & \textbf{Description} \\
\textbf{Argument or Attribute change} & The API call to a function, method, or class has a change in arguments (e.g. name, order, new, deprecated argument) between versions. \\
\textbf{Function Name change} & The name of the API call has changed between versions (e.g. \texttt{pandas.append} to \texttt{pandas.concat}). \\
\textbf{Semantics or Function Behavior change} & The semantic / runtime behavior of the API call changed between versions (e.g. returning a different type). \\
\textbf{New feature or additional dependency-based change} & A feature was introduced in a specific version; therefore, to execute the same functionality, a model using an older version should make use of an additional dependency (e.g. \texttt{torch.special} was introduced in \textsc{torch 1.10}, previously one could use \textsc{numpy} for the same). \\
\bottomrule
\end{tabular}
}
\caption{Categories of API Evolution Changes}
\label{tab:api_evolution_categories}
\end{table}
The main fields of each sample are given in Table~\ref{tab:column-defs}. Additionally, each problem in \GitChameleon{} is associated with metadata to assist in the analysis of the results, as described in Table~\ref{tab:change-metadata}. Each problem is classified with a type of API evolution change among the categories defined in Table~\ref{tab:api_evolution_categories}.

\subsection{Dataset Validation}
To ensure the validity of the dataset examples, we followed the following process: First, we created a clean Docker container for each problem and installed the required dependencies into it. Then, we executed the visible and hidden validation tests to ensure that all are successful. 

\subsection{Hidden Test Construction}
This section presents how we generated the hidden tests for each dataset example. These tests were generated by instructing the \texttt{Zencoder AI Coding Agent}~\footnote{\url{https://zencoder.ai}} to create test files for each example, incorporating the appropriate dependency versions. The Zencoder agent, built on the GPT-4.1 base model, operated with internet search enabled and was granted execution access, allowing it to self-correct outputs that initially failed during runtime. Further errors encountered during verification were resolved by supplying error traces back to Zencoder or through an isolated instance of GPT-4o, supplemented with manual intervention where necessary. This process enabled us to construct a robust and comprehensive test suite, achieving a coverage of \textbf{96.5\%}. The decision to use \textsc{Zencoder} was motivated by limitations observed in alternative unit test generation approaches. Rule-based generators such as Pynguin~\cite{lukasczyk2022pynguin} fail to account for version differences among samples that share the same or similar problem statements. Meanwhile, AI-based unit test generators like Claude Code and EarlyAI\footnote{\url{https://www.startearly.ai/}} were not suitable: the former typically generated test classes where each sub-function was populated only with \texttt{pass()} statements, while the latter was restricted to functional-style problems and could not handle the more complex, class-based structures prevalent in \GitChameleon{}.

\subsection{Additional Dataset Statistics}
\begin{figure}[!htb]
  \centering
  \begin{subfigure}{1\linewidth}
    \centering
    \includegraphics[width=\linewidth]{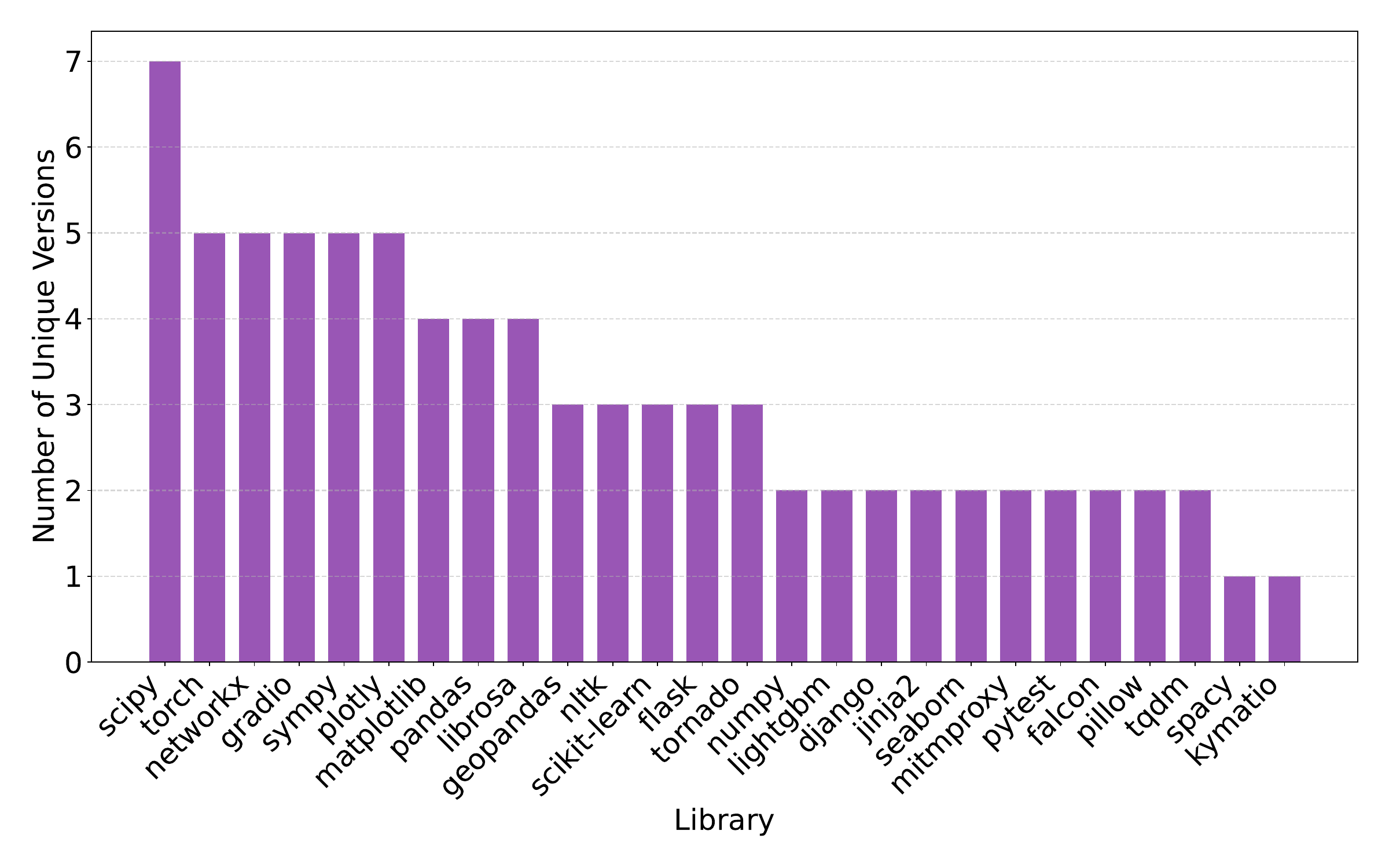}
    \caption{Number of unique versions per library.}
    \label{fig:version_count}
  \end{subfigure}
  \vspace{2mm}
  \begin{subfigure}{1\linewidth}
    \centering
    \includegraphics[width=\linewidth]{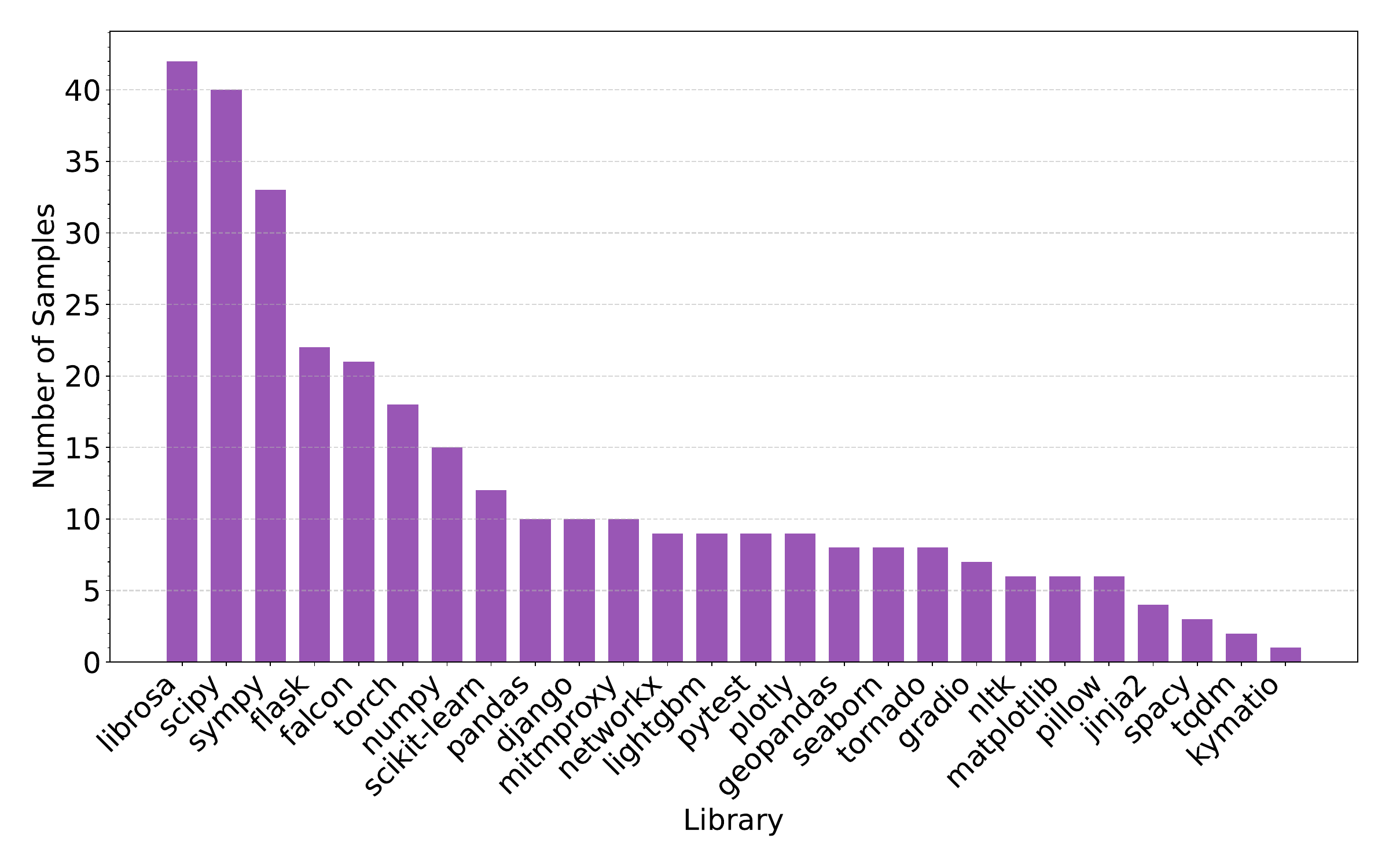}
    \caption{Number of samples per library.}
    \label{fig:samples_by_library}
  \end{subfigure}
  \caption{Dataset library statistics. (a) The count of distinct versions identified for each library, presented in decreasing order of uniqueness. (b) The total frequency of samples containing each library, ordered by their occurrence count.}
  \label{fig:extra_stats}
\end{figure}

Figure~\ref{fig:extra_stats} presents the number of unique versions per library and the number of samples per library.

% \begin{figure}[ht!]
%     \centering
%     \includegraphics[width=\linewidth]{figures/yoy_data.pdf}
%     \caption{Cumulative year-over-year version releases of popular Python-based machine learning libraries show a consistent upward trend, reflecting the rapid pace of development and version updates of code libraries and packages.}
%     \label{fig:fig1}
% \end{figure}

% \newpage
\section{Extra Methodologies: Reasoning, Sampling and Prompting}

This section presents results from additional experimental methodologies:

\begin{itemize}
    \item \textbf{Temperature Sampling:} Results are shown in Table~\ref{tab:temp0.8-performance}. We evaluate sampling at temperature $T = 0.8$ across 10 seeds using both the OpenAI and Gemini model suites. The performance difference compared to greedy decoding is minimal.
    
    \item \textbf{Reasoning Models:} Performance results for the OpenAI o-series reasoning models are provided in Table~\ref{tab:base-model-performance-ext}.
    
    \item \textbf{Self-Explained Keywords (SEK) Prompting:} We evaluate the SEK prompting method proposed by~\citet{fan2024selfexplainedkeywordsempowerlarge}, applied to both OpenAI and Gemini models. SEK involves a two-stage process: (1) \textit{Keyword Extraction}, where the model generates relevant keywords for the coding task, and (2) \textit{Keyword Categorization}, where keywords are ranked and classified into (a) Function, (b) General, and (c) Abstract categories. TF-IDF ranking is performed using a 50{,}000-document subset of the \textsc{evol-codealpaca-v1} corpus~\cite{luo2023wizardcoder}. As shown in our empirical analysis, SEK does not yield significant improvements over greedy sampling, and in several cases underperforms relative to it. \textsc{Note:} Temperature $T = 0$ is used in both stages of SEK prompting.
\end{itemize}

\begin{table*}[!hbt]
  \centering
  \setlength{\tabcolsep}{8pt} % Adjust column spacing
  \renewcommand{\arraystretch}{1.2} % Optional: increase row spacing a bit
  \resizebox{0.85\linewidth}{!}{%
    \begin{tabular}{@{}
                     l
                     >{\columncolor{HighlightGray}}c c c                    
                     >{\columncolor{HighlightGray}}c c c
                     >{\columncolor{HighlightGray}}c c
                     @{}}
      \toprule
      \multirow{3}{*}{\textbf{Model}}
      & \multicolumn{3}{c}{\textbf{Vanilla Decoding}}
      & \multicolumn{3}{c}{\textbf{Vanilla with Self-Debug}}
      & \multicolumn{2}{c}{\textbf{Zero-shot CoT}} \\
      \cmidrule(lr){2-4} \cmidrule(lr){5-7} \cmidrule(lr){7-9}
      & \multicolumn{2}{c}{\shortstack[l]{\textbf{Success}\\\textbf{Rate (\%)}}}
      & \multirow{2}{*}{\shortstack[l]{\textbf{API} \\ \textbf{Hit} \\ \textbf{Rate (\%)}}}
      & \multicolumn{2}{c}{\shortstack[l]{\textbf{Success}\\\textbf{Rate (\%)}}}
      & \multirow{2}{*}{\shortstack[l]{\textbf{API} \\ \textbf{Hit} \\ \textbf{Rate (\%)}}}
      & \multicolumn{1}{c}{\shortstack[l]{\textbf{Success}\\\textbf{Rate (\%)}}}
      & \multirow{2}{*}{\shortstack[l]{\textbf{API} \\ \textbf{Hit} \\ \textbf{Rate (\%)}}} \\
      \cmidrule(lr){2-3} \cmidrule(lr){5-6} \cmidrule(lr){8-8}
      & \cellcolor{HighlightGray}\shortstack[c]{\textbf{Hidden}}
      & \textcolor{DimGrayText}{\shortstack[c]{\textbf{Visible}}}
      &
      & \cellcolor{HighlightGray}\shortstack[c]{\textbf{Hidden}}
      & \textcolor{DimGrayText}{\shortstack[c]{\textbf{Visible}}}
      &
      & \cellcolor{HighlightGray}\shortstack[c]{\textbf{Hidden}}     
      
      & \\
      \midrule
      %\addlinespace[0.5em]      
      \multicolumn{9}{@{}l}{} \\
      \midrule    
o1         & \textbf{51.2}\tiny$\pm$2.8  & \textbf{60.1}\tiny$\pm$2.7  & 42.1\tiny$\pm$2.7  & 57.6\tiny$\pm$2.7  & 68.6\tiny$\pm$2.6  & \textbf{49.2}\tiny$\pm$2.8  & 41.2\tiny$\pm$2.7  & \textbf{41.3}\tiny$\pm$2.7  \\
o3-mini    & 44.5\tiny$\pm$2.7  & 52.7\tiny$\pm$2.8  & 40.6\tiny$\pm$2.7  & \textbf{66.8}\tiny$\pm$2.6  & \textbf{76.5}\tiny$\pm$2.3  & 45.7\tiny$\pm$2.8  & \textbf{50.9}\tiny$\pm$2.8  & 40.7\tiny$\pm$2.7  \\
o4-mini    & 48.2\tiny$\pm$2.8  & 57.0\tiny$\pm$2.7  & \textbf{48.3}\tiny$\pm$2.8  & 63.1\tiny$\pm$2.7  & 75.0\tiny$\pm$2.4  & 45.4\tiny$\pm$2.7  & –                 & –                 \\
codex-mini & 48.5\tiny$\pm$2.8  & 58.2\tiny$\pm$2.7  & 47.5\tiny$\pm$2.8  & –                  & –                  & –                  & 32.0\tiny$\pm$2.6  & 37.9\tiny$\pm$2.7  \\

      \bottomrule
    \end{tabular}%
  }
  \caption{Success rate on visible and hidden tests and API hit rate under the Vanilla, Self-Debug, and Zero-shot CoT settings, for the OpenAI o-series models. Model ranking on the benchmark is determined by \textbf{Hidden Success Rate}. Visible Success Rate figures are for context on Self-Debugging. The best result in each column is in bold. For full model details and citations, please refer to Appendix~\ref{app:artifacts}.}
  \label{tab:base-model-performance-ext}
\end{table*}

\begin{table}[h]
  \centering
  \setlength{\tabcolsep}{8pt} % Adjust column spacing
  \renewcommand{\arraystretch}{1.2} % Increase row spacing
  \resizebox{0.99\linewidth}{!}{%
    \begin{tabular}{@{}l
                     cc  % Hidden Success Rate & API Hit Rate
                     @{}}
      \toprule
      \textbf{Model}
      & \shortstack[c]{\textbf{Hidden Success}\\\textbf{Rate (\%)}}
      & \shortstack[c]{\textbf{API Hit}\\\textbf{Rate (\%)}} \\
      \midrule
      o1               & \textbf{50.5}\tiny$\pm$0.8  & 44.0\tiny$\pm$0.8  \\
      o3-mini          & 46.4\tiny$\pm$1.6  & 42.5\tiny$\pm$0.6  \\
      GPT-4.1          & 48.9\tiny$\pm$1.4  & 48.1\tiny$\pm$1.0  \\
      GPT-4.1-mini     & 45.9\tiny$\pm$1.3  & 46.9\tiny$\pm$0.6  \\
      GPT-4.1-nano     & 33.8\tiny$\pm$1.1  & 43.8\tiny$\pm$0.8  \\
      GPT-4o           & 47.2\tiny$\pm$1.2  & 45.1\tiny$\pm$0.9  \\
      GPT-4o-mini      & 40.2\tiny$\pm$1.2  & 41.0\tiny$\pm$1.1  \\
      \midrule
      Gemini 1.5 Pro   & 45.4\tiny$\pm$1.2  & 45.5\tiny$\pm$0.7  \\
      Gemini 2.5 Pro   & 41.0\tiny$\pm$3.4  & \textbf{48.3}\tiny$\pm$1.7  \\
      Gemini 2.0 Flash & 43.4\tiny$\pm$3.1  & 42.5\tiny$\pm$0.9  \\
      Gemini 2.5 Flash & 46.4\tiny$\pm$0.8  & 46.8\tiny$\pm$1.2  \\
      \bottomrule
    \end{tabular}%
  }
  \caption{\textbf{Hidden Success Rate using temperature sampling ($T=0.8$), averaged over 10 seeds.} A comparison to the greedy decoding baseline in Table~\ref{tab:base-model-performance} reveals that the changes in performance between greedy decoding and temperature sampling are mixed. For most models, the differences are small, but for a few specific models, the changes are big and noteworthy. For the majority of models evaluated (8 out of 11), the performance change is minor, typically within +/- 2 percentage points. For example, \texttt{Gemini-2.5-pro}, shows a notable decrease in success rate (-9.0 points).}
  \label{tab:temp0.8-performance}
\end{table}

\begin{table}[!hbt]
  \centering
  \setlength{\tabcolsep}{8pt} % Adjust column spacing
  \renewcommand{\arraystretch}{1.2} % Increase row spacing slightly
  \resizebox{0.99\linewidth}{!}{%
    \begin{tabular}{@{}l
                     cc  % SEK: Success Rate and API Hit Rate
                     @{}}
      \toprule
      \textbf{Model}
      & \shortstack{\textbf{Hidden Success}\\\textbf{Rate (\%)}}
      & \shortstack{\textbf{API}\\\textbf{Hit Rate (\%)}} \\
      \midrule
GPT-4o              & 29.6\tiny$\pm$2.5  & 43.6\tiny$\pm$2.7  \\
GPT-4o-mini         & 27.7\tiny$\pm$2.5  & 40.3\tiny$\pm$2.7  \\
GPT-4.1             & 43.6\tiny$\pm$2.7  & 49.4\tiny$\pm$2.8  \\
GPT-4.1-mini        & 41.2\tiny$\pm$2.7  & 44.0\tiny$\pm$2.7  \\
GPT-4.1-nano        & 32.9\tiny$\pm$2.6  & 43.8\tiny$\pm$2.7  \\
GPT-4.5             & 33.8\tiny$\pm$2.6  & \textbf{58.0}\tiny$\pm$2.7  \\
\midrule
Gemini 1.5 Pro      & 44.5\tiny$\pm$2.7  & 45.7\tiny$\pm$2.8  \\
Gemini 2.0 Flash    & 41.2\tiny$\pm$2.7  & 43.4\tiny$\pm$2.7  \\
Gemini 2.5 Pro      & 47.3\tiny$\pm$2.8  & 50.0\tiny$\pm$2.8  \\
Gemini 2.5 Flash    & \textbf{48.2}\tiny$\pm$2.8  & 43.4\tiny$\pm$2.7  \\
  \bottomrule
    \end{tabular}%
  }
  \caption{\textbf{Success and API hit rates under the SEK setting.} While SEK, being a two-round prompting scheme, is expected to outperform greedy decoding, we observe that it does not yield significant improvements. For example, with GPT-4.1, the success rate actually drops by 4.9\% when using SEK compared to greedy decoding.} 
  \label{tab:performance-sek}
\end{table}

\section{Extended Experiment Results and Analysis}
This section contains the following additional experimental results: 
\begin{itemize}
    \item  An experiment on Automatic Prompt Optimization of the system prompt for Greedy Decoding is described in Table~\ref{tab:round-prompt-optim}.
    \item An experiment on static analysis based generated solutions fixing to ensure model failures are not attributed to confounding factors like indentation problems and unused imports or variable declarations. Refer to Table~\ref{tab:static} for further details.
    \item Table~\ref{tab:rag_full} contains an extended set of RAG results, including both additional models and the setting where only a single document is retrieved.
\end{itemize}

\begin{figure*}[htb]
    \centering
    \begin{minipage}{\linewidth}
        \centering
        \includegraphics[width=\linewidth]{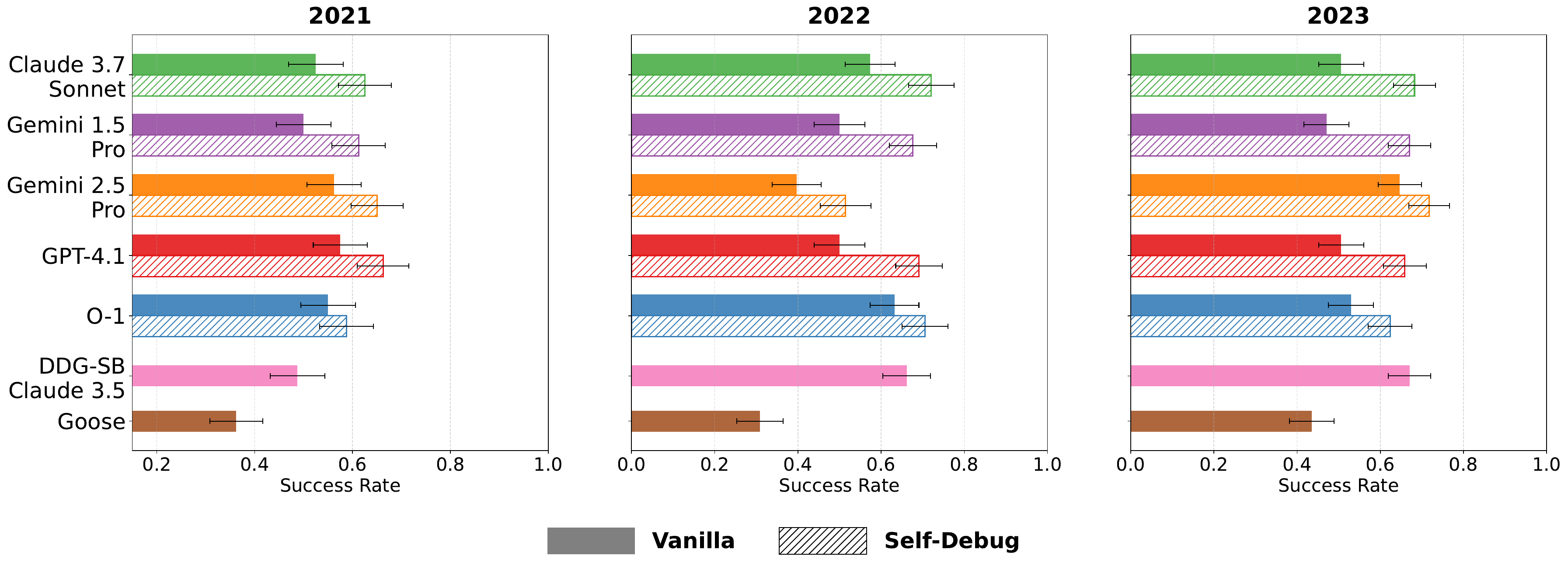}
        \label{fig:model_date_perf}
    \end{minipage}
    \vspace{5mm} % Adjust the spacing between figures as needed
    \caption{\textbf{Success Rate Breakdown by Version Release Year}. Lighter and darker shaded bars represent values obtained with and without Self-Debugging, respectively. Standard error is drawn as a black line. This plot shows that the release year does not significantly impact the results for most evaluated settings. }
    \label{fig:yoy_model_perf}
\end{figure*}

\begin{figure*}[htb]
  \centering
  \includegraphics[width=0.9\textwidth]{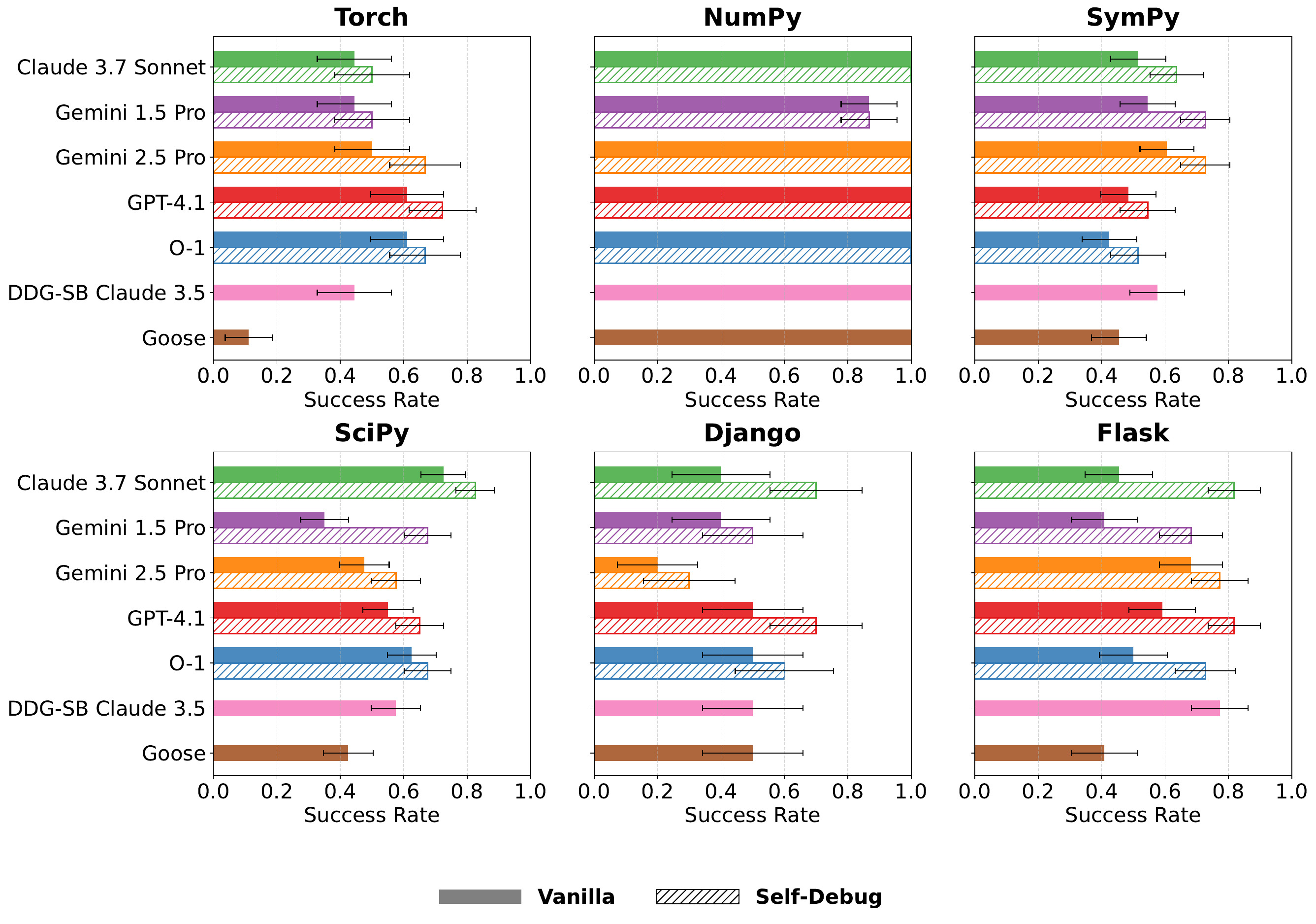}
  \caption{\textbf{Success Rate Breakdown by Library}. This figure shows the differences in success rate between the libraries included in \GitChameleon{}. All evaluated settings do very well on NumPy, which is to be expected given the popularity of the library and the subsequent abundance of code that uses it. The success rates on the web development frameworks are notably lower than on the scientific computing libraries, perhaps due to having more complex abstractions.}
  \label{fig:model_library}
\end{figure*}

\begin{figure*}[!htbp] % 'htbp' suggests placement options: here, top, bottom, page
    \centering % Center the entire figure content

    % Top row: two subfigures side-by-side
    \begin{subfigure}[b]{0.45\textwidth} % Adjust width as needed, 'b' for bottom alignment
        \centering
        \includegraphics[width=\linewidth]{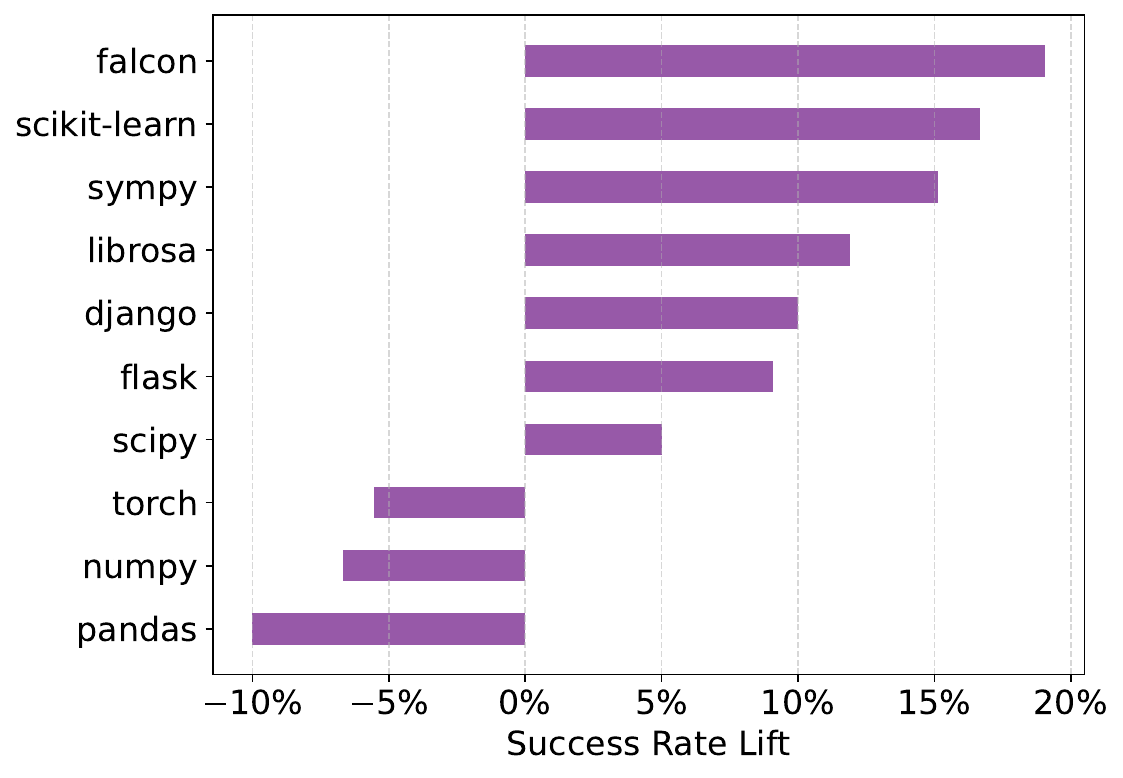}
        \caption{\textbf{GPT-4.1}}
        \label{fig:subfig1}
    \end{subfigure}
    \hspace{0.5cm}
    \begin{subfigure}[b]{0.45\textwidth} % Adjust width as needed
        \centering
        \includegraphics[width=\linewidth]{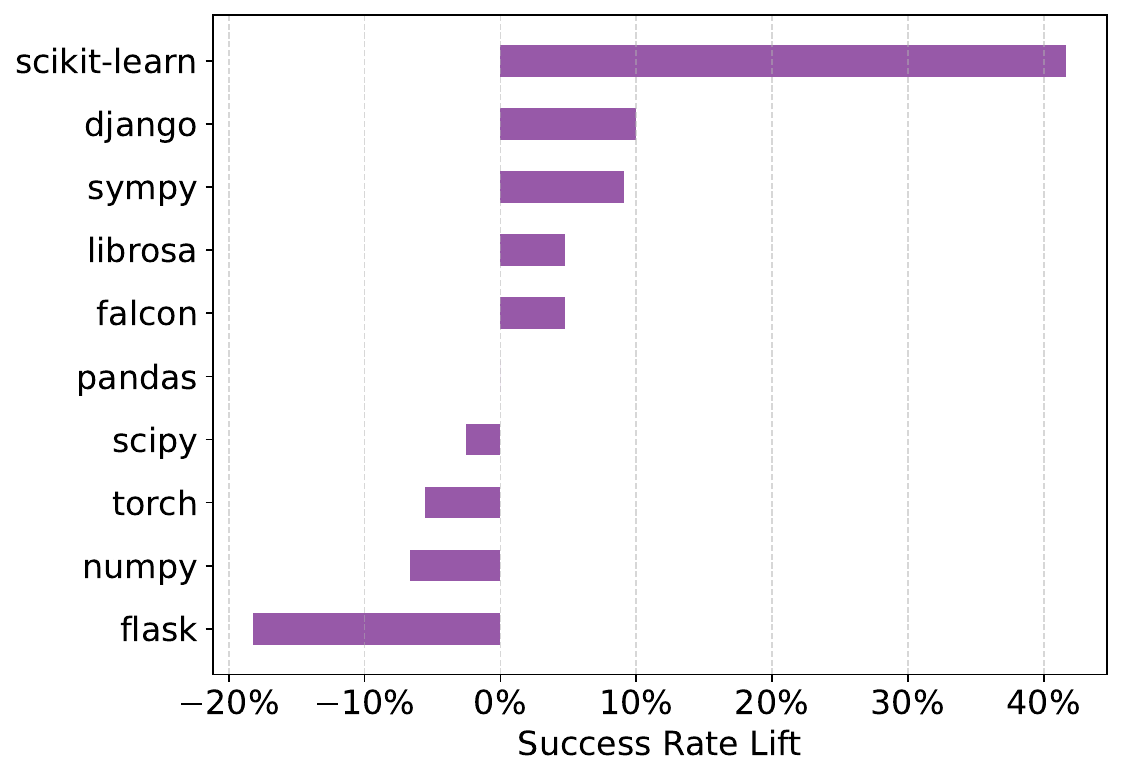}
        \caption{\textbf{GPT-4.1-mini}}
        \label{fig:subfig2}
    \end{subfigure}

    % Bottom row: one subfigure spanning below
    \vspace{0.cm} % Optional: add some vertical space between rows
    \begin{subfigure}[b]{0.45\textwidth} % Adjust width as needed, perhaps wider
        \centering
        \includegraphics[width=\linewidth]{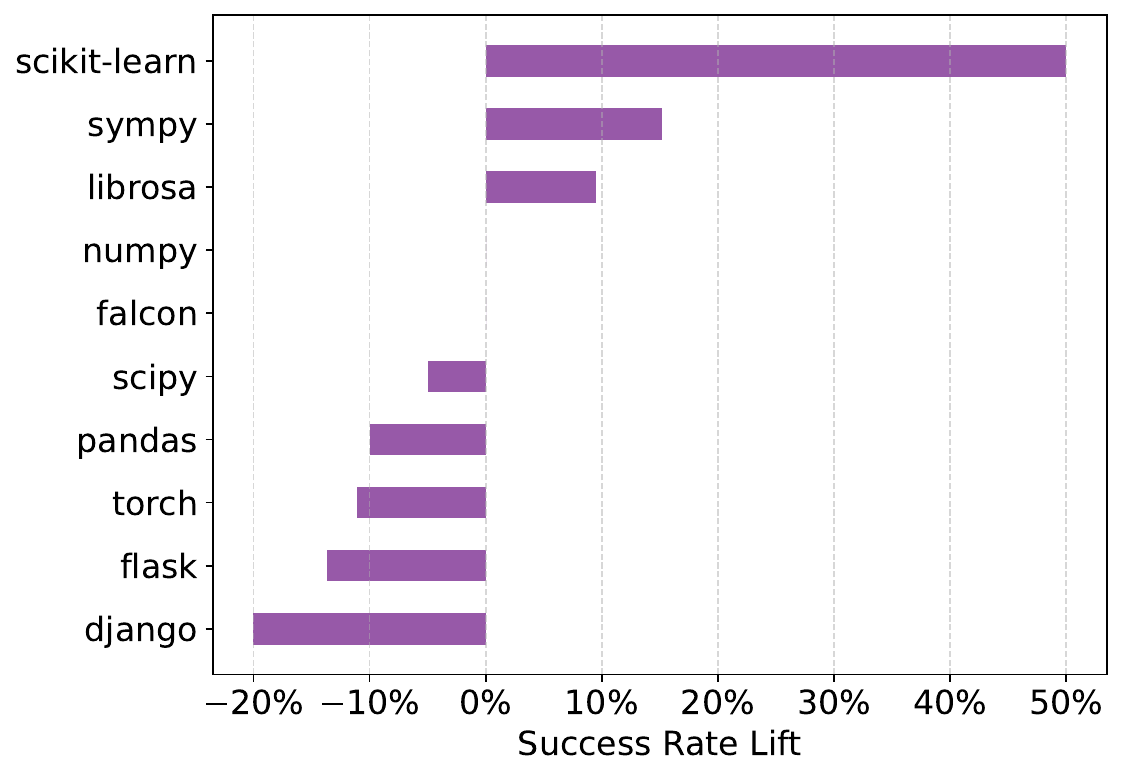}
        \caption{\textbf{GPT-4.1-nano}}
        \label{fig:subfig3}
    \end{subfigure}

    \caption{\textbf{ $\Delta$ Success Rate of RAG over Greedy Decoding, per library}. The 10 most frequent libraries in \GitChameleon{} are shown here. The plots demonstrate a trend where smaller models are less effective at using RAG, with the full-size \texttt{GPT-4.1} improving on 7 libraries, the \texttt{mini} version improving on 5 and the \texttt{nano} version improving only on 3.}
    \label{fig:rag_lib_lift}
\end{figure*}

\begin{figure*}[!htbp]
  \centering
  \begin{subfigure}[b]{0.46\textwidth} % Adjust width as needed, 'b' for bottom alignment
        \centering
          \includegraphics[width=0.99\linewidth]{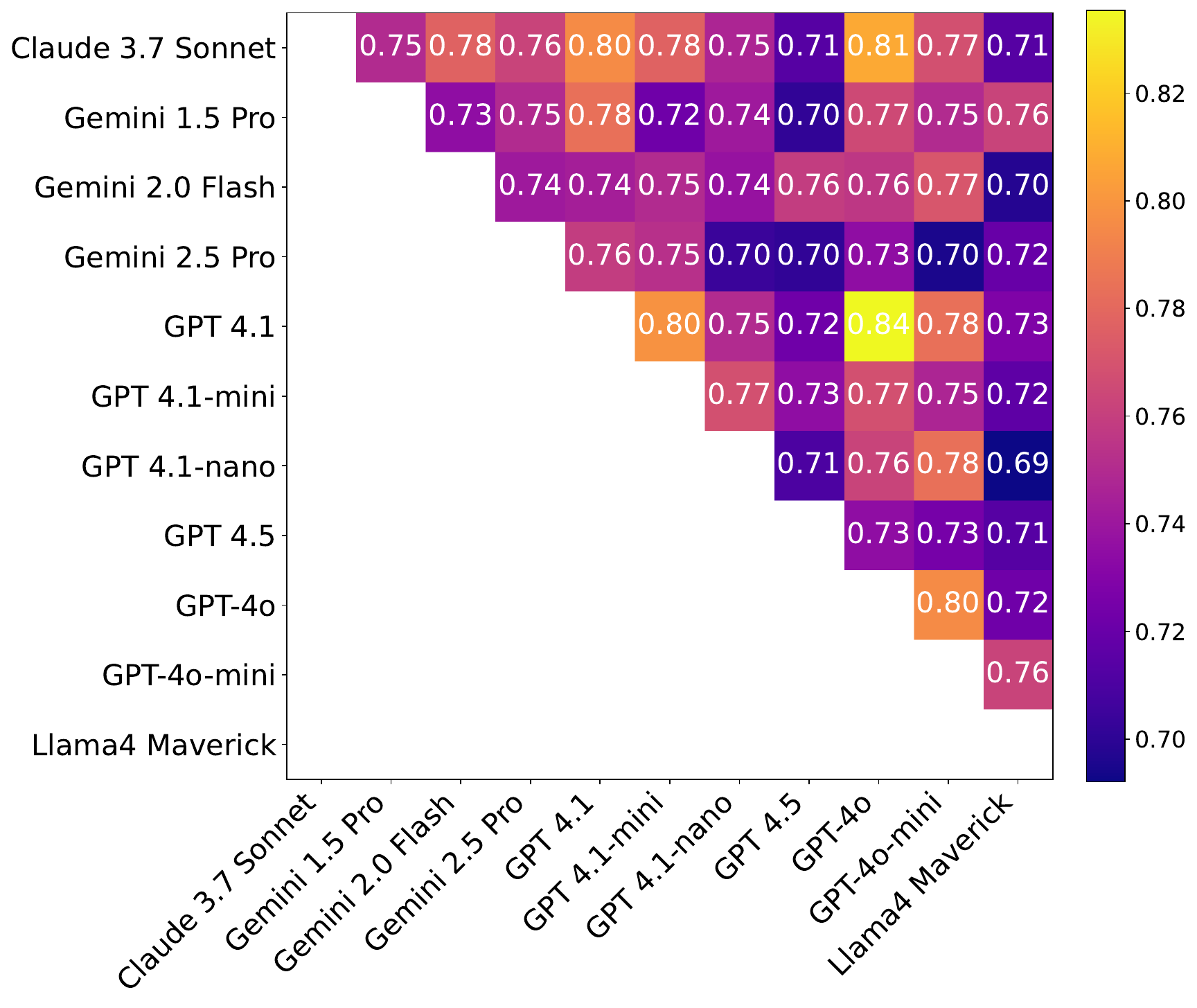}
        \caption{\textbf{Greedy Decoding}}    
    \end{subfigure}
    \hspace{0.2cm}
    \begin{subfigure}[b]{0.46\textwidth} % Adjust width as needed
        \centering
        \includegraphics[width=\linewidth]{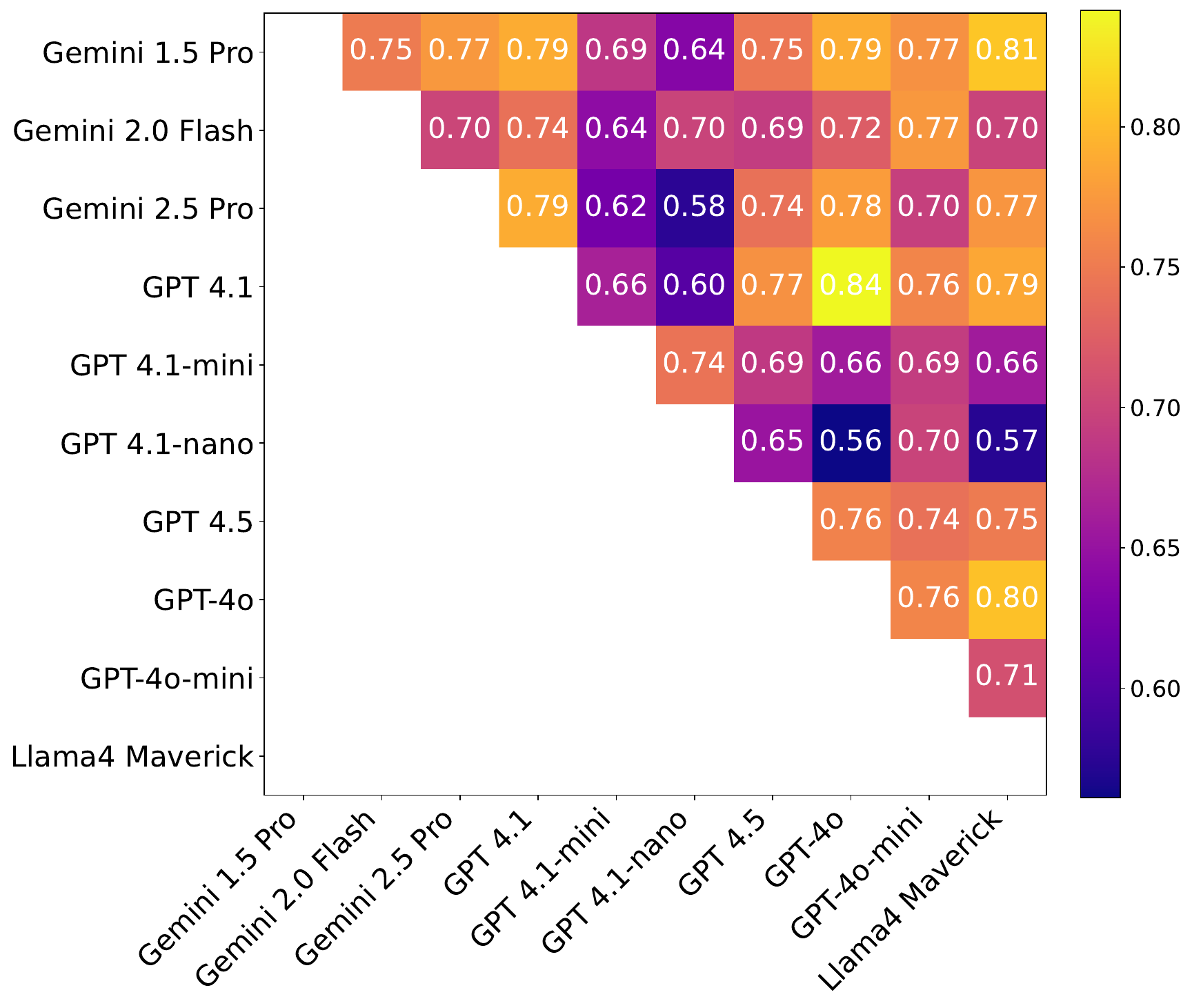}
        \caption{\textbf{Zero-Shot Chain-Of-Thought}}      
    \end{subfigure}
    \vspace{0.1cm} % Optional: add some vertical space between rows
    \begin{subfigure}[b]{0.46\textwidth} % Adjust width as needed, perhaps wider
        \centering
        \includegraphics[width=\linewidth]{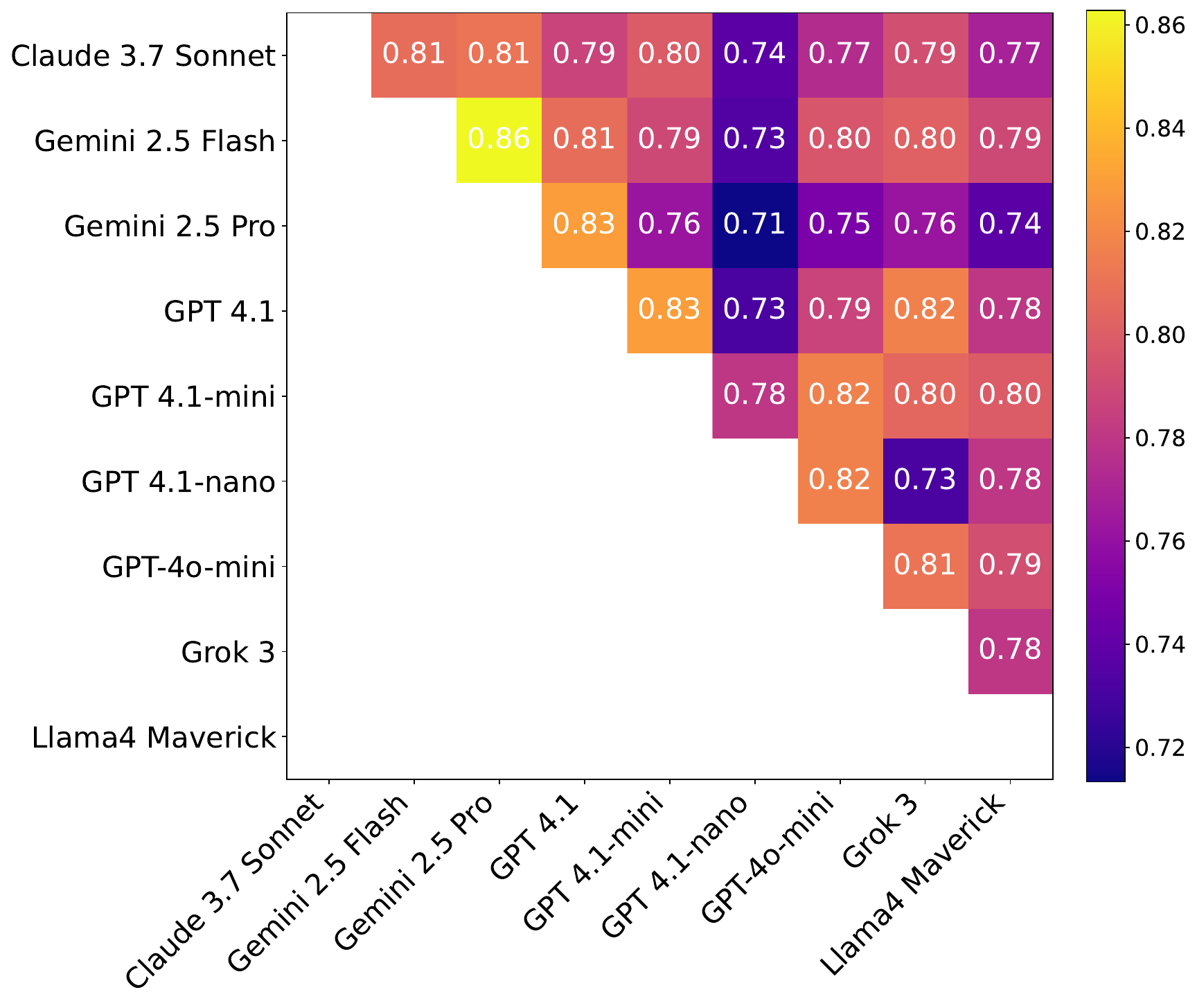}
        \caption{\textbf{RAG (k=3)}}       
    \end{subfigure}

  \caption{\textbf{Intra model sample agreement rates}. These plots show the rate of samples that have the same pass/fail result among all pairs of models, under the Greedy Decoding, Zero-Shot CoT and RAG settings. Each cell in these plots represents the agreement rate of a pair of models, with the rate also being color-coded. The high agreement rates in all three subfigures show that ensembling different models would have a limited effect on the success rates. }
  \label{fig:model_conf_mat}
\end{figure*}

We also present the following additional analyses:
\begin{itemize}
    \item A comparison of success rates between Self-Debug and Greedy Decoding, when broken down by version release year (Figure~\ref{fig:yoy_model_perf}) and by library (Figure~\ref{fig:model_library}).
    \item A comparison of success rates between RAG and Greedy Decoding by library is shown in Figure~\ref{fig:rag_lib_lift}.
    \item Figure~\ref{fig:model_conf_mat} analyzes the intra-model sample agreement rates in the Greedy Decoding, Zero-Shot CoT and RAG settings.
\end{itemize}

% \begin{table}[ht]
% \centering
% \caption{Performance summary for selected LLMs under Greedy Decoding.}
% \label{tab:prompt_optim}
% \begin{tabular}{@{} l l c c c @{}}
% \toprule
% \textbf{Model}      & \textbf{Method}        & \textbf{Hidden Success Rate (\%)} & \textbf{API Hit Rate (\%)} & \textbf{Best Round} \\
% \midrule
% GPT-4o              & Greedy Decoding        & --                                & --                          & --                  \\
% GPT-4.1             & Greedy Decoding        & --                                & --                          & --                  \\
% GPT-4.1-mini        & Greedy Decoding        & --                                & --                          & --                  \\
% GPT-4.1-nano        & Greedy Decoding        & --                                & --                          & --                  \\
% \bottomrule
% \end{tabular}
% \end{table}
\begin{table}[ht]
\centering

  \resizebox{\columnwidth}{!}{%
\begin{tabular}{@{} l c c c @{}}
\toprule
\textbf{Model}      & \textbf{Best Round} & \textbf{Success Rate (\%)} & \textbf{$\Delta$ (\%)} \\
\midrule
GPT-4.1-mini        & 1  & 42.1\tiny$\pm$2.7  & –2.1                \\
GPT-4.1-nano        & 3  & 37.5\tiny$\pm$2.7  & \textbf{+3.7}                \\
GPT-4.1             & 1  & \textbf{50.0}\tiny$\pm$2.8  & +1.5                \\
GPT-4o              & 0  & 49.1\tiny$\pm$2.8  &  0.0                \\
\bottomrule
\end{tabular}
}
\caption{\textbf{Automatic System Prompt Optimization results}. The prompt was optimized for at most 5 rounds using the method described in~\cite{ye2025promptalchemyautomaticprompt}, with early stopping if the improvement over previous round is less than 1.5\%. We used GPT-4.1 as the mutation model and a random fixed 20\% subset of the dataset for the optimization process. For the initial prompt, we use the same system prompt that we had used for our Greedy Decoding experiments, as given in Figure~\ref{fig:greedy-prompts}. 
We report the delta of the hidden test success rate, in comparison to the Greedy Decoding baseline. The results demonstrate the limited utility of further optimizing the prompts we had used in our experiments.}
\label{tab:round-prompt-optim}
\end{table}

\begin{table*}[htbp]
\centering
\label{tab:rag_performance}
\resizebox{\linewidth}{!}{%
\begin{tabular}{@{}lccccccc@{}}
\toprule
\textbf{Model} & \multicolumn{2}{c}{$k=1$} & \multicolumn{5}{c}{$k=3$} \\
\cmidrule(lr){2-3} \cmidrule(lr){4-8} 
& \shortstack[c]{Success\\Rate (\%)} & \shortstack[c]{API Hit\\Rate (\%)} & \shortstack[c]{Success\\Rate (\%)} & \shortstack[c]{API Hit\\Rate (\%)} & \shortstack[c]{Precision\\(\%)} & \shortstack[c]{Recall\\(\%)} & MRR \\
\midrule
\addlinespace[0.5em]
\multicolumn{8}{@{}l}{\textbf{Open-Weights Models}} \\
\midrule
CommandA & 43.6\tiny$\pm$2.7 & 43.9\tiny$\pm$2.7 & \textbf{48.2}\tiny$\pm$2.8 & 45.4\tiny$\pm$2.7 & \textbf{41.9}\tiny$\pm$2.7 & \textbf{50.7}\tiny$\pm$2.8 & \textbf{0.63}\tiny$\pm$0.03 \\
CommandR 7B & 23.2\tiny$\pm$2.3 & 36.3\tiny$\pm$2.7 & 23.2\tiny$\pm$2.3 & 35.6\tiny$\pm$2.6 & 41.6\tiny$\pm$2.7 & 50.4\tiny$\pm$2.8 & 0.62\tiny$\pm$0.03 \\
Deepseek R1 & \textbf{50.9}\tiny$\pm$2.8 & 44.8\tiny$\pm$2.7 & 51.2\tiny$\pm$2.8 & \textbf{47.9}\tiny$\pm$2.8 & 41.5\tiny$\pm$2.7 & 50.1\tiny$\pm$2.8 & 0.62\tiny$\pm$0.03 \\
Reka Flash-3 & 8.5\tiny$\pm$1.5 & 34.5\tiny$\pm$2.6 & 11.6\tiny$\pm$1.8 & 31.9\tiny$\pm$2.6 & 29.9\tiny$\pm$2.5 & 39.6\tiny$\pm$2.8 & 0.47\tiny$\pm$0.03 \\
Jamba 1.6 Mini & 18.0\tiny$\pm$2.1 & 35.4\tiny$\pm$2.6 & 29.3\tiny$\pm$2.5 & 40.4\tiny$\pm$2.7 & 41.6\tiny$\pm$2.7 & 50.1\tiny$\pm$2.8 & 0.62\tiny$\pm$0.03 \\
OpenHands LM 32B v0.1 & 34.8\tiny$\pm$2.6 & 41.0\tiny$\pm$2.7 & 28.9\tiny$\pm$2.5 & 36.5\tiny$\pm$2.7 & 25.9\tiny$\pm$2.4 & 33.7\tiny$\pm$2.7 & 0.42\tiny$\pm$0.03 \\
Llama 4 Scout & 38.7\tiny$\pm$2.7 & \textbf{45.1}\tiny$\pm$2.7 & 39.3\tiny$\pm$2.7 & 43.6\tiny$\pm$2.7 & 41.3\tiny$\pm$2.7 & 50.4\tiny$\pm$2.8 & 0.62\tiny$\pm$0.03 \\
\midrule
\addlinespace[0.5em]
\multicolumn{8}{@{}l}{\textbf{Enterprise Models}} \\
\midrule
Arcee CoderL & 46.3\tiny$\pm$2.8 & 47.3\tiny$\pm$2.8 & 36.6\tiny$\pm$2.7 & 40.4\tiny$\pm$2.7 & 31.1\tiny$\pm$2.6 & 41.0\tiny$\pm$2.8 & 0.49\tiny$\pm$0.03 \\
Claude 3.5 Haiku & 43.6\tiny$\pm$2.7 & 47.9\tiny$\pm$2.8 & 43.0\tiny$\pm$2.7 & 47.5\tiny$\pm$2.8 & 41.9\tiny$\pm$2.7 & 50.7\tiny$\pm$2.8 & 0.62\tiny$\pm$0.03 \\
Claude 3.5 Sonnet & 8.5\tiny$\pm$1.5 & 18.6\tiny$\pm$2.1 & 49.4\tiny$\pm$2.8 & 51.5\tiny$\pm$2.8 & 41.9\tiny$\pm$2.7 & 50.7\tiny$\pm$2.8 & 0.62\tiny$\pm$0.03 \\
Codestral & 44.2\tiny$\pm$2.7 & 47.3\tiny$\pm$2.8 & 46.0\tiny$\pm$2.8 & 48.5\tiny$\pm$2.8 & 41.9\tiny$\pm$2.7 & 50.7\tiny$\pm$2.8 & 0.62\tiny$\pm$0.03 \\
CommandR+ & 32.0\tiny$\pm$2.6 & 43.0\tiny$\pm$2.7 & 36.6\tiny$\pm$2.7 & 41.9\tiny$\pm$2.7 & 41.6\tiny$\pm$2.7 & 50.4\tiny$\pm$2.8 & 0.62\tiny$\pm$0.03 \\
Gemini 2.5 Flash & \textbf{54.3}\tiny$\pm$2.8 & \textbf{50.5}\tiny$\pm$2.8 & \textbf{55.2}\tiny$\pm$2.8 & \textbf{51.2}\tiny$\pm$2.8 & \textbf{41.9}\tiny$\pm$2.7 & \textbf{50.7}\tiny$\pm$2.8 & \textbf{0.62}\tiny$\pm$0.03 \\
GPT-4.1-mini & 46.9\tiny$\pm$2.8 & 50.0\tiny$\pm$2.8 & 48.8\tiny$\pm$2.8 & 50.0\tiny$\pm$2.8 & 41.3\tiny$\pm$2.7 & 50.4\tiny$\pm$2.8 & 0.62\tiny$\pm$0.03 \\
GPT-4.1-nano & 38.1\tiny$\pm$2.7 & 45.1\tiny$\pm$2.7 & 37.8\tiny$\pm$2.7 & 45.0\tiny$\pm$2.7 & 41.3\tiny$\pm$2.7 & 50.4\tiny$\pm$2.8 & 0.62\tiny$\pm$0.03 \\
GPT-4o-mini & 41.5\tiny$\pm$2.8 & 45.4\tiny$\pm$2.7 & 43.3\tiny$\pm$2.8 & 46.8\tiny$\pm$2.8 & 41.0\tiny$\pm$2.7 & 50.1\tiny$\pm$2.8 & 0.62\tiny$\pm$0.03 \\
GPT-4o & 48.2\tiny$\pm$2.8 & 47.0\tiny$\pm$2.7 & 52.1\tiny$\pm$2.8 & 49.4\tiny$\pm$2.8 & 40.6\tiny$\pm$2.7 & 49.5\tiny$\pm$2.8 & 0.61\tiny$\pm$0.03 \\
Inflection 3 Productivity & 24.7\tiny$\pm$2.8 & 42.0\tiny$\pm$2.6 & 21.9\tiny$\pm$2.7 & 44.2\tiny$\pm$2.7 & 41.9\tiny$\pm$2.7 & 50.7\tiny$\pm$2.8 & 0.62\tiny$\pm$0.03 \\
LFM 40B MoE & 30.8\tiny$\pm$2.7 & 38.3\tiny$\pm$2.7 & 20.7\tiny$\pm$2.7 & 34.0\tiny$\pm$2.7 & 33.8\tiny$\pm$2.7 & 44.8\tiny$\pm$2.8 & 0.53\tiny$\pm$0.03 \\
\bottomrule
\end{tabular}%
}
\caption{RAG performance of additional models when retrieving $k=1$ and $k=3$ most relevant documents. Precision is shown only for $k=3$ as it is equivalent to Recall in the $k=1$ case. This table shows that retrieving three documents is better in almost all cases than retrieving a single document, despite the incurred false positives that arise due to most of the examples having less than three relevant documents. }
\label{tab:rag_full}
\end{table*}

\clearpage
\begin{table}[ht]
\centering
\resizebox{\columnwidth}{!}{%
\begin{tabular}{@{} l l l c c @{}}
\toprule
\textbf{Assistant} & \textbf{Model} & \textbf{Linter} & \textbf{\shortstack[l]{Pylint\\Score $\uparrow$}} & \textbf{\shortstack[l]{Success\\Rate (\%)} }\\
\midrule
\multirow{3}{*}{Cline (IDE)}   & \multirow{3}{*}{GPT-4.1}            & N/A              & 1.06 & 54.6\tiny$\pm$2.8 \\
                               &                                     & Black + Isort    & 1.69 & 54.6\tiny$\pm$2.8 \\
                               &                                     & Ruff             & \textbf{2.64} & 54.6\tiny$\pm$2.8 \\
\midrule
\multirow{3}{*}{Goose (CLI)}   & \multirow{3}{*}{GPT-4o}             & N/A              & 0.53 & 36.3\tiny$\pm$2.7 \\
                               &                                     & Black + Isort    & 1.82 & 36.3\tiny$\pm$2.7 \\
                               &                                     & Ruff             & \textbf{2.92} & 36.3\tiny$\pm$2.7 \\
\midrule
\multirow{3}{*}{\shortstack[l]{Claude\\Code (CLI)}}  & \multirow{3}{*}{\shortstack[l]{Claude\\3.7 Sonnet}}  & N/A              & 0.00 & 48.8\tiny$\pm$2.8 \\
                               &                                     & Black + Isort    & 1.92 & 48.8\tiny$\pm$2.8 \\
                               &                                     & Ruff             & \textbf{2.60} & 48.8\tiny$\pm$2.8 \\

\bottomrule
\end{tabular}
}
\caption{\textbf{Static Analysis and Auto-linting/Formatting.} Pylint\tablefootnote{https://pylint.pycqa.org/en/latest/index.html} scores are averaged across code samples and are scored out of 10. The success rate numbers presented are the same as in Table~\ref{tab:cli-ide-assistants} wherein Goose has no access to problem statement while Cline and Claude are provided with the same. We observe that the original generated solutions via coding assistants do not meet minimum quality standard requirements, however when improved via auto-linters like Black\tablefootnote{https://black.readthedocs.io/en/stable/}, ISort\tablefootnote{https://pycqa.github.io/isort/} and Ruff\tablefootnote{https://docs.astral.sh/ruff/}, their code quality improves but with no impact to the success rate. This demonstrates that there are no confounding errors like indentation issues, unused imports and other formatting issues influencing our evaluation results observed. \textsc{Note}: For Ruff formatting, we used the already formatted/ linted solutions via Black and ISort.}
\label{tab:static}
\end{table}

\begin{table*}[ht!]
\centering
\resizebox{\linewidth}{!}{%
\begin{tabular}{@{} l l l p{4.5cm} p{4cm} p{5cm} @{}}
\toprule
\textbf{Benchmark} & \textbf{Language} & \textbf{Evaluation Method} & \textbf{Core Task} & \textbf{Source of Changes} & \textbf{Key Differentiator from \GitChameleon{}} \\
\midrule
\rowcolor{gray!10}
\textbf{\GitChameleon{}} & Python & Execution-Based & \textbf{Generation for a static version:} Writes new code for a specific, often older, library version. & Real, documented historical breaking changes. & \textit{(Baseline for comparison)} \\
\addlinespace
CodeUpdateEval & Python & Execution-Based & \textbf{Code Updating}: Modifies existing code to work with a newer library version. & Real-world software update commits. & Focuses on migrating code \textbf{forward} to a newer version, not generating for a static one. \\
\addlinespace
JavaVersionGenBench & Java & Execution-Based & \textbf{Code Updating}: Modifies existing Java code to handle version updates. & Real-world Java projects. & Focuses on the \textbf{Java} ecosystem and its specific language/tooling challenges. \\
\addlinespace
LLM-Deprecated-APl & Python & Non-Executable  & \textbf{Deprecation Fixing}: Identifies and replaces specific deprecated API calls. & A curated list of deprecated APIs. & Uses a \textbf{non-executable} evaluation method and has a narrow scope focused only on API deprecation. \\
\addlinespace
LibEvolutionEval & Python & Non-Executable  & \textbf{Code Completion}: Fills in a missing part of a code snippet based on context. & API documentation and release notes. & Is a \textbf{completion-based} task that does not test functional correctness through execution. \\
\addlinespace
RustEvo2 & Rust & Execution-Based & \textbf{Code Repair}: Fixes existing code that fails to compile after a dependency update. & Real breaking changes from Rust libraries ("crates"). & Focuses on the \textbf{Rust} ecosystem and a reactive, compiler-error-driven repair task. \\
\addlinespace
CODEMENV & Python & Execution-Based & \textbf{Environment Compatibility:} Generates code that is compatible with a complex environment specification. & A broad set of environment configurations. & Has a broader focus on overall environment compatibility, not specifically on historical breaking changes. \\
\bottomrule
\end{tabular}%
}
\caption{Detailed comparison of \GitChameleon{} with related benchmarks across several key dimensions, highlighting differences in evaluation methodology, core task, and primary programming language.}
\label{tab:benchmark-comparison-appendix}
\end{table*}
\vspace{-3cm}
\section{Related Work}
\vspace{-1cm}
\subsection{Code Evolution Datasets}
\vspace{-1cm}
While the main text provides a high-level overview of the most similar benchmarks, this section offers a more detailed differentiation between \GitChameleon{} and other relevant works. We categorize these benchmarks based on several key dimensions, including their evaluation method (execution-based vs. non-executable) and, most importantly, their core \textbf{task format (instruction-based generation vs. completion- or repair-based tasks)}. This distinction is critical as it tests different capabilities of language models.

\subsubsection{Task Format: Instruction-Based Generation}
\GitChameleon{} is fundamentally an \textbf{instruction-based} benchmark. For each problem, the model is given a natural language "Problem Statement" and starter code. The core challenge is to comprehend the user's intent and generate a new, functionally correct solution that adheres to specific version constraints. This tests a model's ability to translate human requirements into code.

\subsubsection{Task Format: Code Update, Repair, and Completion}
In contrast, many other benchmarks focus on tasks where the primary input is existing code, not a natural language instruction. The model's goal is to modify, repair, or complete a given code snippet.

\paragraph{Code Update and Repair Benchmarks}
A significant body of work evaluates a model's ability to modify or repair existing code. 
\begin{itemize}
    \item \textbf{CodeUpdateEval}~\citep{liu2024automaticallyrecommendcodeupdates} and \textbf{JavaVersionGenBench}~\citep{ciniselli2024generalizabilitydeeplearningbasedcode} are code modification benchmarks for Python and Java, respectively. They provide a model with a working piece of code and require it to be updated to a newer library version. 
    \item \textbf{RustEvo2}~\citep{liang2025rustevo2evolvingbenchmarkapi} is a code repair benchmark for Rust. It provides a model with code that is broken due to a dependency update and asks it to generate a fix based on compiler errors. 
\end{itemize}
These tasks are distinct from \GitChameleon{}'s, as they test a reactive, corrective capability rather than the proactive generation of new code from a specification.

\paragraph{Completion-Based and Non-Executable Benchmarks}
Another category of benchmarks uses non-executable metrics or focuses on code completion.
\begin{itemize}
    \item \textbf{LibEvolutionEval}~\citep{kuhar2024libevolutionevalbenchmarkstudyversionspecific} is a non-executable benchmark structured as a "fill-in-the-middle" \textbf{completion-based task}. Its evaluation is based on textual similarity metrics (e.g., F1 score), not the functional correctness of the code.
    \item \textbf{LLM-Deprecated-APl}~\citep{wang2025llmsmeetlibraryevolution}, which we note in our introduction, focuses on replacing deprecated APIs. This is a specific type of repair task that is evaluated using non-executable string matching. 
    \item \textbf{CODEMENV}~\citep{cheng2025codemenvbenchmarkinglargelanguage} evaluates a model's ability to generate code compatible with a complex environment specification. While execution-based, its task is primarily driven by satisfying technical constraints rather than implementing a distinct, high-level natural language instruction.
\end{itemize}

For a detailed breakdown, Table~\ref{tab:benchmark-comparison-appendix} contrasts \GitChameleon{} with these related benchmarks across several key methodological dimensions.

\subsection{Specialized Frameworks and Repair Techniques}
\label{app:ext_related_work}

Recognizing the unique challenges of library evolution, researchers and practitioners are developing specialized frameworks and automated repair techniques that often combine LLMs with other methods.

\subsubsection{DepsRAG}
This framework utilizes a multi-agent system built around RAG and Knowledge Graphs specifically for reasoning about software dependencies~\cite{arxiv_depsrag_v2}. It employs distinct agents managed by an LLM: one to construct and query the dependency KG, another for web searches, and a critic agent to review and refine the generated responses, aiming for higher accuracy in complex dependency analysis tasks.

\subsubsection{Dr.Fix}
This tool represents a family of approaches using LLMs, often combined with program analysis and RAG, for automated program repair. It focuses on fixing API misuse in LLM-generated code based on the taxonomy of misuse types. It employs a detect-reason-fix pipeline and demonstrates substantial improvements in repair accuracy metrics such as BLEU and Exact Match~\cite{behrang2025drfixautomaticallyfixingdata}.

\subsubsection{ReplaceAPI / InsertPrompt}
These are lightweight, targeted techniques designed specifically to mitigate the use of deprecated APIs in LLM-based code completion. ReplaceAPI performs a direct, post-generation substitution of known deprecated API calls with their replacements, achieving high fix rates in evaluations~\cite{wang2025llmsmeetlibraryevolution}. InsertPrompt modifies the input prompt to discourage the generation of deprecated APIs in the first place. They serve as valuable baseline approaches for this specific problem~\cite{llmlibeval}.

\subsubsection{Conclusion}
These works indicate a trend towards hybrid and agentic systems, moving beyond single LLM calls to more sophisticated architectures that integrate LLMs with other methods for handling library evolution. \GitChameleon{} serves as an essential resource for evaluating such systems.

In the subsequent sections we present qualitative sample analyses and model generation differences. 

% Define Python style for listings
% Literate programming feature to replace _ with \_ for display
\lstdefinestyle{python}{
    language=Python,
    basicstyle=\ttfamily\footnotesize,
    keywordstyle=\color{blue}\bfseries,
    commentstyle=\color{green!40!black}\itshape,
    stringstyle=\color{purple},
    showstringspaces=false,
    breaklines=true,
    frame=tb, % Adds a frame at the top and bottom
    captionpos=b, % Puts caption below the listing
    literate=*{_}{\_}1 % Crucially escapes underscores
}

% Define style for traceback
\lstdefinestyle{traceback}{
    basicstyle=\ttfamily\scriptsize, % Smaller font for traceback
    breaklines=true,
    frame=tb,
    captionpos=b,
    literate=*{_}{\_}1 % Escape underscores in traceback
             {.}{.}1 % Ensure periods are treated as literals
             {>}{\textgreater}1 % Escape >
             {<}{\textless}1 % Escape <
}

\section{Case Study: Code Assistant Failure With Search}

Here we present an example where a coding assistant (Zencoder) fails at solving a problem in our benchmark even when it can search the internet for the correct API call. The task required using the \texttt{torch.special} module or equivalent \texttt{scipy.special} to calculate the the modified Bessel function of the first kind of order 0. This can be computed using \texttt{torch.special.i0} or \texttt{scipy.special.i0} if the version pre-dates the introduction of \texttt{torch.special}.

\subsection{Inputs}
\textbf{Problem Statement:}
Calculate the modified Bessel function of the first kind, order 0 using pytorch's special functions if available in this version, otherwise you may use another library.

\textbf{Environment}
\begin{itemize}
    \item Python Version: 3.7
    \item Pytorch Version: 1.10
\end{itemize}

\textbf{Starting Code}
\begin{lstlisting}[style=python, caption={Initial Python function signature provided.}, label={lst:starting_code_i0}]
import torch
def bessel_i0(input_tensor: torch.Tensor) -> torch.Tensor:
\end{lstlisting}

\subsection{Model Attempt and Failure}
The model's attempt involved manually implementing the modified Bessel function, handling small and large inputs as separate cases.

\textbf{Assistant Trace}
\begin{lstlisting}[style=python, caption={The assistant's trace while designing the solution.}, label={lst:model1_code_i0}]
[truncated]
**Web search**
Done
Let me check if the torch.special module with Bessel functions is specifically available in PyTorch 1.10.0.


**Web search**
Done
The search results dont specifically confirm if torch.special.i0 is available in PyTorch 1.10.0. Let me search more specifically.


**Web search**
Done
Since Im not getting clear information about the availability of torch.special.i0 in PyTorch 1.10.0, Ill implement the function manually using the mathematical definition of the modified Bessel function of the first kind of order 0.
Now, lets implement the solution:
\end{lstlisting}

\textbf{Proposed Solution}
\begin{lstlisting}[style=python, caption={The solution proposed by the assistant.}, label={lst:prop_sol}]
def bessel_i0(input_tensor: torch.Tensor) -> torch.Tensor:    
    result = torch.zeros_like(input_tensor)
    abs_x = torch.abs(input_tensor)
    
    # For small values (|x| < 15), use the series expansion
    small_mask = abs_x < 15.0
    small_x = abs_x[small_mask]
    
    if small_x.numel() > 0:
        small_result = torch.ones_like(small_x)
        term = torch.ones_like(small_x)
        x_squared_over_4 = (small_x / 2.0) ** 2
        
        # Sum the series until convergence or max iterations
        for k in range(1, 30):  # 30 terms should be enough for good precision
            term = term * x_squared_over_4 / (k ** 2)
            small_result = small_result + term
            
            # Check for convergence
            if torch.all(term < 1e-10 * small_result):
                break
                
        result[small_mask] = small_result
    
    # For large values (|x| $\geq$ 15), use the asymptotic approximation
    large_mask = ~small_mask
    large_x = abs_x[large_mask]
    
    if large_x.numel() > 0:
        large_result = torch.exp(large_x) / torch.sqrt(2 * math.pi * large_x)
        result[large_mask] = large_result
    
    return result}
\end{lstlisting}

\textbf{Error Traceback:}
\begin{lstlisting}[style=python, caption={Error trace for the implementation of \texttt{bessel\_i0}.}, label={lst:error1_code_i0}]
TestBesselI0.test_large_values ________________________
def test_large_values(self):
    """Test bessel_i0 with large positive and negative values."""
    input_tensor = torch.tensor([10.0, -10.0, 20.0, -20.0], dtype=torch.float32)
    result = bessel_i0(input_tensor)
    expected = torch.from_numpy(scipy_i0(input_tensor.numpy()))
    
>       torch.testing.assert_close(result, expected)
E       AssertionError: Tensor-likes are not close!
\end{lstlisting}

\textbf{Explanation:}
The failure in the manual implementation arises because the asymptotic approximation  
\[
I_0(x)\approx\frac{e^x}{\sqrt{2\pi x}}
\]
has only \(O(1/x)\) accuracy and omits higher-order correction terms, so for \(x\gtrsim20\) its relative error exceeds the test tolerance of \(10^{-5}\). To fix this, either use the built-in  
\(\mathtt{torch.special.i0}(x)\)  
or include at least the first correction term in the expansion:  
\[
I_0(x)\approx\frac{e^x}{\sqrt{2\pi x}}\Bigl(1 - \tfrac{1}{8x}\Bigr),
\]  
which reduces the large-\(x\) error below the required threshold.  

\lstdefinestyle{python}{
    language=Python,
    basicstyle=\ttfamily\footnotesize,
    keywordstyle=\color{blue}\bfseries,
    commentstyle=\color{green!40!black}\itshape,
    stringstyle=\color{purple},
    showstringspaces=false,
    breaklines=true,
    frame=tb, % Adds a frame at the top and bottom
    captionpos=b, % Puts caption below the listing
    literate=*{_}{\_}1 % Crucially escapes underscores
}

% Define style for traceback
\lstdefinestyle{traceback}{
    basicstyle=\ttfamily\scriptsize, % Smaller font for traceback
    breaklines=true,
    frame=tb,
    captionpos=b,
    literate=*{_}{\_}1 % Escape underscores in traceback
             {.}{.}1 % Ensure periods are treated as literals
             {>}{\textgreater}1 % Escape >
             {<}{\textless}1 % Escape <
}

\section{Case Study: Self-Debugging in Batched Matrix Exponential Computation}

This appendix details an instance where self-debugging helped rectify an initially incorrect solution from GPT-4.5 for computing the matrix exponential of batched matrices. The task required leveraging the \texttt{scipy.linalg.expm} function while ensuring compatibility with batched inputs.

\subsection{Inputs}
\textbf{Problem Statement:} Complete the function that computes the matrix exponential of batched matrices, non specified parameters should use the default value

\textbf{Environment:}
\begin{itemize}
    \item Python Version: 3.10
    \item SciPy Version: 1.8.1
\end{itemize}

\textbf{Starting Code}
\begin{lstlisting}[style=python, caption={Initial Python function signature provided.}, label={lst:initial_code_scipy}]
from scipy import linalg
import numpy as np

def compute_matrix_exponential(A: np.ndarray) -> np.ndarray:
    # Model needs to complete this function
    pass
\end{lstlisting}

\subsection{First Model Attempt and Failure}
The model's first attempt involved reshaping the input array and applying \texttt{scipy.linalg.expm} directly.

\textbf{Initial Model Output}
\begin{lstlisting}[style=python, caption={First attempt by the model to define \texttt{compute\_matrix\_exponential}.}, label={lst:model_output_1_scipy}]
from scipy import llinalg
import numpy as np

def compute_matrix_exponential(A: np.ndarray) -> np.ndarray:
    return linalg.expm(A)
\end{lstlisting}

\textbf{Model's Explanation (Initial):}
\textit{The function directly returns linalg.expm(A) because SciPy's expm computes the matrix exponential for a given matrix. Given that A is a batch of matrices, SciPy should naturally handle this batch input and apply the exponential operation to each matrix independently, producing the correct batched result.}

\textbf{Result:} \texttt{Failure}

\textbf{Error Traceback}
\textbf{Failure:} Calling \texttt{linalg.expm(A)} on a 3D array $A\in\mathbb R^{b\times n\times n}$ makes SciPy treat it as one “matrix,” and raises
  \[
    \texttt{ValueError: expected a square matrix}
  \]
  because it only supports 2D inputs.

\begin{lstlisting}[style=traceback, caption={Error traceback from the initial model output.}, label={lst:traceback_scipy}]
Traceback (most recent call last):
  File "/app/repo/eval_venvs/gcham_venv_111/lib/python3.10/site-packages/scipy/sparse/linalg/_matfuncs.py", line 601, in _expm
    raise ValueError('expected a square matrix')
ValueError: expected a square matrix
\end{lstlisting}

\subsection{Self-Debugging Process and Correction}
The self-debugging process recognized that simply calling linalg.expm was insufficient. Instead, \texttt{linalg.expm} must be applied individually to each 2D matrix within the batch. Additionally, a crucial pre-condition is that each matrix within the batch must be square.

\textbf{Second Model Attempt (Post-Debugging)}
The model regenerated the solution, incorporating input validation and iterating over the individual matrices in the batch.

\textbf{Regenerated Model Output}
\begin{lstlisting}[style=python, caption={Corrected \texttt{compute\_matrix\_exponential} after self-debugging.}, label={lst:model_output_2_scipy}]
from scipy import linalg
import numpy as np

def compute_matrix_exponential(A: np.ndarray) -> np.ndarray:
    return np.stack([linalg.expm(A[i]) for i in range(A.shape[0]) if A[i].shape[0] == A[i].shape[1]], axis=0)
\end{lstlisting}

\textbf{Model's Explanation (Regenerated):}
\textit{The corrected solution applies linalg.expm separately to each 2D square matrix in the batch using a list comprehension, then re-stacks the results into a 3D array. This ensures each call to expm receives a valid input, resolving the error.}

\textbf{Result:} \texttt{Success}

\subsection{Analysis of the Correction}
The regenerated output effectively addressed the shortcomings of the initial attempt:
\begin{itemize}
     \item \textbf{Fix:} Apply \texttt{expm} to each 2D slice $A_i\in\mathbb R^{n\times n}$ and re-stack,
  ensuring every call gets a proper square matrix.
\end{itemize}
This revised approach correctly computes the matrix exponential for each matrix in the batch and handles the data dimensions appropriately, thus resolving the \texttt{ValueError} and fulfilling the problem requirements.

\section{Qualitative Analysis}
This section presents additional qualitative studies of attempted solutions in the Greedy Decoding and the Chain-of-Thought settings.

\lstdefinestyle{python}{
    language=Python,
    basicstyle=\ttfamily\footnotesize,
    keywordstyle=\color{blue}\bfseries,
    commentstyle=\color{green!40!black}\itshape,
    stringstyle=\color{purple},
    showstringspaces=false,
    breaklines=true,
    frame=tb, % Adds a frame at the top and bottom
    captionpos=b, % Puts caption below the listing
    literate=*{_}{\_}1 % Crucially escapes underscores
}

% Define style for traceback
\lstdefinestyle{traceback}{
    basicstyle=\ttfamily\scriptsize, % Smaller font for traceback
    breaklines=true,
    frame=tb,
    captionpos=b,
    literate=*{_}{\_}1 % Escape underscores in traceback
             {.}{.}1 % Ensure periods are treated as literals
             {>}{\textgreater}1 % Escape >
             {<}{\textless}1 % Escape <
}

\subsection{Greedy Decoding}

\subsubsection{Example 1: (PyTorch)}

\noindent\textbf{Problem Statement:} Calculate the logarithm of the cumulative distribution function of the standard normal distribution using available functions. If not available in PyTorch, use another library.

\noindent\textbf{Environment:}
\begin{itemize}
    \item Python Version: 3.7
    \item Torch Version: 1.9.0
    \item Extra Dependencies: scipy==1.7.3, numpy==1.21.6
\end{itemize}

\noindent\textbf{Starting Code}
\begin{lstlisting}[style=python, caption={Greedy Example 1 Starting Code}, label={lst:GreedyE1_SC}]
import torch
def log_ndtr(input_tensor: torch.Tensor) -> torch.Tensor:
    pass
\end{lstlisting}

\noindent\textbf{Solution}
\begin{lstlisting}[style=python, caption={Greedy Example 1 Solution}, label={lst:GreedyE1_SO}]

import numpy as np
    from scipy.stats import norm
    output = torch.from_numpy(norm.logcdf(input_tensor.numpy()))
    return output

\end{lstlisting}

\noindent\textbf{Claude 3.7 Failure Output}
\begin{lstlisting}[style=python, caption={Claude 3.7 Output}, label={lst:GreedyE1_Out_cla}]
import torch
import math
from torch.distributions import Normal

def log_ndtr(input_tensor: torch.Tensor) -> torch.Tensor:
    normal = Normal(0, 1)
    return normal.cdf(input_tensor).log()
\end{lstlisting}

\noindent\textbf{Gemini 2.5 Pro Failure Output}
\begin{lstlisting}[style=python, caption={Gemini 2.5 Pro Output}, label={lst:GreedyE1_Out_gemi2}]
import torch
from torch.distributions import Normal

def log_ndtr(input_tensor: torch.Tensor) -> torch.Tensor:
  loc = torch.tensor(0.0, device=input_tensor.device, dtype=input_tensor.dtype)
  scale = torch.tensor(1.0, device=input_tensor.device, dtype=input_tensor.dtype)

  standard_normal = Normal(loc=loc, scale=scale)

  log_cdf_values = standard_normal.log_cdf(input_tensor)

  return log_cdf_values
\end{lstlisting}

\noindent\textbf{Analysis of Model Outputs}

Both Claude 3.7 and Gemini 2.5 Pro incorrectly attempt to compute the log-CDF of the standard normal distribution using PyTorch's \texttt{Normal} distribution. Claude 3.7 computes \texttt{normal.cdf(input\_tensor).log()}, which is numerically unstable for large negative inputs since the CDF approaches zero and its logarithm can underflow to $-\infty$, leading to invalid results. Gemini 2.5 Pro mistakenly calls a non-existent \texttt{log\_cdf} method on the PyTorch \texttt{Normal} object, causing an immediate \texttt{AttributeError} in PyTorch 1.9. This reflects a confusion between PyTorch and other libraries (such as SciPy or TensorFlow Probability) that do provide this method. In summary, neither model provides a robust or correct implementation; the only reliable approach in this environment is to use \texttt{scipy.stats.norm.logcdf} as shown in the reference solution, or to implement a numerically stable approximation in PyTorch when SciPy is unavailable.

\subsubsection{Greedy Example 2 (SciPy)}

\noindent\textbf{Problem Statement:} complete the following function that check if all the batch of matrices are invertible, using numpy 1.25.1.

\noindent\textbf{Environment:}
\begin{itemize}
    \item Python Version: 3.10
    \item Scipy Version: 1.11.1
\end{itemize}

\noindent\textbf{Starting Code}
\begin{lstlisting}[style=python, caption={Greedy Example 3 Starting Code}, label={lst:GreedyE3_SC}]
import warnings
from scipy.linalg import det
import numpy as np
warnings.filterwarnings('error')

def check_invertibility(matrices: np.ndarray) -> np.bool_:
    pass
\end{lstlisting}

\noindent\textbf{Solution}
\begin{lstlisting}[style=python, caption={Greedy Example 3 Solution}, label={lst:GreedyE3_SO}]
    return np.all(det(matrices))
\end{lstlisting}

\noindent\textbf{Gemini 2.5 Correct Output}
\begin{lstlisting}[style=python, caption={Gemini 2.5 Pro Output}, label={lst:GreedyE3_out_gem25}]
import warnings
from scipy.linalg import det
import numpy as np
warnings.filterwarnings('error')

def check_invertibility(matrices: np.ndarray) -> np.bool_:
    try:
        # Check if matrices is a batch of matrices
        if matrices.ndim < 3:
            # If it's a single matrix, add a batch dimension
            matrices = np.expand_dims(matrices, axis=0)
        
        # Check if all matrices are square
        batch_size, rows, cols = matrices.shape
        if rows != cols:
            return np.bool_(False)
        
        # Calculate determinants for all matrices in the batch
        determinants = np.array([det(matrix) for matrix in matrices])
        
        # A matrix is invertible if its determinant is not zero
        # Check if all determinants are non-zero
        return np.bool_(np.all(determinants != 0))
    
    except Exception:
        # If any error occurs during computation, return False
        return np.bool_(False)
\end{lstlisting}

\noindent\textbf{GPT 4.1 Failure Output}
\begin{lstlisting}[style=python, caption={GPT 4.1 Output}, label={lst:GreedyE3_out_41}]
import warnings
from scipy.linalg import det
import numpy as np
warnings.filterwarnings('error')

def check_invertibility(matrices: np.ndarray) -> np.bool_:
    for matrix in matrices:
        try:
            d = det(matrix)
            if np.isclose(d, 0):
                return np.bool_(False)
        except Warning:
            return np.bool_(False)
    return np.bool_(True)
\end{lstlisting}

\noindent\textbf{Analysis of Model Outputs}

Gemini 2.5 Pro produces a robust solution that handles both single matrices and batches, verifies that matrices are square, computes the determinant for each matrix, and correctly checks if all determinants are non-zero to determine invertibility, returning \texttt{False} on any exception to ensure safe failure. In contrast, GPT 4.1's output contains a subtle but critical error in exception handling: the code attempts to catch \texttt{Warning} as an exception, but Python treats warnings and exceptions differently, so this approach is ineffective—especially since \texttt{warnings.filterwarnings('error')} escalates warnings to exceptions that should be caught as \texttt{Exception}. As a result, GPT 4.1’s function may fail or produce unpredictable results in the presence of numerical issues or shape mismatches, despite using the correct logic for checking determinant values.

\subsubsection{Greedy Example 3 (SymPy)}

\noindent\textbf{Problem Statement:} Write a custom jacobi symbols function that compute the Jacobi symbol.

\noindent\textbf{Environment:}
\begin{itemize}
    \item Python Version: 3.9
    \item Sympy Version: 1.13
\end{itemize}

\noindent\textbf{Starting Code}
\begin{lstlisting}[style=python, caption={Greedy Example 4 Starting Code}, label={lst:GreedyE4_SC}]
import sympy

def custom_jacobi_symbols(a: int, n: int) -> int:
    pass
\end{lstlisting}

\noindent\textbf{Solution}
\begin{lstlisting}[style=python, caption={Greedy Example 4 Solution}, label={lst:GreedyE4_SO}]
    return sympy.jacobi_symbol(a, n)

\end{lstlisting}

\noindent\textbf{GPT 4.1 Correct Output}
\begin{lstlisting}[style=python, caption={GPT 4.1 Output}, label={lst:GreedyE4_out_41}]
import sympy

def custom_jacobi_symbols(a: int, n: int) -> int:
    return sympy.ntheory.jacobi_symbol(a, n)
\end{lstlisting}

\noindent\textbf{O1 Failure Output}
\begin{lstlisting}[style=python, caption={O1 Output}, label={lst:GreedyE4_out_o1}]
import sympy

def custom_jacobi_symbols(a: int, n: int) -> int:
    return sympy.ntheory.residues.jacobi_symbol(a, n)
\end{lstlisting}

\noindent\textbf{Analysis of Model Outputs}

GPT-4.1 provides the correct solution by calling \texttt{sympy.ntheory.jacobi\_symbol}, which is the proper and documented API for computing the Jacobi symbol in SymPy 1.13. In contrast, O1 incorrectly uses \texttt{sympy.ntheory.residues.jacobi\_symbol}, assuming that the function is nested in the \texttt{residues} submodule; however, in this version of SymPy, \texttt{jacobi\_symbol} is not found there, resulting in an \texttt{AttributeError} on all tests. This error highlights a lack of up-to-date API knowledge or reliance on outdated documentation.

\subsubsection{Greedy Example 4 (Flask)}
\noindent\textbf{Problem Statement:} Complete the app set-up so that, when given a batch of matrix, the json encoding compute the determinants of each matrix,  before flattening and converting the result to a list, we are using scipy 1.11.1. Do not run the app in your code.

\noindent\textbf{Environment:}
\begin{itemize}
    \item Python Version: 3.10
    \item Flask Version: 3.0.0
    \item Extra Dependencoes: scipy==1.11.1
\end{itemize}

\noindent\textbf{Starting Code}
\begin{lstlisting}[style=python, caption={Greedy Example 5 Starting Code}, label={lst:GreedyE5_SC}]
import flask
import numpy as np
from scipy import linalg

app = flask.Flask('test1')
@app.route('/data')
def data(num_list):
    return flask.jsonify({'numbers': num_list})
def eval_app(app, data_fn, num_arr):
    with app.test_request_context():
        response = data_fn(num_arr)
        return response.get_data(as_text=True)

class MyCustomJSONHandler(flask.json.provider.DefaultJSONProvider):
    def default(self, obj):
        if isinstance(obj, np.ndarray) and len(obj.shape)==3 and obj.shape[-1]==obj.shape[-2] :
            pass
\end{lstlisting}

\noindent\textbf{Solution}
\begin{lstlisting}[style=python, caption={Greedy Example 5 Solution}, label={lst:GreedyE5_SO}]
    res = linalg.det(obj)
                return res.tolist()
            return super().default(obj)
    
    app.json_provider_class = MyCustomJSONHandler
    app.json = app.json_provider_class(app)
\end{lstlisting}

\noindent\textbf{Gemini 2.5 Pro Correct Output}
\begin{lstlisting}[style=python, caption={Gemini 2.5 Pro Output}, label={lst:GreedyE5_out_gemini25}]
import flask
import numpy as np
from scipy import linalg

app = flask.Flask('test1')
@app.route('/data')
def data(num_list):
    return flask.jsonify({'numbers': num_list})
def eval_app(app, data_fn, num_arr):
    with app.test_request_context():
        response = data_fn(num_arr)
        return response.get_data(as_text=True)

class MyCustomJSONHandler(flask.json.provider.DefaultJSONProvider):
    def default(self, obj):
        if isinstance(obj, np.ndarray) and len(obj.shape)==3 and obj.shape[-1]==obj.shape[-2] : 
            # Calculate determinants for each matrix in the batch
            determinants = linalg.det(obj)
            # Convert the resulting 1D array of determinants to a list
            return determinants.tolist()
        # Fallback to the default serialization for other types
        return super().default(obj)

# Apply the custom JSON provider to the app
app.json = MyCustomJSONHandler(app)
\end{lstlisting}

\noindent\textbf{Claude 3.7 Failure Output}
\begin{lstlisting}[style=python, caption={Claude 3.7 Output}, label={lst:GreedyE5_out_claude}]
import flask
import json

app = flask.Flask('test')
@app.route('/data')
def data(num_set):
    return flask.jsonify({'numbers': num_set})

def eval(app, data_fn, num_set):
    with app.test_request_context():
        response = data_fn(num_set)
        return response.get_data(as_text=False)

def app_set_up(app: flask.Flask) -> None:
    def set_encoder(obj):
        if isinstance(obj, set):
            return sorted(list(obj))
        raise TypeError(f"Object of type {type(obj)} is not JSON serializable")
    
    app.json.encoder = lambda: json.JSONEncoder(default=set_encoder)
\end{lstlisting}

\noindent\textbf{Analysis of Model Outputs}

Gemini 2.5 Pro provides a solution that correctly subclasses \texttt{flask.json.provider.DefaultJSONProvider} and overrides the \texttt{default} method to check for batches of square matrices, compute their determinants using \texttt{scipy.linalg.det}, and serialize the result as a list, fully aligning with both the requirements and Flask 3.0.0’s JSON provider API. In contrast, Claude 3.7’s output incorrectly tries to customize JSON serialization by modifying \texttt{app.json.encoder}—a property that is not available in Flask 3.0.0—and focuses on encoding sets rather than NumPy arrays, neglecting the determinant computation altogether. Consequently, Claude’s approach fails with an \texttt{AttributeError} and does not satisfy the task requirements even if the encoder logic were correct, while Gemini’s implementation is robust and functional for the specified environment.

\subsection{Zero-Shot Chain-Of-Thought}

\subsubsection{CoT Example 1 (Torch)}

\noindent\textbf{Problem Statement:} Calculate the modified Bessel function of the first kind, order 1 using PyTorch's special functions if available in this version, otherwise you may use another library.

\noindent\textbf{Environment:}
\begin{itemize}
    \item Python Version: 3.7
    \item Torch Version: 1.9.0
    \item Extra Dependencies: scipy==1.7.3, numpy==1.21.6
\end{itemize}

\noindent\textbf{Starting Code}
\begin{lstlisting}[style=python, caption={CoT Example 1 Starting Code}, label={lst:CotE1_SC}]
import torch
def bessel_i1(input_tensor: torch.Tensor) -> torch.Tensor:
    pass
\end{lstlisting}

\noindent\textbf{Solution}
\begin{lstlisting}[style=python, caption={CoT Example 1 Solution}, label={lst:CotE1_SO}]
    import numpy as np
    from scipy.special import i1 as scipy_i1
    output = torch.from_numpy(scipy_i1(input_tensor.numpy()))
    return output
\end{lstlisting}

\noindent\textbf{Claude 3.7 Correct Output}
\begin{lstlisting}[style=python, caption={Claude 3.7 Output}, label={lst:CotE1_Out_cla}]
import torch
from scipy import special

def bessel_i1(input_tensor: torch.Tensor) -> torch.Tensor:
    # Convert PyTorch tensor to NumPy array
    input_numpy = input_tensor.cpu().numpy()
    
    # Apply the modified Bessel function of the first kind, order 1
    result_numpy = special.i1(input_numpy)
    
    # Convert back to PyTorch tensor
    result_tensor = torch.from_numpy(result_numpy).to(input_tensor.device)
    
    # Ensure the output has the same dtype as the input
    result_tensor = result_tensor.to(dtype=input_tensor.dtype)
    
    return result_tensor
\end{lstlisting}

\noindent\textbf{Gemini 2.5 Pro Failure Output}
\begin{lstlisting}[style=python, caption={Gemini 2.5 Pro Output}, label={lst:CotE1_Out_gemi25}]
import torch
def bessel_i1(input_tensor: torch.Tensor) -> torch.Tensor:
    """Calculates the modified Bessel function of the first kind, order 1.

    Args:
        input_tensor: The input tensor.

    Returns:
        A tensor containing the element-wise modified Bessel function of the first kind, order 1.
    """
    # Check if torch.special.i1 is available (it is in torch 1.9.0)
    if hasattr(torch.special, 'i1'):
        return torch.special.i1(input_tensor)
    else:
        # Fallback or error handling if needed for other versions
        # For torch 1.9.0, this else block should not be reached.
        raise NotImplementedError("torch.special.i1 is not available in this PyTorch version.")
\end{lstlisting}

\noindent\textbf{Analysis of Model Outputs}

Claude 3.7 provides a robust solution by correctly identifying that \texttt{torch.special.i1} is not available in PyTorch 1.9.0 and instead leveraging \texttt{scipy.special.i1} to compute the modified Bessel function, converting the result back to a PyTorch tensor and ensuring proper device and dtype handling; this approach is fully compatible with the given environment and passes all tests. In contrast, Gemini 2.5 Pro mistakenly assumes that \texttt{torch.special.i1} exists in PyTorch 1.9.0, checks for its presence using \texttt{hasattr}, and raises a \texttt{NotImplementedError} when unavailable; however, since this function is absent in the specified version, the code fails every test due to the raised exception.

\subsubsection{CoT Example 2 (Scikit-learn)}

\noindent\textbf{Problem Statement:} Adapt the use of `manhattan distances` to obtain a pairwise distance matrix.

\noindent\textbf{Environment:}
\begin{itemize}
    \item Python Version: 3.10
    \item Scikit-learn Version: 1.1
    \item Extra Dependencies: numpy==1.23.5
\end{itemize}

\noindent\textbf{Starting Code}
\begin{lstlisting}[style=python, caption={CoT Example 2 Starting Code}, label={lst:CotE2_SC}]
from sklearn.metrics.pairwise import manhattan_distances
import numpy as np
def get_pairwise_dist(X: np.ndarray,Y: np.ndarray) -> np.ndarray:
    distances = manhattan_distances(X, Y, sum_over_features=False)
    pass
\end{lstlisting}

\noindent\textbf{Solution}
\begin{lstlisting}[style=python, caption={CoT Example 2 Solution}, label={lst:CotE2_SO}]
    return np.sum(distances, axis=1)
\end{lstlisting}

\noindent\textbf{GPT 4.1 Correct Output}
\begin{lstlisting}[style=python, caption={GPT 4,1 Output}, label={lst:CotE2_Out_gpt41}]
from sklearn.metrics.pairwise import manhattan_distances
import numpy as np
def get_pairwise_dist(X: np.ndarray, Y: np.ndarray) -> np.ndarray:
    distances = manhattan_distances(X, Y)
    return distances
\end{lstlisting}

\noindent\textbf{Claude 3.7 Failure Output}
\begin{lstlisting}[style=python, caption={Claude 3.7 Pro Output}, label={lst:CotE2_Out_claude37}]
from sklearn.metrics.pairwise import manhattan_distances
import numpy as np
def get_pairwise_dist(X: np.ndarray,Y: np.ndarray) -> np.ndarray:
    distances = manhattan_distances(X, Y, sum_over_features=False)
    return np.sum(distances, axis=2)
\end{lstlisting}

\noindent\textbf{Analysis of Model Outputs}

In this example, GPT 4.1 provides a correct implementation by calling \texttt{manhattan\_distances(X, Y)} without the \texttt{sum\_over\_features} argument, which by default returns the desired pairwise distance matrix of shape \((n_{\text{samples}_X}, n_{\text{samples}_Y})\). In contrast, Claude 3.7 incorrectly sets \texttt{sum\_over\_features=False}, causing the output to be a 3D array of shape, and then applies \texttt{np.sum(distances, axis=2)}, which does not match the problem's intent and produces incorrect results if further summed incorrectly. The misuse of the \texttt{sum\_over\_features} parameter and the erroneous axis for reduction demonstrates a misunderstanding of the scikit-learn API and leads to a failure, while GPT 4.1's output is concise and correct for all valid input shapes.

\subsubsection{CoT Example 3 (Falcon)}

\noindent\textbf{Problem Statement:} Create a class named CustomRouter to manage your application's routes. The class should maintain an internal dictionary named routes for storing the mapping between URI templates and their associated resources. Implement an add\_route method that accepts three arguments: a URI template, a resource, and additional keyword arguments. This method should use Falcon’s routing utility—specifically, the map\_http\_methods function from the falcon.routing module—to generate a mapping of HTTP methods to resource handlers. If a 'fallback' parameter is provided in the keyword arguments, it should be passed to map\_http\_methods. The method should then store a tuple consisting of the resource and the generated method mapping in the routes dictionary, using the URI template as the key, and finally return the generated method mapping. You must not modify the provided starting code, which includes the empty CustomRouter class. Instead, complete the solution by dynamically adding the required method in the solution() function.

\noindent\textbf{Environment:}
\begin{itemize}
    \item Python Version: 3.10
    \item Falcon Version: 2.0
\end{itemize}

\noindent\textbf{Starting Code}
\begin{lstlisting}[style=python, caption={CoT Example 3 Starting Code}, label={lst:CotE3_SC}]
class CustomRouter:
    def __init__(self):
        self.routes = {}

        
def solution() -> None:
    
    def add_route(
\end{lstlisting}

\noindent\textbf{Solution}
\begin{lstlisting}[style=python, caption={CoT Example 3 Solution}, label={lst:CotE3_SO}]
self, uri_template, resource, **kwargs):
        from falcon.routing import map_http_methods
        method_map = map_http_methods(resource, kwargs.get('fallback', None))
        self.routes[uri_template] = (resource, method_map)
        return method_map
    
    CustomRouter.add_route = add_route
\end{lstlisting}

\noindent\textbf{O1 Correct Output}
\begin{lstlisting}[style=python, caption={O1 Output}, label={lst:CotE3_Out_O1}]
import falcon
from falcon.routing import map_http_methods

class CustomRouter:
    def __init__(self):
        self.routes = {}

def solution() -> None:
    def add_route(self, uri_template, resource, **kwargs):
        fallback = kwargs.pop('fallback', None)
        if fallback is not None:
            method_map = map_http_methods(resource, fallback=fallback)
        else:
            method_map = map_http_methods(resource)
        self.routes[uri_template] = (resource, method_map)
        return method_map
    CustomRouter.add_route = add_route
\end{lstlisting}

\noindent\textbf{GPT 4.1 Failure Output}
\begin{lstlisting}[style=python, caption={GPT 4.1 Output}, label={lst:CotE3_Out_gpt41}]
from falcon.routing import map_http_methods

def solution() -> None:
    def add_route(self, uri_template, resource, **kwargs):
        fallback = kwargs.pop('fallback', None)
        if fallback is not None:
            method_map = map_http_methods(resource, fallback=fallback)
        else:
            method_map = map_http_methods(resource)
        self.routes[uri_template] = (resource, method_map)
        return method_map

    CustomRouter.add_route = add_route
\end{lstlisting}

\noindent\textbf{Analysis of Model Outputs}

In this case, the O1 output provides a fully correct solution by dynamically attaching an \texttt{add\_route} method to the \texttt{CustomRouter} class, properly handling the optional \texttt{fallback} argument, and updating the \texttt{routes} dictionary according to the problem statement. The method is inserted via the \texttt{solution()} function without modifying the provided class definition, and it uses Falcon’s \texttt{map\_http\_methods} utility to construct the method mapping as required. In contrast, the GPT 4.1 output omits the explicit definition of the \texttt{CustomRouter} class in its solution, violating the requirement to use the existing starting code. Although the logic within the \texttt{solution()} function is correct, the absence of a \texttt{CustomRouter} definition in the completed module would lead to a \texttt{NameError} or otherwise prevent the expected dynamic method attachment. The critical distinction is that O1 respects all constraints including not modifying the class definition directly, while GPT 4.1 provides an incomplete module, failing to meet the initialization requirements set by the problem.

\section{Logic vs. Knowledge Retention}

\begin{figure}[htb]
  \centering
  \includegraphics[width=0.9\linewidth]{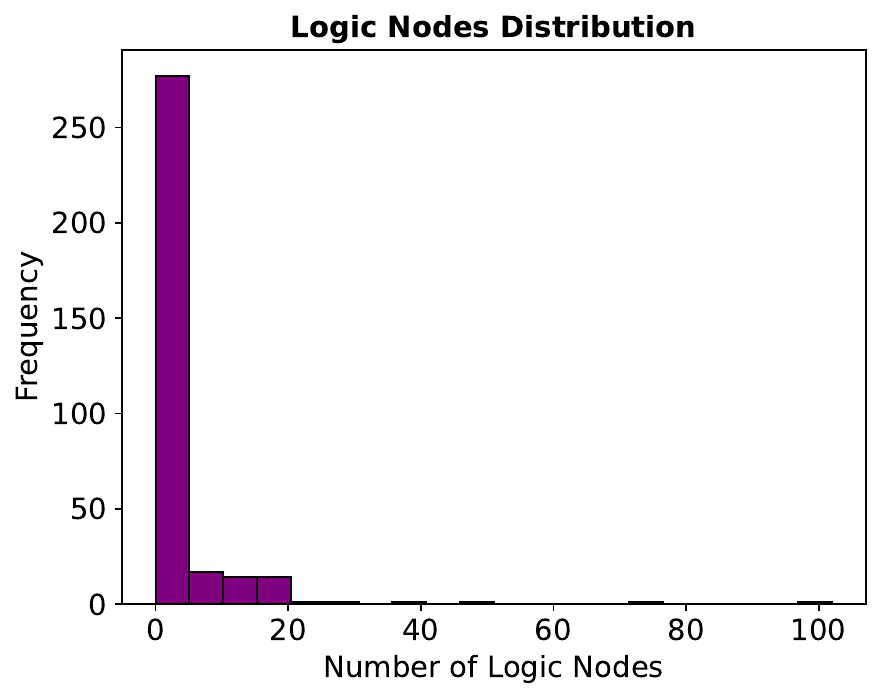}
  \caption{\textbf{Logic Nodes Distribution over samples' ground truth solutions' ASTs.} Most ground truth solutions have less than \textbf{five} logic nodes.}
  \label{fig:logic_count}
\end{figure}

The goal of our proposed benchmark, \textbf{\textcolor{violet}{GitChameleon}}, is to evaluate a model’s ability to retain version-specific knowledge—specifically, whether it can recall the functionalities associated with particular library versions it has been trained on. Notably, this capability is distinct from the ability to generate logically correct code. While we do not explicitly disentangle whether model failures on our evaluation suite stem from incorrect logic generation or incorrect API version usage, our benchmark is intentionally designed so that most problems primarily test knowledge retention rather than complex logic reasoning. For each problem in our dataset, we compute the number of logic-related nodes in the Abstract Syntax Tree (AST) of the ground-truth solution and present their distribution in Figure~\ref{fig:logic_count}. As shown, most ground-truth solutions contain fewer than \textbf{five} logic-related AST nodes. This supports our claim that the benchmark is primarily designed to assess version-specific knowledge retention rather than complex logic-based code generation. 

\begin{table}[h]
\small
\centering
\caption{Criteria for classifying AST nodes as logic-related.}
\label{tab:logic_criteria}
\begin{tabular}{@{}p{0.72\linewidth}>{\centering\arraybackslash}p{0.22\linewidth}@{}}
\toprule
\textbf{Condition} & \textbf{Classification} \\ \midrule
Calling a user-defined function & \checkmark \\
Calling built-in Python operators (e.g., \texttt{+}) & \checkmark \\
Calling a math or utility function with non-obvious purpose & \checkmark \\
Calling a library method (e.g., \texttt{torch.from\_numpy}) & \xmark \\
Composing multiple calls together & \checkmark \\
\bottomrule
\end{tabular}
\end{table}

\begin{figure*}[htb]
  \centering
  \includegraphics[width=0.9\textwidth]{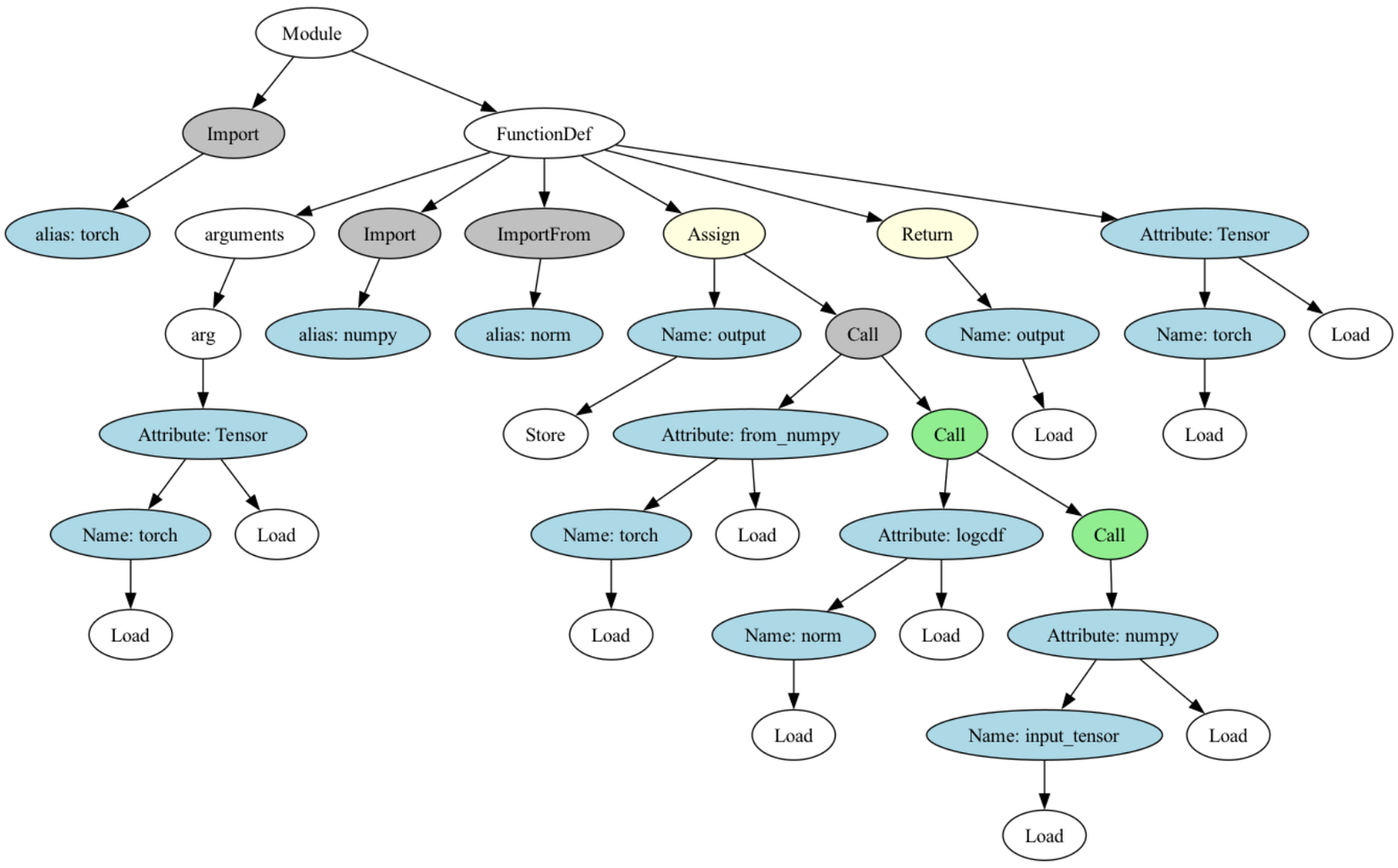}
  \caption{AST visualization for the ground-truth solution of Sample ID 0. The three color-coded \texttt{call} nodes (in grey and green) represent the logic-related components, classified under the “composing multiple calls together” category. The corresponding ground-truth code is shown in Code block~\ref{lst:GT_0} for reference.}
  \label{fig:logic_0}
\end{figure*}

\begin{figure*}[!h]
  \centering
  \includegraphics[width=0.9\textwidth]{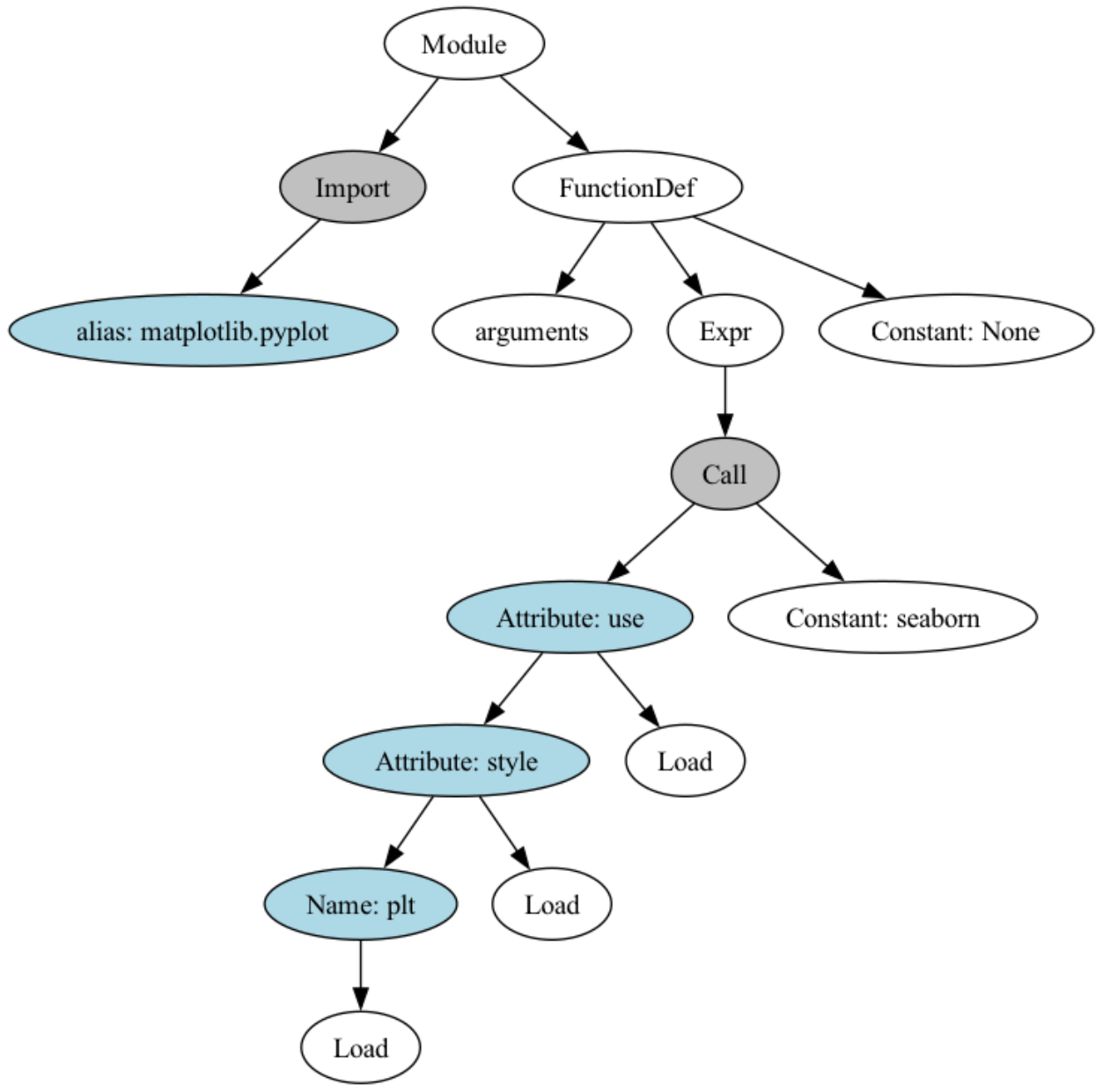}
  \caption{AST visualization for the ground-truth solution of Sample ID 329. No logic nodes are present, as the only \texttt{call} node corresponds to the “calling a library method” category. The ground-truth solution is provided for reference in Code block~\ref{lst:GT_329}.}
  \label{fig:logic_329}
\end{figure*}

The criteria for classifying AST nodes as logic-related are provided in Table~\ref{tab:logic_criteria}, and we include visualizations of the ASTs for two example ground-truth solutions for further illustration in Figures \ref{fig:logic_0} and \ref{fig:logic_329} respectively.

\begin{enumerate}
    \item Sample ID: 0, Logic Nodes: 3
    \begin{lstlisting}[style=python, caption={Sample 0 Ground Truth Solution}, label={lst:GT_0}]
import torch
def log_ndtr(input_tensor: torch.Tensor) -> torch.Tensor:
    import numpy as np
    from scipy.stats import norm
    output = torch.from_numpy(norm.logcdf(input_tensor.numpy()))
    return output
\end{lstlisting}
    \item Sample ID: 329, Logic Nodes: 0
    \begin{lstlisting}[style=python, caption={Sample 329 Ground Truth Solution}, label={lst:GT_329}]
import matplotlib.pyplot as plt 
def use_seaborn() -> None:
    plt.style.use("seaborn")
\end{lstlisting}
\end{enumerate}

\section{Prompt Templates}
\label{sec:prompt_template}
This appendix contains all the prompts we had used for our experiments:
\begin{itemize}
    \item The prompts for greedy sampling are given in Figure~\ref{fig:greedy-prompts}.
    \item The prompts for self-debugging are given in Figure~\ref{fig:self-debugging-prompts}.
    \item The prompt for the multi-step agent is given in Figure~\ref{fig:agent-prompt}.
    \item The prompt for RAG is given in Figure~\ref{fig:rag-prompt}.
    \item The prompt and file format for Coding Assistants are given in Figure~\ref{fig:code-assist}.
    \item The prompt for SEK is given in Figure~\ref{fig:sek_prompt} (for keywords generation) and Figure~\ref{fig:sek_prompt_2} (for code generation).
\end{itemize}

\begin{figure*}[htbp]
  \centering
  \caption{Prompts for Greedy Sampling}
  \label{fig:greedy-prompts}
  \begin{subfigure}[t]{0.45\linewidth}
    \caption*{(a) System Prompt for Zero-Shot Prompting}
    \begin{promptcode}
You are a skilled Python programmer tasked with solving a coding problem. Your goal is to provide a clear, efficient, and correct solution that meets all the specified requirements.

Please provide your solution following these guidelines:

1. Use the required library in your solution.
2. Incorporate the provided starter code correctly.
3. Write your solution in Python.
4. Format your solution within a markdown code block.
5. Ensure your code is clean, efficient, and well-commented.
6. Output only the code block and nothing else.

Example output format:

```python
# [Your code here, incorporating the starter code]

# [Additional code and comments as needed]
```

After writing your solution, please review it to ensure all requirements are met and the code is correct and efficient.

Here are the key elements for this task: 
    \end{promptcode}
  \end{subfigure}
  \hfill
  \begin{subfigure}[t]{0.45\linewidth}
    \caption*{(b) System Prompt for Chain-Of-Thought Prompting}
    \begin{promptcode}
You are a skilled Python programmer tasked with solving a coding problem. Your goal is to provide a clear, efficient, and correct solution that meets all the specified requirements.

First, let's think step-by-step. Then, please provide your solution following these guidelines:

1. Use the required library in your solution.
2. Incorporate the provided starter code correctly.
3. Write your solution in Python.
4. Format your solution within a markdown code block.
5. Ensure your code is clean, efficient, and well-commented.
6. Output nothing else after the code block.


Example output format:

[Step-by-step thinking]
```python
# [Your code here, incorporating the starter code]

# [Additional code and comments as needed]
```

After writing your solution, please review it to ensure all requirements are met and the code is correct and efficient.

Here are the key elements for this task: 
    \end{promptcode}
  \end{subfigure}
  \vspace{1em}
  \begin{subfigure}[t]{0.30\linewidth}
    \caption*{(c) User Prompt}
    \begin{promptcode}
1. Required Library:
<library>
{{library}}
</library>

2. Python version:
<python>
{{python_version}}
</python>

2. Coding Problem:
<coding_problem>
{{coding_problem}}
</coding_problem>

3. Starter Code:
<starter_code>
{{starter_code}}
</starter_code>
    \end{promptcode}
  \end{subfigure}
\end{figure*}

\begin{figure*}[ht]
  \centering
  \caption{Prompts for Self-Debugging}
  \label{fig:self-debugging-prompts}
  \begin{subfigure}[t]{0.48\linewidth}
    \caption*{(a) System Prompt}
    \begin{promptcode}
You are an expert programming assistant. Your task is to fix issues in a generated Python solution for a given programming problem. You are provided with:

- A problem statement
- Starter code
- A previously generated incorrect solution
- A top-level execution trace or error message
- Dependencies information (versions, libraries).

Please generate a corrected Python solution by following these strict guidelines:

1. Use the required libraries explicitly in your code.
2. Correctly incorporate the provided starter code - do not remove or alter its structure.
3. Write in standard Python syntax.
4. Wrap your entire solution within a single Markdown code block.
5. Do not include any text outside the code block - no explanations, comments, docstrings, or usage examples.
6. Ensure the code is clean, efficient, and syntactically valid.
7. Avoid interactive, stateful, or environment-dependent constructs (e.g., Django projects, web servers).
8. Your output must be executable in a non-interactive environment (e.g., a test harness or script runner).


Example output format:

```python
# [Your corrected code here]
```

Before submitting, carefully review your code for correctness, completeness, and adherence to all constraints.

    \end{promptcode}
  \end{subfigure}
  \hfill
  \begin{subfigure}[t]{0.48\linewidth}
    \caption*{(b) User Prompt}
    \begin{promptcode}
<Problem>
{problem}
</Problem>

<Python Version>
{python_version}
</Python Version>

<Library>
{library}
</Library>

<Version>
{version}
</Version>

<Extra Dependencies>
{additional_dependencies}
</Extra Dependencies>

<Starting Code>
{starting_code}
</Starting Code>

<Generated Solution>
{solution}
</Generated Solution>

<Trace>
{top_level_trace}
</Trace>
    \end{promptcode}
  \end{subfigure}
\end{figure*}

\begin{figure}[htbp]
  \centering
  \caption{Tool-Calling Agent Prompt}
  \label{fig:agent-prompt}
    \begin{promptcode}
You are to solve a coding problem in Python.

# Instructions:

* The coding problem requires using the library {library}=={version}. Try using the problem with only this library and the standard Python libraries.

* Do a thorough research on the web about how to solve the coding problem for the given library version. Repeat multiple times if needed.

* BEFORE FINISHING YOUR WORK, YOU MUST check your solution to the coding problem by running the `docker_problem_sandbox` tool.

* Use the `final_answer` tool to return a self-contained Python script that solves the problem. DO NOT INCLUDE ANY TEXT BESIDES FOR THE CODE IN THE FINAL ANSWER.

* The solution needs to be in a markdown code block.

* The solution needs to start with the starter code provided below.


# Coding Problem:

{problem}


# Starter Code:

```python
{starting_code}
```
    \end{promptcode}
 \end{figure}

\begin{figure}[htbp]
  \centering
  \caption{RAG Prompt}
  \label{fig:rag-prompt}
    \begin{promptcode}
You are an AI assistant specialized in solving Python programming problems using information derived from documentation.

Each query may specify particular libraries and version constraints. Your task is to generate a correct, efficient, and minimal Python solution that adheres strictly to these requirements.

Please follow these rules when crafting your response:

1. Use only the specified libraries and respect the given version constraints.
2. Incorporate any provided starter code as required.
3. Write only Python code- no in- line comments or usage examples. Do not provide anything in the response but the code.
4. Ensure the code is clean, minimal, and adheres to best practices.
5. The code must be executable in a non-interactive environment (e.g., avoid frameworks like Django or code requiring a web server).Context:
{context}

Based on the above, respond to the user query below.

Query: {query}
\end{promptcode}
\caption*{Here, \texttt{\{context\}} refers to the context of the top-\texttt{k} retrieved documents from the vectorized database for that query and \texttt{\{query\}} is the same as the User Prompt given in Figure~\ref{fig:greedy-prompts}(c).}
\label{app:prompt_rag}
\end{figure}

\begin{figure*}[ht]
  \centering
  \caption{Prompt and File Format for Coding Assistants}
  \label{fig:code-assist}
  \begin{subfigure}[t]{0.48\linewidth}
    \caption*{(a) Prompt}
    \begin{promptcode}
 Solve each sample_{i}.py in this folder then subsequently save your solutions as py files with the same name in a separate subfolder called "{assistant name}" that just completes the starting code provided in the sample and uses the instructions written in the comments at the start of each file.
    \end{promptcode}
  \end{subfigure}
  \hfill
  \begin{subfigure}[t]{0.48\linewidth}
    \caption*{(b) Input File Format}
    \begin{promptcode}
# Complete using the following libraries and/or extra dependencies and their versions:
# problem statement: {problem}
# library: {library}
# version: {version}
# extra_dependencies: {extra_dependencies}
{starting_code}
    \end{promptcode}
  \end{subfigure}
\caption*{(a) presents the prompt template we had used for our Coding Assistant experiments. (b) shows the format of the example files referenced in the prompt.}
\label{app:assistant_query_template}
\end{figure*}

\begin{figure*}[ht]
  \centering
  \caption{Prompts for SEK (Keyword Generation Stage)}
  \label{fig:sek_prompt}
  \begin{subfigure}[t]{0.48\linewidth}
    \caption*{(a) System Prompt}
    \begin{promptcode}
You are a seasoned Python developer at a Fortune 500 company who excels at analyzing complex code. Analyze the given code problem from the problem statement and starter code provided. Try to extract the keywords from the code problem. For each identified keyword:
1. Provide the keyword.
2. Give a formalized explanation of the keyword using technical languages.

Provided Format:
Keywords:[Keywords]
Explainations:[Formalized explanations]

Guidelines:
- Prioritize keywords that are crucial to understanding the input parameters, return content or supplementary information.
- Use precise languages in explanations and provide formalized definitions where appropriate.
- Ensure explanations are consistent with the behaviors expected based on the problem description.
- Limit to the top 1-3 important keywords to focus on core concepts.
- You are supposed to output a structured JSON output containing the extracted keywords and their corresponding formalized explanations in individual lists of strings. The keys for this JSON must be Keywords and Explainations.
- Strictly adhere to the provided format, do not output anything else.

    \end{promptcode}
  \end{subfigure}
  \hfill
  \begin{subfigure}[t]{0.48\linewidth}
    \caption*{(b) User Prompt}
    \begin{promptcode}
<Problem Statement>
{problem}
</Problem Statement>

<Starting Code>
{starting_code}
</Starting Code>
    \end{promptcode}
  \end{subfigure}

\end{figure*}

\begin{figure*}[ht]
  \centering
  \caption{Prompts for SEK (Code Generation Stage)}
  \label{fig:sek_prompt_2}
\begin{subfigure}[t]{0.48\linewidth}
    \caption*{(a) System Prompt}
    \begin{promptcode}
You are a skilled Python programmer tasked with solving a coding problem. Your goal is to provide a clear, efficient, and correct solution that meets all the specified requirements.

Please provide your solution following these guidelines:

1. Use the required library in your solution.
2. Incorporate the provided starter code correctly.
3. Write your solution in Python.
4. Format your solution within a markdown code block.
5. Ensure your code is clean and efficient.
6. Output only the code block and nothing else. Do not add any in-line comments, documentations, references or usage examples.
7. Make sure your code is executable in a non-interactive environment. For example, do not write code which requires building a Django project or deploying a web-app.

Example output format:

```python
# [Your code here, incorporating the starter code]
```

After writing your solution, please review it to ensure all requirements are met and the code is correct and efficient.

Here are the key elements for this task: 

    \end{promptcode}
  \end{subfigure}
  \hfill
  \begin{subfigure}[t]{0.48\linewidth}
    \caption*{(b) User Prompt}
    \begin{promptcode}
<Python Version>
{python_version}
</Python Version>

<Library>
{library}
</Library>

<Version>
{version}
</Version>

<Extra Dependencies>
{extra_dependencies}
</Extra Dependencies>

<Problem Statement>
{problem}
</Problem Statement>

<Keywords>
Analyze the following key terms and their relationships within the problem context:
{General_Keywords}
{Abstract_Keywords}
</Keywords>

<Starting Code>
{starting_code}
</Starting Code>
    \end{promptcode}
  \end{subfigure}
\end{figure*}
\section{Artifacts and Model Details}
\label{app:artifacts}
This appendix provides citations for various artifacts and models mentioned in the paper.

\subsection{Libraries}
This is the full list of libraries included in \GitChameleon{}.
\begin{itemize}
    \item \texttt{PyTorch} \citep{paszke2019pytorch}
    \item \texttt{Geopandas} \citep{kelsey_jordahl_2020_3946761}
    \item \texttt{NLTK} \citep{nltk}
    \item \texttt{NetworkX} \citep{hagberg2008exploring}
    \item \texttt{GeoPy}\footnote{\url{https://pypi.org/project/geopy/}}
    \item \texttt{Gradio} \citep{abid2019gradiohasslefreesharingtesting}
    \item \texttt{Scikit-Learn} \citep{sklearn_api}
    \item \texttt{Matplotlib} \citep{Hunter:2007}
    \item \texttt{PyCaret}\footnote{\url{https://pycaret.org/}}
    \item \texttt{Pandas} \citep{reback2020pandas,mckinney-proc-scipy-2010}
    \item \texttt{NumPy} \citep{harris2020array}
  \item \texttt{LightGBM}\footnote{\url{https://lightgbm.readthedocs.io/}}
    \item \texttt{spaCy} \footnote{\url{https://spacy.io/}}
    \item \texttt{Django}\footnote{\url{https://www.djangoproject.com/}}
    \item \texttt{SciPy} \citep{virtanen2020scipy}
    \item \texttt{Flask}\footnote{\url{https://flask.palletsprojects.com/}}
    \item \texttt{Jinja2}\footnote{\url{https://jinja.palletsprojects.com/}}
    \item \texttt{SymPy}\footnote{\url{https://www.sympy.org/en/index.html}}
    \item \texttt{Seaborn}\footnote{\url{https://seaborn.pydata.org/}}
    \item \texttt{mitmproxy}\footnote{\url{https://mitmproxy.org/}}
    \footnote{\url{https://mitmproxy.org/}}
    \item \texttt{pytest} \footnote{\url{https://pytest.org/}}
    \item \texttt{Falcon} web framework\footnote{\url{https://falconframework.org/}}
    \item \texttt{Tornado} web server\footnote{\url{https://www.tornadoweb.org/}}
    \item \texttt{Plotly}\footnote{\url{https://plotly.com/python/}}
    \item \texttt{Librosa}\footnote{\url{https://librosa.org/doc/latest/index.html}}
    \item \texttt{Pillow} \footnote{\url{https://python-pillow.org/}}
    \item \texttt{tqdm} \footnote{\url{https://github.com/tqdm/tqdm}}
    \item \texttt{Kymatio}\footnote{\url{https://librosa.org/doc/latest/index.html}}
\end{itemize}

\subsection{Models}
\subsubsection*{Open-Weights Models}
The following open-weights models were evaluated:
\begin{itemize}
    \item \texttt{Llama 3.1 Instruct Turbo}: \cite{llama3_1_tech_report}
    \item \texttt{Llama 3.3 Instruct Turbo 70B}: \cite{llama_llamacon_2025}
    \item \texttt{Llama 4 Maverick 400B}: \cite{llama_llamacon_2025}
    \item \texttt{Qwen 2.5-VL Instruct 72B}: \cite{qwen2025qwen25technicalreport}
    \item \texttt{Qwen 3 235B}:\cite{yang2025qwen3technicalreport}
    \item \texttt{Command A 111B}: \cite{cohere2025commandaenterprisereadylarge}
    \item \texttt{DeepSeek R1 685B}: \cite{deepseekai2025deepseekr1incentivizingreasoningcapability}
    \item \texttt{DeepSeek v3}: \cite{deepseekai2025deepseekv3technicalreport}
    \item \texttt{Openhands LM 32B v0.1}: \cite{wang2025-introducing-openhands-lm}
    \item \texttt{Reka Flash-3}: \cite{huggingfaceRekaAIrekaflash3Hugging}
    \item \texttt{Jamba 1.6 Mini, Large}: \cite{lieber2024jambahybridtransformermambalanguage}
\end{itemize}

\subsubsection*{Enterprise Models}
The following enterprise models were evaluated:
\begin{itemize}
    \item \texttt{Arcee CoderL}: \cite{arceeModelSelection}
    \item \texttt{Claude 3.5 Haiku}\footnote{https://www.anthropic.com/claude/haiku}
    \item \texttt{Claude 3.5 Sonnet}\footnote{https://www.anthropic.com/news/claude-3-5-sonnet}
    \item \texttt{Claude 3.7 Sonnet}: \cite{claude3_7_sonnet_anthropic_2025}
    \item \texttt{Claude 4 Sonnet}\footnote{https://www.anthropic.com/claude/sonnet}
    \item \texttt{CommandR+}\footnote{\url{https://cohere.com/blog/command-r-plus-microsoft-azure}}
    \item \texttt{Gemini 1.5 Pro}: \cite{geminiteam2024gemini15unlockingmultimodal}
    \item \texttt{Gemini 2.0 Flash}: \cite{gemini2_0_flash_image_2025}
    \item \texttt{Gemini 2.5 Pro}: \cite{gemini2_5_vertex_ai_2025}
    \item \texttt{Gemini 2.5 Flash}: \cite{gemini2_5_vertex_ai_2025}
    \item \texttt{GPT-4.1}: \citep{gpt4_1_openai_2025}
    \item \texttt{GPT-4.1-mini}: \citep{gpt4_1_openai_2025}
    \item \texttt{GPT-4.1-nano}: \citep{gpt4_1_openai_2025}
    \item \texttt{GPT-4o}: \cite{gpt4o_system_card_2024}
    \item \texttt{GPT-4o-mini}: \cite{gpt4o_system_card_2024}
    \item \texttt{GPT-4.5}: \cite{gpt4_5_openai_2025}
    \item \texttt{o1}: \citep{openai_o1_system_card}
    \item \texttt{o3-mini}: \cite{openai_o1_system_card}
    \item \texttt{codex-mini}\footnote{\url{https://platform.openai.com/docs/models/codex-mini-latest}}
    \item \texttt{Grok 3}: \cite{xAIgrok3}
    \item \texttt{Mistral Medium 3}: \cite{mistralMedium3}
    \item \texttt{Devstral Small}\footnote{\url{https://mistral.ai/news/devstral}}
    \item \texttt{Inflection 3 Productivity}\footnote{\url{https://openrouter.ai/inflection/inflection-3-productivity}}
    \item \texttt{Liquid LFM 40B MoE}\footnote{\url{https://www.liquid.ai/blog/liquid-foundation-models-our-first-series-of-generative-ai-models}}
    \item \texttt{Nova Pro}:\cite{Intelligence2024}
\end{itemize}

\subsection{Coding Assistants (CLI/IDE)}
The following coding assistants were studied as part of the experimentation pipeline:
\begin{itemize}
    \item \texttt{Claude Code}\footnote{\url{https://docs.anthropic.com/en/docs/claude-code/overview}} (CLI)
    \item \texttt{Goose}\footnote{\url{https://block.github.io/goose/}} (CLI)
    \item \texttt{Cline}\footnote{\url{https://cline.bot/}} (IDE-VSCode)
    \item \texttt{RooCode}\footnote{\url{https://roocode.com/}} (IDE-VSCode)
    \item \texttt{KiloCode}\footnote{\url{https://kilocode.ai/}} (IDE-VSCode)
\end{itemize}
\end{document}